\documentclass[ejs]{imsart}

\RequirePackage[OT1]{fontenc}
\RequirePackage{amsthm,amsmath,amssymb}
\RequirePackage[numbers]{natbib}
\RequirePackage[colorlinks,citecolor=blue,urlcolor=blue]{hyperref}
\RequirePackage{epsfig}
\RequirePackage{graphicx}
\RequirePackage[mathscr]{eucal}
\usepackage{bbm}
\usepackage{multirow}

\doi{10.1214/154957804100000000}
\pubyear{0000}
\volume{0}
\firstpage{0}
\lastpage{0}

\startlocaldefs

\def\argmin{\mathop{\rm argmin}}

\newcommand{\real}{\ensuremath{\mathbb{R}}}

\newcommand{\ltwo}{\ensuremath{\mathbb{L}^2}}

\newtheorem{defn}{Definition}

\newtheorem{algorithm}{Algorithm}

\endlocaldefs

\begin{document}

\begin{frontmatter}
\title{A Geometric Approach to Pairwise Bayesian Alignment of Functional Data Using Importance Sampling}
\runtitle{Geometric Bayesian Alignment of Functional Data}

\begin{aug}
\author{\fnms{Sebastian} \snm{Kurtek}\corref{}\ead[label=e1]{kurtek.1@stat.osu.edu}}

\address{Department of Statistics\\
The Ohio State University\\
Columbus, OH\\
\printead{e1}}

\runauthor{S. Kurtek}

\affiliation{The Ohio State University}

\end{aug}

\begin{abstract}
We present a Bayesian model for pairwise nonlinear registration of functional data. We use the Riemannian geometry of the space of warping functions to define appropriate prior distributions and sample from the posterior using importance sampling. A simple square-root transformation is used to simplify the geometry of the space of warping functions, which allows for computation of sample statistics, such as the mean and median, and a fast implementation of a $k$-means clustering algorithm. These tools allow for efficient posterior inference, where multiple modes of the posterior distribution corresponding to multiple plausible alignments of the given functions are found. We also show pointwise $95\%$ credible intervals to assess the uncertainty of the alignment in different clusters. We validate this model using simulations and present multiple examples on real data from different application domains including biometrics and medicine.
\end{abstract}


\tableofcontents
\end{frontmatter}

\section{Introduction}

The problem of registration of functional data is important in many branches of science. In simple terms,
it deals with deciding how points on one function match in some optimal way with points on
another function. In contrast to landmark-based matching, such an approach matches the entire domains of the functions in a general registration problem. The study of registration problems is popular in image analysis where pixels or voxels across images are matched, and in shape analysis of objects where points across shapes are matched.
One can broadly classify registration problems into two main groups: (1) pairwise registration and (2) groupwise
registration. In pairwise registration, one solves for an optimal matching between two objects, while in groupwise
registration multiple ($>2$) objects are simultaneously registered. In this paper, we focus on the
problem of pairwise registration. This problem has been referred to in many different ways, some of
which are alignment, warping, deformation matching, amplitude-phase separation, and so on.
While registration can be studied for many types of objects, from simple functions to complex high-dimensional
structures, the fundamental issues in registration are often similar. We will focus on perhaps the
simplest objects for studying registration problems, $\real$-valued functions on $[0,1]$. More specifically, we will take
a Bayesian approach to this problem, motivated by geometrical considerations; the method will be characterized by the definition of a geometric prior on a suitable function space, representing the parameter space of interest. We also compare the proposed method to past ideas that often take an optimization-based approach.

\begin{figure}[!t]
\begin{center}
\begin{tabular}{|c|c|c|c|}
\hline
(a)&(b)&(c)&(d)\\
\hline
\includegraphics[width=1.1in]{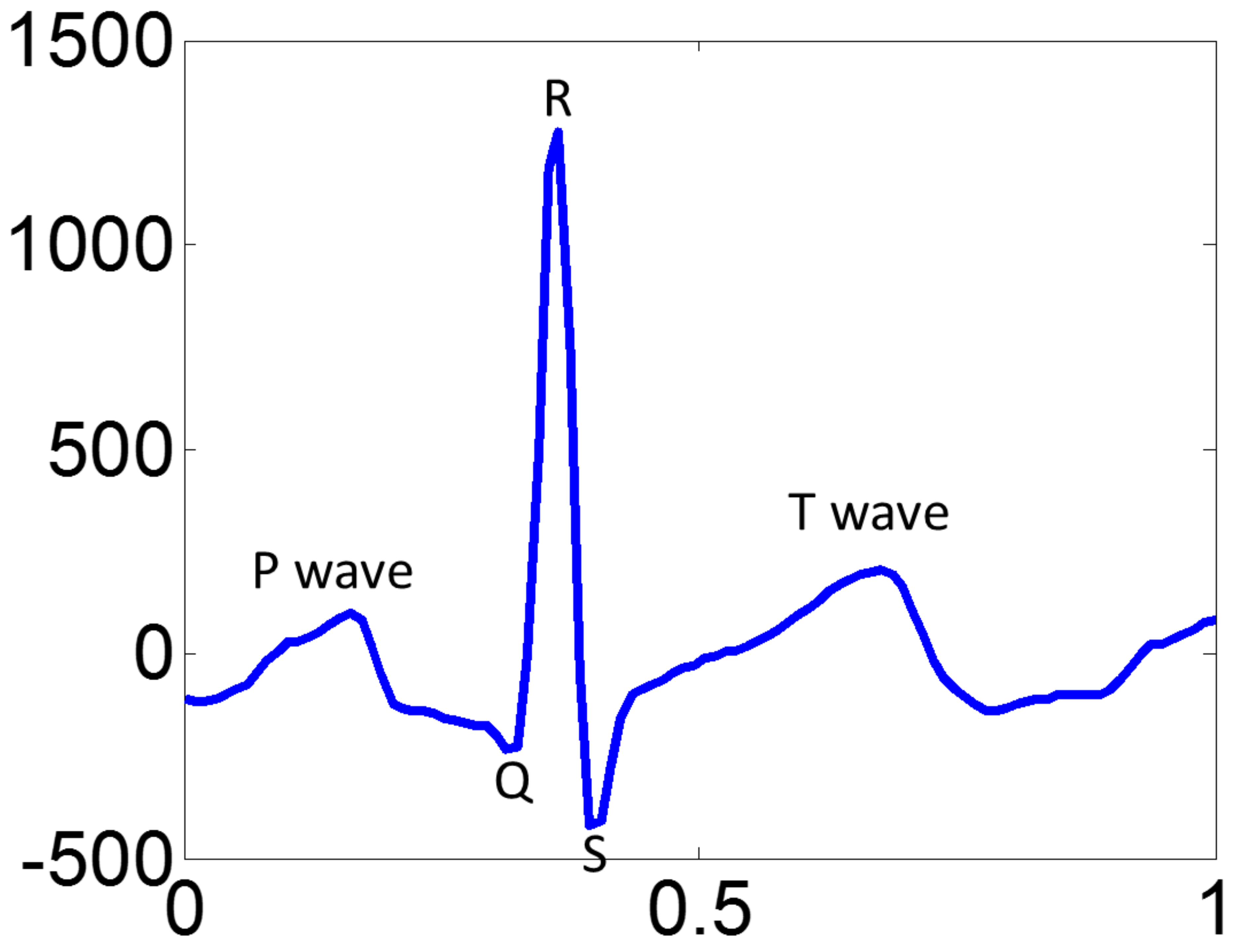}&\includegraphics[width=1.1in]{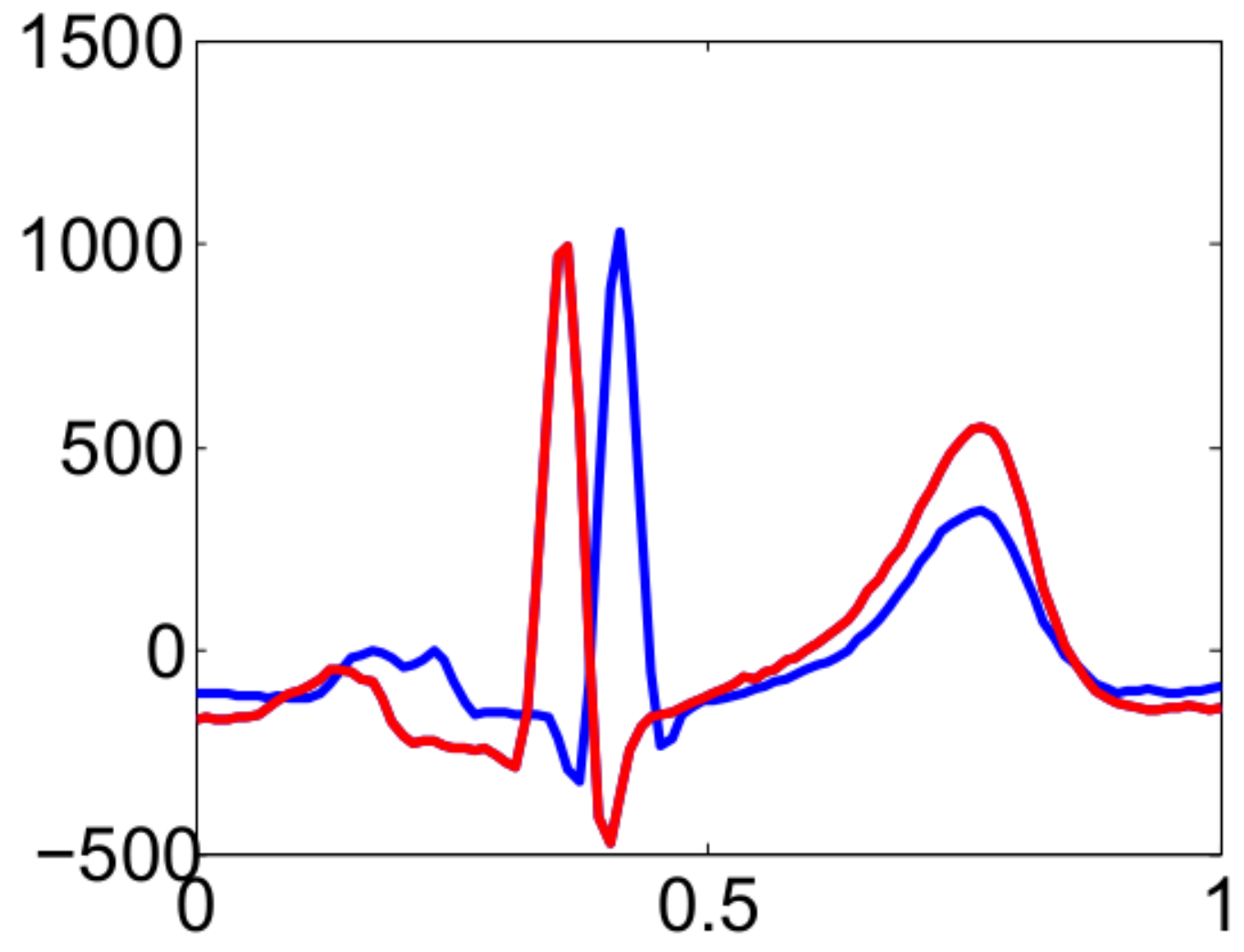}&\includegraphics[width=.7in]{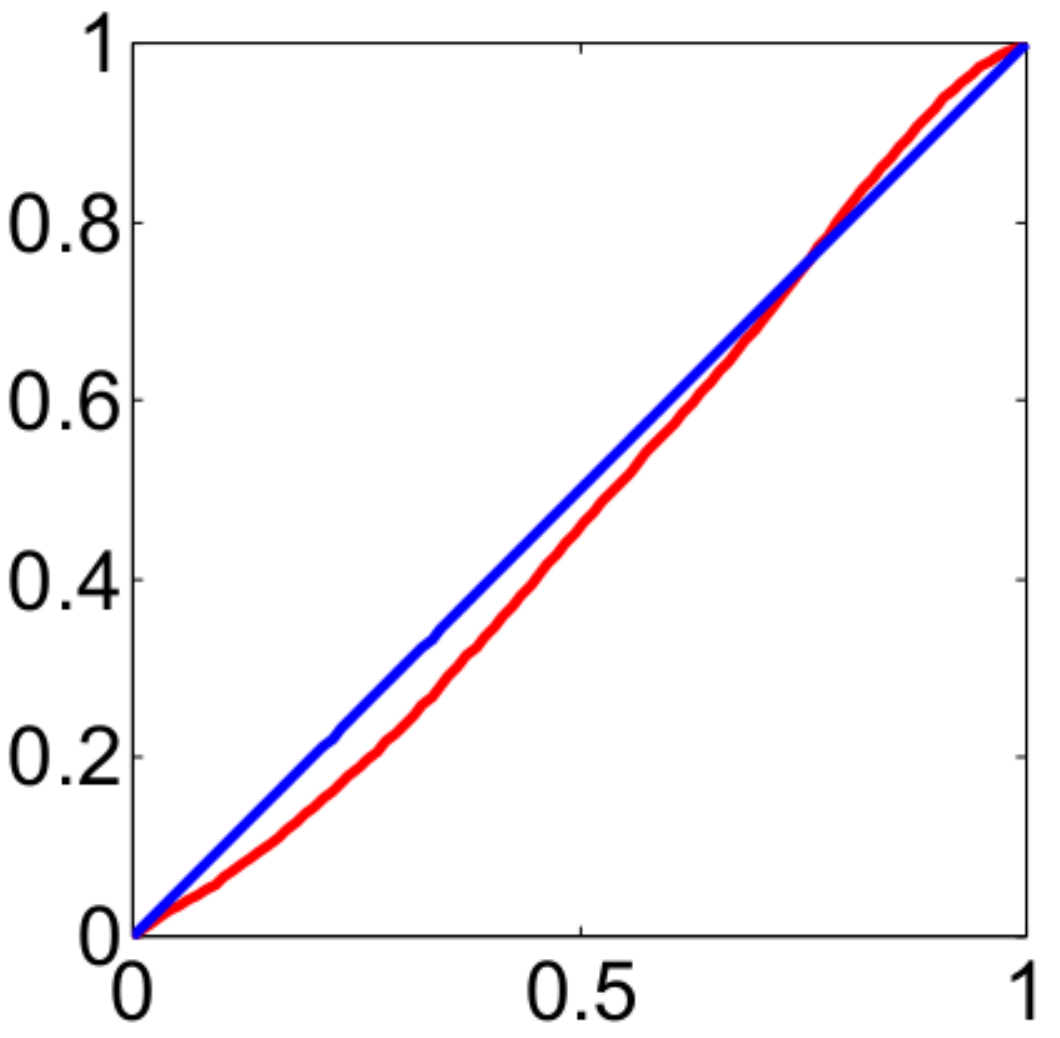}&\includegraphics[width=1.1in]{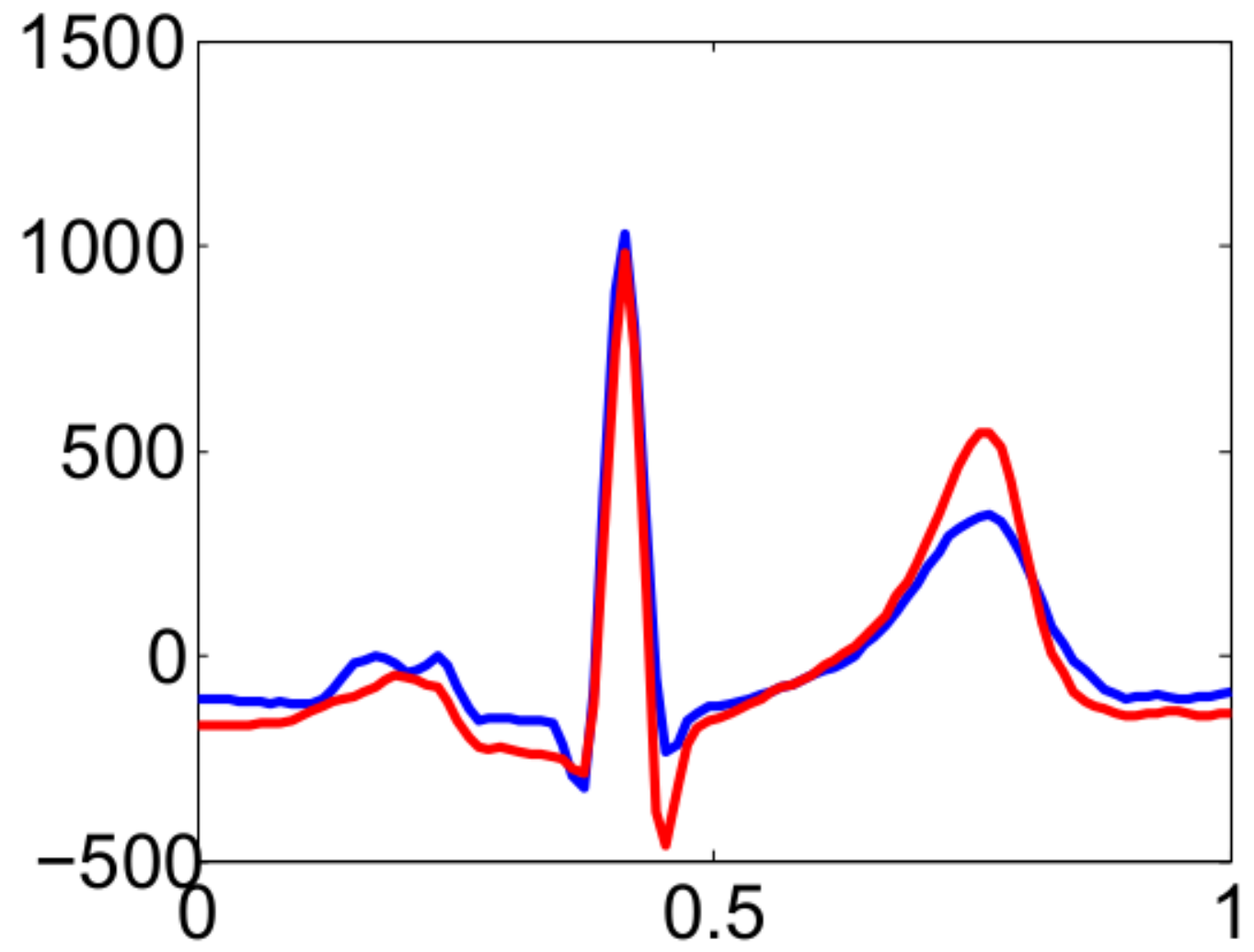}\\
\hline
\end{tabular}
\caption{(a) A PQRST complex with labeled salient features: the P wave, QRS complex and T wave. (b) Two unaligned PQRST complexes. (c) The estimated warping function in red with identity warping in blue as a reference. (d) The same PQRST complexes as in panel (b) but aligned using the estimated warping.}\label{fig:motex}
\end{center}
\end{figure}

To motivate the function alignment problem, consider the example shown in Figure \ref{fig:motex}. In panel (a), we display an example of a PQRST complex with labeled structures (P wave, QRS complex, T wave). This function represents a full heartbeat cycle and can be extracted from long electrocardiogram (ECG) signals for the purposes of diagnosing heart diseases such as myocardial infarction. The difficulty with using such objects for diagnosis is highlighted in panel (b). As given, the P wave and QRS complex on the red function occur earlier than on the blue one. This is usually due to natural variability in nonlinear heartbeat dynamics. In general, given two PQRST complexes, their important salient features are often not in correspondence. This presents a major challenge when modeling these functions. Even simple statistics such as the cross-sectional mean can be meaningless (see Figure \ref{fig:ex1ECG} and row 3 in Figure \ref{fig:exma1}). Aligning the functions prior to subsequent statistical analyses is thus required. The purpose of pairwise alignment is to estimate a warping function, and additionally the uncertainty in this estimate, that aligns the prominent features across two functions. In panel (c), we display the estimated warping function in red, and in panel (d) we show the resulting alignment of the two PQRST complexes. Now, the P wave, QRS complex and T wave occur at the same time across both functions.

There exists a large literature on statistical analysis of
functions, in part due to the pioneering efforts of Ramsay and Kneip \cite{ramsay-silverman-2005,kneip-gasser-annals:92}, and several others
\citep{muller-JASA:2004,muller-biometrika:2008}. When restricting
to the analysis of elastic functions (functions that are temporally aligned) the literature is relatively
recent \citep{ramsay-li-RSSB:98,gervini-gasser-RSSB:04,muller-JASA:2004,james:07,muller-biometrika:2008,kneip-ramsay:2008,Raket20141}.
The general approach in most of these methods is to use an energy function to compute optimal registrations and perform subsequent analysis on the aligned functions using standard tools from
functional data analysis such as the cross-sectional mean, covariance and functional Principal Component Analysis (fPCA). The importance of registration in functional data is undeniable as evidenced in a recent Special Section of the Electronic Journal of Statistics titled \textit{Statistics of Time Warpings and Phase Variations} \cite{marron2014}; this section contained a set of applied papers that analyzed four different datasets, including mass spectrometry functions \cite{koch2014}, neural spike trains \cite{wu2014}, juggling trajectories \cite{ramsay2014} and internal carotid arteries \cite{sangalli2014}.

Recently, it has been argued that a Bayesian approach rather than pure optimization is a better option for many situations. The advantages of a model-based Bayesian approach include:
\begin{enumerate}
\item A comprehensive exploration of the warping variable space resulting in potential multimodal solutions to the registration problem;
\item Assessment of uncertainty, via credible intervals, associated with the computed estimates.
\end{enumerate}
The literature on registration methods that are based on Bayesian principles is fairly limited. Telesca and Inoue \cite{telesca} proposed a semi-parametric model for groupwise alignment of functional data. These models were further extended in the context of analyzing microarray data in \cite{telesca1}. A nonparameteric approach to the groupwise registration problem was also proposed recently in \cite{hayashi}. A different Bayesian model was proposed for registering liquid chromatography-mass spectrometry data in \cite{tsai}. The main difficulty in specifying Bayesian registration models lies in defining an appropriate prior on the space of warping functions, or some relevant subset, to enable efficient inference. In \cite{10.1109/TPAMI.2008.223}, Srivastava and Jermyn defined a Gaussian-type prior distribution on the space of warping functions, via the geodesic distance, in the context of detecting shapes in two-dimensional point clouds. The recent model of Cheng et al. \cite{Dryden:13} used the square-root slope function (SRSF) representation of functional data and utilized the fact that
the derivative of a warping function is a probability density function. In this way, they constructed a Dirichlet process to impose a prior model implicitly on the space of warping functions, and sampled from the posterior distribution using Markov chain Monte Carlo (MCMC) techniques. The SRSF representation of functional data has many desirable properties related to the registration problem, which we emphasize in Section \ref{probback}.

In the current paper, we describe a convenient geometric structure, a unit sphere, using the square-root density (SRD) representation of warping functions and use
its geometry to impose the prior. In this setup, we develop a Bayesian registration model and utilize importance sampling from the posterior to compute posterior functionals such as the mean, median or maximum a posteriori (MAP) estimate. We also provide pointwise standard deviations and credible intervals to assess alignment uncertainty. We show that these tools are especially effective when two or more registrations are plausible. Thus, the main contributions of this paper are the following:
\begin{enumerate}
\item We use the spherical geometry of the space of warping functions to define a class of truncated wrapped normal prior distributions for the purpose of Bayesian alignment of functional data; 
\item We define a sampling importance re-sampling approach to sample from the marginal posterior distribution of warping functions;
\item We use the Riemannian geometry of the space of warping functions to define an efficient $k$-means clustering algorithm, which can be used to identify multiple modes in the posterior representing different plausible alignments of the observed functions.
\end{enumerate}

The rest of this paper is organized as follows. In Section \ref{probback}, we give a detailed description of the registration problem and describe tools for statistical analysis on the space of warping functions. In Section \ref{sec:hrm}, we introduce our registration model and in Section \ref{sec:impsamp} we describe an importance sampling approach for sampling from the posterior distribution of warping functions. Finally, in Sections \ref{sec:sim} and \ref{sec:expres}, we present simulation studies and different applications of the proposed framework. We emphasize examples where the posterior distribution is multimodal. Finally, we close with a brief summary and directions for future work in Section \ref{sec:conc}.

\section{Problem Background}
\label{probback}

Before we describe our Bayesian framework, we first setup the registration problem mathematically.
Let ${\cal F}$ be an appropriate subset (made precise later) of real-valued functions on the interval $[0,1]$.
For any two functions $f_1, f_2 \in {\cal F}$, the registration problem is defined as finding the mapping $\gamma$
such that point $t \in [0,1]$ on the domain of $f_1$ is matched to the point $\gamma(t) \in [0,1]$ on the domain of $f_2$. In other
words, the functions $f_1(t)$ and $f_2(\gamma(t))$ are optimally matched under the chosen optimality
criterion. The main question that arises is: What should be the criterion for optimal registration? A natural tendency is to choose
an $\mathbb{L}^p$-norm between $f_1$ and $f_2 \circ \gamma$, but there are some known limitations
of that approach. For instance, if we choose the $\ltwo$ norm, defined as $\|f_1-f_2\|=\sqrt{\int_0^1 |f_1(t)-f_2(t)|^2dt}$ ($|\cdot|$ is the standard Euclidean norm), we
obtain the following optimization problem:
\begin{equation}
\gamma^* = \arg \inf_{\gamma} \| f_1 - f_2 \circ \gamma\|.\label{eq:fdist}
\end{equation}
This setup can lead to a degenerate solution, termed the {\it pinching effect} demonstrated in \cite{marron2015}. In this case, one can pinch the entire function $f_2$ to get arbitrarily close to $f_1$ in $\ltwo$ norm. To
avoid this situation, one often adds a roughness penalty on $\gamma$,  denoted by ${\cal R}(\gamma)$, leading to the optimization
problem given by $\gamma^*=\arg \inf_{\gamma}\left( \| f_1 - f_2 \circ \gamma\|^2 + \lambda {\cal R}(\gamma) \right)$. Although this avoids
the pinching effect, it introduces some other issues. First, the choice of $\lambda$ is not obvious in general cases.
Second and more important is the fact that this solution is not symmetric. That is, the optimal registration of $f_1$ to $f_2$ can be
quite different from that of $f_2$ to $f_1$. Another related issue is that this criterion is not a proper
metric and this leads to additional problems in later analysis. Most papers on registration of functional data involve this setup and inherit the above-mentioned limitations.

To avoid these issues, Srivastava et al. \cite{EFA:10,NIPS} proposed an approach that has its foundations in differential
geometry. First, let the set of all registration or warping functions be defined as $\Gamma=\{\gamma:[0,1]\to [0,1]\ |\ \gamma(0)=0,\ \gamma(1)=1,\ 0<\dot{\gamma}<\infty\}$. $\Gamma$ forms a Lie group under composition, i.e., for any $\gamma_1, \gamma_2 \in \Gamma$ their composition $\gamma_1 \circ \gamma_2$ is also in $\Gamma$, and for any $\gamma \in \Gamma$ there is a unique $\gamma^{-1} \in \Gamma$. The $\gamma_{id}(t) = t$
is the identity element of this group. The next item is to represent the given functions by their square-root slope
functions (SRSFs): $q(t) = \mbox{sign}(\dot{f}(t)) \sqrt{|\dot{f}(t)|}$. Note that the SRSF is the one-dimensional version of the square-root velocity function used for shape analysis of higher-dimensional curves \cite{srivastava_etal_PAMI:10,KurtekJASA}.

For registration under this approach, each $f \in {\cal F}$ is represented
by its SRSF $q$. One sets ${\cal F}$ to be the space of all absolutely continuous functions and
the resulting space of all SRSFs is $\ltwo([0,1],\real)$ henceforth referred to simply as $\ltwo$.
For every $q\in \ltwo$ there exists a function $f$ (unique up to a constant) such that the given $q$ is the SRSF of that $f$. In fact, this function can be obtained precisely using $f(t) = f(0) + \int_0^t q(s)|q(s)|ds$. Note that if a function $f$ is warped by
$\gamma$ to $f \circ \gamma$, its SRSF changes from $q$ to $(q, \gamma) = (q \circ \gamma) \sqrt{\dot{\gamma}}$; this last term
involving $\sqrt{\dot{\gamma}}$ is an important departure from previous solutions. To setup the registration problem, we define an equivalence class of an SRSF as $[q]=\{(q,\gamma)|\gamma\in\Gamma\}$.
Finally, the pairwise registration between any two functions $f_1$ and $f_2$ is performed by solving an optimization problem over equivalence classes of their SRSF representations:
\begin{equation}
\gamma_{DP}=\arg \inf_{\gamma \in \Gamma} \| q_1 - (q_2, \gamma) \|.\label{eq:SRSFregist}
\end{equation}
The solution to this problem is computed using the dynamic programming (DP) algorithm. The resulting distance between the aligned $f_1$ and $f_2$ is given by $d([q_1],[q_2])=\| q_1 - (q_2, \gamma_{DP}) \|$.

As described in \cite{EFA:10}, this framework has many advances: it avoids the pinching problem, its
registration solution is symmetric, it does not require an additional regularization term and the choice of $\lambda$ that goes with it, and it is actually a proper metric on the
quotient space ${\cal F}/\Gamma$, which provides important tools for ensuing analysis. The most important reason why
this setup avoids many problems of Equation \ref{eq:fdist} is that $\| q_1 - q_2 \| = \| (q_1, \gamma) - (q_2, \gamma)\|$ for any $\gamma
\in \Gamma$. In mathematical terms, it means that the action of $\Gamma$ on $\ltwo$ is by isometries. The original method was later extended to apply to statistical analysis of cyclostationary biosignals \citep{biosignals}, and was shown to perform well in different applications \citep{tucker1,tucker2,Wu1,marron2014}.

While the framework of Srivastava et al. \cite{EFA:10} is precise in mathematically defining the function registration problem, it solves for optimal warping functions via energy optimization. In this paper, we argue that a model-based Bayesian approach has many additional advantages. Thus, to preserve the nice properties, such as the isometric action of $\Gamma$ under the $\ltwo$ metric, we build our Bayesian model using the SRSF representation of functional data.

\subsection{Representation Space of Warping Functions}
\label{sec:warpfun}

The proposed Bayesian model defines prior distributions and importance functions on the space of warping functions $\Gamma$. Thus, we are faced with defining statistics and probability distributions on this space. In order to do this we use the Fisher-Rao Riemannian metric on $\Gamma$, which is given by (for $w_1,\ w_2\in T_{\gamma}(\Gamma)$ and $\gamma\in\Gamma$) \citep{srivastava-etal-Fisher-Rao-CVPR:2007,srivastava_etal_PAMI:10,KurtekJASA}:
\begin{equation}
\langle\langle w_1,w_2 \rangle\rangle_\gamma = \int_0^1 \dot{w}_1(t)\dot{w}_2(t)\frac{1}{\dot{\gamma}(t)}dt,
\end{equation}where $\dot{w}$ and $\dot{\gamma}$ represent derivatives. An important property of the Fisher-Rao metric is that it is invariant to re-parameterizations of probability density functions \citep{Cencov82}. While this is not the only metric that achieves this property, it is important to note that there is no invariant metric that does not include derivatives. It is possible to define statistics and probability models directly on $\Gamma$ under the Fisher-Rao metric, but this proves to be very complicated due to the non-trivial Riemannian geometry of this space. We use the Fisher-Rao Riemannian geometry in our Bayesian setup because the desirable properties of this metric (i.e., parameterization invariance) will naturally translate to the prior distributions on $\Gamma$.

Inference on $\Gamma$ is greatly simplified using a convenient transformation, which is similar to the definition of the SRSF for general functions \cite{bhattacharya-43}.
\begin{defn}
Define the mapping $\phi:\Gamma\to\Psi$. Then, given an element $\gamma\in\Gamma$, define a new representation $\psi:[0,1]\to \real_{>0}$ using the square-root of its derivative as $\phi(\gamma)=\psi=\sqrt{\dot{\gamma}}$.
\end{defn}
\noindent This is the same as the SRSF defined earlier for functions and takes this form because $\dot{\gamma} > 0\ \forall\ t$. For simplicity and to distinguish it from the SRSF representation of observed functions, we refer to this representation as the square-root density (SRD). The identity map $\gamma_{id}(t)=t$ maps to a constant function with value $\psi_{id}(t) = 1$. An important advantage of this transformation is that the $\ltwo$ norm of a function $\psi$ is 1. Thus, the set of all such $\psi$s, denoted by $\Psi$, is a subset of the unit sphere in $\ltwo$. Furthermore, as shown in \citep{bhattacharya-43,srivastava-etal-Fisher-Rao-CVPR:2007,srivastava_etal_PAMI:10,KurtekJASA}, the Fisher-Rao metric on the space of warping functions simplifies to the $\ltwo$ metric on $\Psi$, which in turn greatly simplifies all computation. Given a function $\psi$ one can easily compute the corresponding warping function via integration using $\gamma(t)=\int_0^t\psi(s)^2 ds$; this provides the inverse mapping $\phi^{-1}:\Psi\to\Gamma$. Thus, the geodesic path between two warping functions, $\gamma_1,\ \gamma_2\in\Gamma$ represented using their SRDs $\psi_1,\ \psi_2\in\Psi$, is simply the great circle connecting them ($\alpha:[0,1]\to\Psi$), $\alpha(\tau)=\frac{1}{\sin(\theta)}[\sin(\theta-\theta \tau)\psi_1+\sin(\theta \tau)\psi_2]$, where $\theta$ represents the length of this path (geodesic distance between warping functions $\gamma_1$ and $\gamma_2$ under the Fisher-Rao metric) and is simply the arc-length between $\psi_1$ and $\psi_2$:
\begin{equation}
d_{FR}(\gamma_1,\gamma_2)=d(\psi_1,\psi_2)=\theta=\cos^{-1}(\langle \psi_1,\psi_2 \rangle),\label{eqn:dpsi}
\end{equation} where $\langle \cdot,\cdot\rangle$ is the standard $\ltwo$ inner product.

Since the differential geometry of the sphere is well known, this transformation also simplifies the problem of defining probability distributions of warping functions. The general approach will be to define wrapped probability distributions, and perform random sampling and probability calculations on tangent spaces of $\Psi$; the tangent space for all $\psi\in\Psi$ is defined as $T_{\psi}(\Psi)=\{v:[0,1]\to\real |\langle v,\psi\rangle=0\}$. In order to achieve this goal, we must first define some standard tools from differential geometry for this space:
\begin{enumerate}
\item \textbf{Exponential map:} For $\psi\in\Psi$ and $v\in T_{\psi}(\Psi)$, the exponential map is defined as $\exp: T_{\psi}(\Psi)\to\Psi$ by $\exp_{\psi}(v)=\cos(\|v\|)\psi+\frac{\sin(\|v\|)}{\|v\|}v$.
\item \textbf{Inverse exponential map:} For $\psi_1,\ \psi_2\in\Psi$, the inverse exponential map is defined as $\exp^{-1}:\Psi\to T_{\psi}(\Psi)$ by $\exp^{-1}_{\psi_1}(\psi_2)=\frac{\theta}{\sin(\theta)}(\psi_2-\cos(\theta)\psi_1)$.
\item \textbf{Parallel transport:} For $\psi_1,\ \psi_2\in\Psi$, the shortest geodesic path $\alpha:[0,1]\to\Psi$ such that $\alpha(0)=\psi_1$ and $\alpha(1)=\psi_2$, and a vector $v\in T_{\psi_1}(\Psi)$, its parallel transport along $\alpha$ to $\psi_2$ is defined as $\kappa: T_{\psi_1}(\Psi)\to T_{\psi_2}(\Psi)$ by $\kappa(v)=v-\frac{2\langle v,\psi_2 \rangle}{\|\psi_1+\psi_2\|}(\psi_1+\psi_2)$.
\end{enumerate}
The exponential and inverse exponential maps provide a natural way of mapping points from the representation space $\Psi$ to the tangent space (at a particular element of $\Psi$) and vice versa. Parallel transport long geodesic paths allows translation of tangent vectors from one tangent space to another. An important property of parallel transport is that the mapping $\kappa$ is an isometry between the two tangent spaces, i.e., for $v_1,\ v_2\in T_{\psi_1}(\Psi)$, $\langle v_1,v_2 \rangle=\langle \kappa(v_1),\kappa(v_2) \rangle$. This tool from differential geometry is useful in defining probability models on the space of warping functions. In particular, we define an orthonormal basis in the tangent space at any point on $\Psi$ by transporting a standard basis defined on the tangent space at the identity element, $T_{1}(\Psi)$.

\textbf{Summary statistics on $\Psi$:} In addition to defining prior distributions on the space of warping functions, we would like to be able to compute summary statistics such as the mean or median. These tools are especially useful in inference based on samples generated from the posterior distribution. Suppose that we have a sample of warping functions $\gamma_1,\dots,\gamma_p$. To begin, we are interested in defining a mean and median of these functions. To do this we again exploit the geometry of $\Psi$. We begin by mapping all of the functions $\gamma$ to their corresponding SRD representations resulting in $\psi_1,\dots,\psi_p$. Once this is done, all of our data is on the subset of the unit sphere, where the geodesic distance is used to compute their intrinsic mean and median as follows. The sample Karcher mean is given by $\bar{\psi} = \argmin_{\psi \in \Psi} \sum_{i=1}^p d(\psi,\psi_i)^2$ while the sample geometric median is defined as $\tilde{\psi} = \argmin_{\psi \in \Psi} \sum_{i=1}^p d(\psi,\psi_i)$. Gradient-based approaches for finding the Karcher mean and geometric median are given in several places \citep{le-mean-shape,dryden-mardia_book:98,FletcherMedian,KurtekCVIU} and are omitted here for brevity.

\textbf{K-means clustering on $\Psi$:} One of the motivations behind this work is the discovery and analysis of multiple modes in the posterior distribution of warping functions. For this purpose, we introduce a $k$-means clustering approach on $\Psi$. In the previous section, we defined a procedure to compute the Karcher mean of warping functions and we will use it to specify the $k$-means clustering algorithm. Let $\gamma_1,\dots,\gamma_p$ be a sample from the posterior distribution and $\psi_1,\dots,\psi_p$ be their corresponding SRDs. The $k$-means clustering approach computes a partition of the sample space such that the within cluster sum of squared distances is minimized. This is achieved using the following standard algorithm \citep{kmeans}:
\begin{algorithm}
\textbf{($k$-Means):} Initialize using $k$ unique functions $\bar{\psi}_{1,0},\dots,\bar{\psi}_{k,0}$ as cluster centers and set $j=0$. \\
(1) For each $i=1,\dots,p$ and $m=1,\dots,k$, compute $d_{i,m} = d(\bar{\psi}_{m,j},\psi_i)$ using Equation \ref{eqn:dpsi}.\\
(2) Assign each function $\psi_i,\ i=1,\dots,p$, to the cluster which minimizes $d_{i,\cdot}$ .\\
(3) Update cluster means $\bar{\psi}_{1,j+1},\dots,\bar{\psi}_{k,j+1}$ using the Karcher mean.\\
(4) Set $j=j+1$.\\
(5) Repeat Steps 1-4 until cluster assignments remain unchanged.
\end{algorithm}
\noindent A major benefit of this algorithm is its flexibility. One can easily replace the $k$-means formulation by, for example, $k$-medians. This is especially useful when the mean may not be a good estimate of the posterior mode of interest.

There are two main limitations of this algorithm: (1) the solution strongly depends on the initialization of the $k$ cluster means, and (2) the number of clusters $k$ must be specified a priori (usually the expected number of posterior modes is unknown). We address the first issue using hierarchical distance-based clustering as follows. To overcome limitation (1), we compute all pairwise distances between the given samples using Equation \ref{eqn:dpsi} and perform hierarchical clustering using the maximum linkage criterion. We then initialize the $k$-means clustering algorithm using the $k$ clusters provided by hierarchical clustering. To address the second issue, we use the following procedure to determine the ``correct" number of clusters or posterior modes $k$. First, we compute the pooled total variance across all clusters for $k=1$ and $k=2$. To decide whether the posterior has multiple modes, we examine the percentage decrease in the pooled variance due to the additional second cluster. If the percentage decrease is greater than $30\%$, we proceed to cluster the posterior samples. While this cutoff value seems ad-hoc, we have found through many simulations and real data examples that it works well in practice. Then, to decide on the final number of clusters, we use the silhouette measure of Rousseeuw \cite{rousseeuw}. To construct the silhouette for warping function $i$, we require the following two values: (1) $a(i)$, which is the average dissimilarity of warping function $i$ to all other warping functions in the same cluster, and (2) $b(i)$, which is the minimum average dissimilarity of warping function $i$ to any of the other clusters; we use the Fisher-Rao distance (Equation \ref{eqn:dpsi}) as the dissimilarity measure. Then, the silhouette can be calculated as $s(i)=\frac{b(i)-a(i)}{\max\{a(i),b(i)\}}$. The silhouette for a given warping function measures the appropriateness of its cluster assignment. The average of the silhouette measures over all posterior warping function samples can take values between -1 and 1, which represent very poor and very good clusterings, respectively. The number of modes in the posterior is chosen as the number of clusters, which maximizes the average silhouette measure.

\textbf{Discretization:} To define the Bayesian registration model, we first discretize the observed functional data using a dense sampling of $N$ points: $[t]=\{t_1,\dots,t_N\}\in [0,1]$, where $N$ depends on the application of interest. We study the effects of different values of $N$ on the posterior inference in Section \ref{sec:sim1}. This allows us to model differences between SRSFs using multivariate normal distributions. Note that the function $f$ evaluated at the $N$ discrete points is denoted by $f([t])$ (similarly $q([t])$ for the SRSFs). As will be seen later, the warping functions do not require an explicit discretization in the given model. But, in order to compute the action of $\Gamma$ on the observed functions (SRSFs), we also discretize them with the same $N$ points in the implementation. Finally, we use discrete approximations to compute the quantities defined in this section.

\section{Bayesian Registration Model}
\label{sec:hrm}

Given two functions $f_1,\ f_2$ and their corresponding SRSFs $q_1,\ q_2$, we introduce a novel Bayesian model for function registration. Let $q_2^*$ denote $(q_2\circ\phi^{-1}(\psi))\psi$. At the first stage, we model the difference $q_1-q_2^*|\psi$ using a zero-mean Gaussian process. After discretization of the observed functions, we model the $N$ differences $q_1[t]-q_2^*[t]|\psi$ using the multivariate normal distribution as follows:
\begin{equation}
q_1[t]-q_2^*[t]\ |\ \psi,\kappa\sim MVN(0_N,\frac{1}{2\kappa} I_{N})\ \mbox{(likelihood denoted by $L$)}.
\end{equation}This part of our model is exactly the same as that proposed in \cite{Dryden:13}.

The second stage of our model places a truncated wrapped normal (TWN) prior distribution on the space of warping functions $\Gamma$ by using their SRD representation:
\begin{equation}
\psi\sim TWN_{\Psi}(\mu_{\psi},\Sigma_{\psi})\ \mbox{(denoted by $\pi_\psi$)}.
\end{equation}
We set the mean of the prior to be the identity mapping $\mu_\psi=1$, which provides natural regularization toward $\gamma_{id}$ (i.e., no warping). We also truncate the support of the prior to the valid space of warping functions given by $\Psi$. Thus, the prior distribution $\pi_{\psi}$ is a truncated wrapped normal distribution defined and evaluated in $T_{1}(\Psi)$. This definition is similar to that presented in Kurtek et al. \cite{KurtekJASA}; an alternative construction of Gaussian distributions on high-dimensional spheres is given in \cite{dryden2005}.

To define the covariance structure in the prior on warping functions, we require an orthonormal basis in the tangent space $T_{1}(\Psi)$. We begin by defining a set of basis elements, which are orthogonal to the representation space and have unit $\ltwo$ norm: $\tilde{B}_1=\{\sqrt{3}(1-2t),\ \sqrt{2}\sin(2\pi lt),\ \sqrt{2}\cos(2\pi lt)\ |\ l=1,2,\dots\,n\}$. Then, to form an orthonormal basis for the tangent space $T_1(\Psi)$, we use the Gramm-Schmidt procedure under the $\ltwo$ metric. Notice that this orthonormal basis, denoted by $\tilde{B}$, is truncated by choosing a maximum number $l=n$, which yields $2n+1$ basis elements denoted by $\tilde{b}$. The truncation of the basis is important for additional regularization (smoothness of the warping functions) and computational efficiency. Given an orthonormal basis in the tangent space $T_{1}(\Psi)$, one can approximate any warping function using a set of basis coefficients given by $\psi\approx c=\{c_j=\langle \exp_1^{-1}(\psi),\tilde{b}_j\rangle,\ j=1,\dots,2n+1\}$. Using this notation, we can write the truncated wrapped normal prior on warping functions as follows:
\begin{equation}
\pi_{\psi}(\psi|1,K)\propto \exp(-\frac{1}{2}c^T K^{-1}c)\mathbbm{1}_{\Psi}, \label{eqn:priorgen1}
\end{equation}where $\mathbbm{1}$ is the indicator function. We specify $K$ as a diagonal covariance matrix with $\sigma^2/j^4$ as the $j$th diagonal element with a large value for $\sigma^2=1000$. Thus, we assume a weakly informative prior distribution on the directions given by the basis $\tilde{B}$. We choose quadratic decay of the standard deviation with respect to the degree of the basis functions based on simulations presented in Section \ref{sec:sim1}. We require at least a linear decay for the eigenvalues of the covariance operator to be summable \cite{cotter2013}. In practice, we want to favor smoother warping functions; thus, we weigh the low frequency basis elements (corresponding to low values of $j$) higher than the high frequency basis elements; the variance of the additional linear basis element is not penalized.

To model the concentration parameter in the likelihood, $\kappa$, we use a vague gamma prior with parameters $\alpha=1$ and $\beta=0.01$ ($E(\kappa)=100$, $V(\kappa)=10000$). This prior is denoted by $\pi_{\kappa}$. We assume that the registration variable, $\psi$, and the concentration in the likelihood, $\kappa$, are independent. This is a reasonable assumption due to the fact that the alignment of two functions does not depend on their scale as shown in \cite{EFA:10}.

Under this specification of the model, the marginal posterior distribution of $\psi$ becomes:
\begin{align}
\label{postdist}
p(\psi|q_1,q_2) &\propto \int_0^{\infty}L(q_1[t]-q_2^*[t]|\psi,\kappa)\pi_{\psi}(\psi)\pi_{\kappa}(\kappa)d\kappa\nonumber\\
&\propto \frac{\tilde{\Gamma}(N/2+1)}{(0.01+|q_1[t]-q_2^*[t]|^2)^{N/2+1}}\pi_{\psi}(\psi),
\end{align}where $\tilde{\Gamma}$ denotes the gamma function. We will use importance sampling to sample form the posterior distribution and perform Bayesian inference.

\subsection{Model Justification}

Here, we give a brief justification for each component of the proposed Bayesian registration model. In particular, we focus on the advantages of the given model over other possible choices.

\textbf{Likelihood:} In the current work, we specify the likelihood as a multivariate normal distribution on the pointwise differences between two SRSFs representing the observed functions. An alternative approach that is common in current literature is to model the pointwise differences between the observed functions themselves. Unfortunately, this suffers from the drawbacks discussed in detail in Secton \ref{probback}. In particular, it is clear that, under that setup, the likelihood changes depending on whether one is aligning $f_2$ to $f_1$ or vice versa. This is a direct result of the lack of isometry of the $\ltwo$ metric under the action of $\Gamma$, i.e., $\|f_1\circ\gamma-f_2\circ\gamma\|\neq \|f_1-f_2\|$. See \cite{Dryden:13} for further justification of the given likelihood.

\textbf{Prior on $\Gamma$:} We model the warping functions using a truncated wrapped normal distribution on the SRD space. This allows us to avoid discretizing the warping functions in the specification of the model (we only discretize at the final implementation stage), which is in contrast to the method presented by Cheng et al. \cite{Dryden:13}. In that work, the authors observe that warping functions are akin to cumulative distribution functions. Thus, they place a Dirichlet prior on increments of the discretized warping functions. In contrast, we use a basis on the SRD tangent space, which allows us to model the full warping function up to the level of basis truncation (the warping function can be easily evaluated at any point on the domain $[0,1]$ using the given basis). The proposed approach also permits one to easily incorporate prior knowledge into the model. First, the prior can be defined in a tangent space centered at any warping function. The given basis can be parallel translated using the simple expression in Section \ref{sec:warpfun} to aid in this definition (see the next section for details). This can be especially useful if the observed functions are annotated with landmarks. Second, the prior knowledge about smoothness of the warping functions can be incorporated through the level of basis truncation. For smooth warping functions, the basis can be truncated at a relatively small number (and vice versa for ``rougher" warpings). Finally, we are able to control the variance and decay in the diagonal covariance $K$, allowing further flexibility in the model.

\textbf{Prior on $\kappa$:} We choose the standard gamma prior on the concentration parameter $\kappa$. The main advantage of this choice is that we are able to analytically marginalize the posterior over this parameter. This simplifies the importance sampling approach discussed in the next section.

\section{Importance Sampling}
\label{sec:impsamp}

We begin by briefly introducing the concept of importance sampling and then provide some details of how this can be applied to our problem. Importance sampling is a variance reduction technique in Monte Carlo estimation where instead of directly sampling from a distribution of interest, which may be inefficient, one first samples from an importance function and then re-samples based on appropriate weights.

Suppose that we are interested in estimating the value of the following integral: $\theta=\int_{\cal X} g(x)p(x)dx$, where $p$ is a probability density function. The classical Monte Carlo estimate of this integral is given by $\hat{\theta}=\sum_{i=1}^Sg(x_i)$, where $\{x_1,\dots,x_S\}$ are $iid$ samples from $p$. If the variance of the classical Monte Carlo estimate is large it may be beneficial to introduce a new function $h$, termed the {\it importance function}, which can be used to generate the samples instead of $p$. One can then rewrite the integral as $\theta=\int_{\cal X}\frac{g(x)p(x)}{h(x)}h(x)dx$. The improved Monte Carlo estimate becomes $\tilde{\theta}=\sum_{i=1}^Sg(x_i)w(x_i)$, where $\{x_1,\dots,x_S\}$ are $iid$ samples from $h$ and $w(x)=\frac{p(x)}{h(x)}$. We use this idea to generate samples from the posterior distribution represented by $p$ as follows. Given a large sample $\{x_1,\dots,x_S\}$ from $h$, we compute the associated weights as $\{\frac{p(x_i)}{h(x_i)},\ i=1,\dots,S\}$. Then, to obtain $s$ samples from $p$ (where $s\ll S$), we re-sample the set $\{x_1,\dots,x_S\}$ with the corresponding (normalized) weights. This provides a flexible and efficient method for sampling from the posterior distribution. This process is also called sampling importance re-sampling (SIR). In the current work, we use an improved SIR method without replacement given in \cite{oivind}.

For our problem, we are faced with defining an importance function $h$ that allows us to efficiently sample from the posterior $p$. The main requirement on $h$ is that its support is the same as that of $p$. One option is to use the prior as the importance function directly, and generate the weights using the likelihood. But, in other cases, one may want to ``upsample" a different part of the space, e.g., near the dynamic programming solution. Thus, we provide a general recipe for constructing wrapped normal importance functions similar to the definition of the prior on $\Psi$.

\begin{figure}[!t]
\begin{center}
\includegraphics[width=4in]{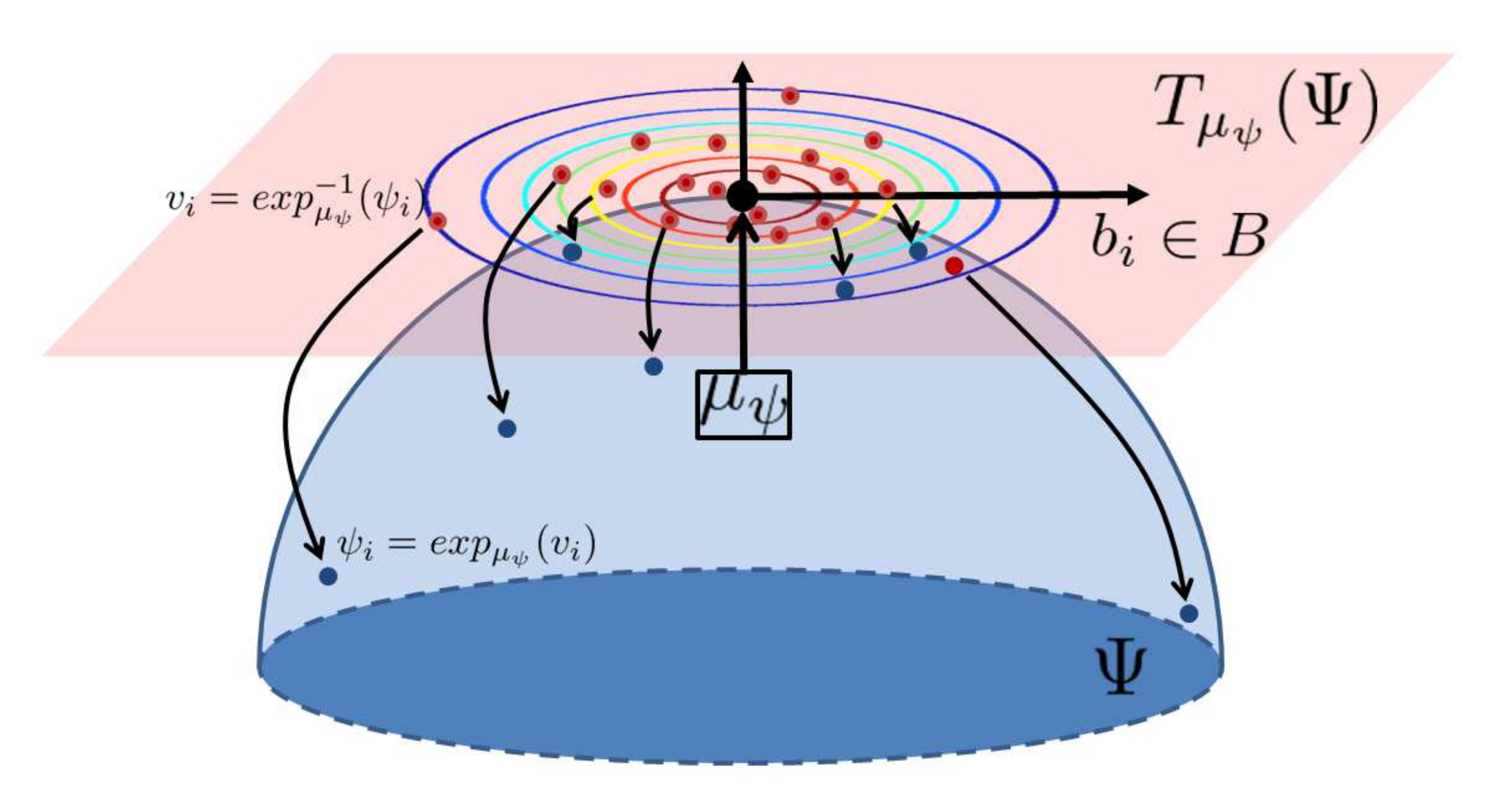}
\caption{We define wrapped normal importance functions in the tangent space at a pre-specified mean. One can generate random samples from these models on the tangent space and then use the exponential map to get a random warping function.}\label{fig:gaussmodel}
\end{center}
\end{figure}

In order to do this, we require a method for defining an orthonormal basis in the tangent space at any point on $\Psi$. Given the truncated basis $\tilde{B}$ in $T_{1}(\Psi)$ defined in the previous section, we can define an orthonormal basis in the tangent space at an arbitrary point $T_{\mu_{\psi}}(\Psi)$ using parallel transport, which was defined in Section \ref{sec:warpfun}. Parallel transport defines an isometric mapping between tangent spaces, and thus preserves the lengths of the basis vectors and the angles between them. We refer to the orthonormal basis in $T_{\mu_{\psi}}(\Psi)$ as $B$ (with elements $\{b_k,\ k=1,\dots,m\}$), and use it to define a coordinate system in that space. Thus, we can again approximate any warping function using a set of basis coefficients given by $\psi\approx d=\{d_k=\langle \exp_{\mu_{\psi}}^{-1}(\psi),b_k\rangle,\ k=1,\dots,m\}$. In this way, we can define a general version of the importance function as:
\begin{equation}
h(\psi|\mu_{\psi},K_h)\propto \exp(-\frac{1}{2}d^T K_h^{-1} d),
\end{equation}
where $K_h$ is a diagonal matrix ($K_h$ can be specified in the same was as in the prior). Note that there is no need to truncate the importance function. Figure \ref{fig:gaussmodel} provides a pictorial explanation of our definition of the wrapped normal importance function in the tangent space. Under this setup, we can generate random samples from $h$ in using the following procedure:
\begin{enumerate}
\item For $k=1,\dots,m$, sample $z_k\stackrel{iid}{\sim}N(0,1)$;
\item For $k=1,\dots,m$, compute the random basis coefficients $d_k^{rnd}=z_k\sqrt{K_h(k)}$, where $K_h(k)$ denotes the $k$th diagonal element of the matrix $K_h$; 
\item Form the random tangent vector using the basis $B$ as $v^{rnd}=\sum_{k=1}^{m} d_kb_k$;
\item Map $v^{rnd}$ to $\Psi$ using $\psi^{rnd}=\exp_{\mu_{\psi}}(v^{rnd})$;
\item Compute the random warping function using $\gamma^{rnd}=\phi^{-1}(\psi^{rnd})$.
\end{enumerate}

Using the idea of importance sampling, we can re-write the posterior distribution in Equation \ref{postdist} as follows:
\begin{equation}
p(\psi|q_1,q_2)\propto \frac{\tilde{\Gamma}(N/2+1)\pi_{\psi}(\psi)h(\psi)}{(0.01+|q_1[t]-q_2^*[t]|^2)^{N/2+1}h(\psi)}.\label{eqn:impsampfin}
\end{equation}
It is obvious from the expression in Equation \ref{eqn:impsampfin}, that in the special case when the importance function is the same as the prior, one can simply sample from the prior distribution and weight each sample using the integrated likelihood. Thus, our approach is to generate a large sample $\{\psi_1,\dots,\psi_S\}$ from $h$ and evaluate a weight for each sampled warping function using:
\begin{equation*}
\eta_i=\frac{\tilde{\Gamma}(N/2+1)}{(0.01+|q_1[t]-q_2^*[t]|^2)^{N/2+1}}\exp(-\frac{1}{2}(c^T K^{-1}c-d^T K_h^{-1} d))\mathbbm{1}_{\Psi},\ i=1,\dots,S,
\end{equation*}
where $c=\{c_j=\langle \exp_1^{-1}(\psi),\tilde{b}_j\rangle,\ \tilde{b}_j\in\tilde{B},\ j=1,\dots,2n+1\}$ and $d=\{d_j=\langle \exp_{\mu_{\psi}}^{-1}(\psi),b_j\rangle,\ b_j\in B,\ j=1,\dots,m\}$ as before. Once all of the weights have been computed, we re-sample a small number $s$ of $\psi$s from the original set using the methods proposed in \cite{oivind}. The re-sampled functions $\psi_1,\dots,\psi_s$ are samples from the posterior distribution $p$, and can be mapped to their corresponding warping functions using $\phi^{-1}$. Posterior functionals can be mapped to $\Gamma$ in the same way.

\section{Simulation Studies}
\label{sec:sim}

In this section, we present warping results using simulated scenarios. In all examples, we fix the original sample size to $S=500000$, the posterior sample size to $s=200$, and the number of basis elements in the prior and importance function to $N-1$, where $N$ is the sampling density of the observed functions. The importance function used throughout the simulation studies and the real applications is a wrapped normal centered at the identity element with the same covariance structure as the prior.

\subsection{Simulation 1}
\label{sec:sim1}

\begin{figure}[!t]
\begin{center}
\begin{tabular}{|c|c|c|c|}
\hline
(a)&(b)&(c)&(d)\\
\hline
\includegraphics[width=1in]{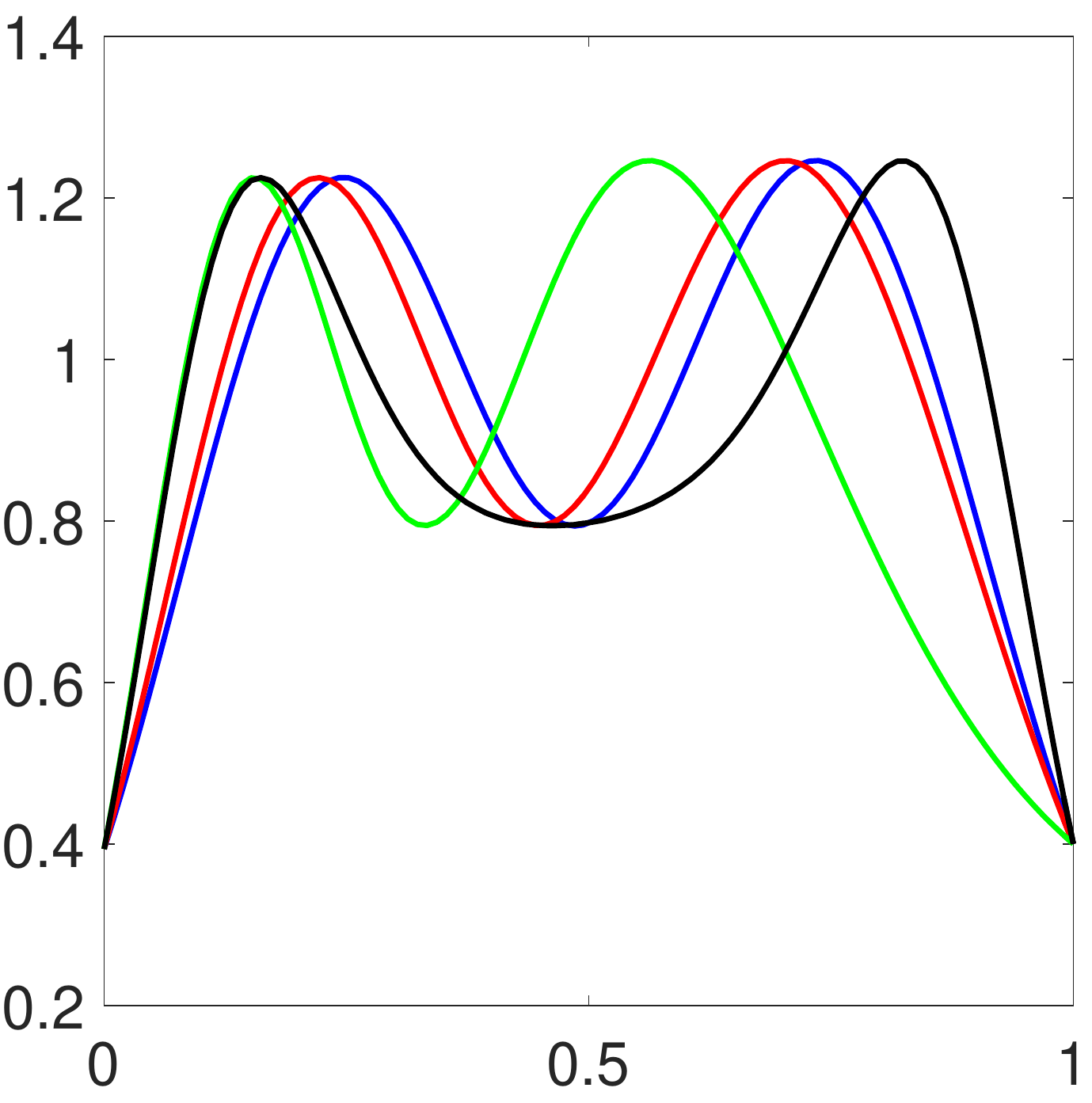}&\includegraphics[width=1in]{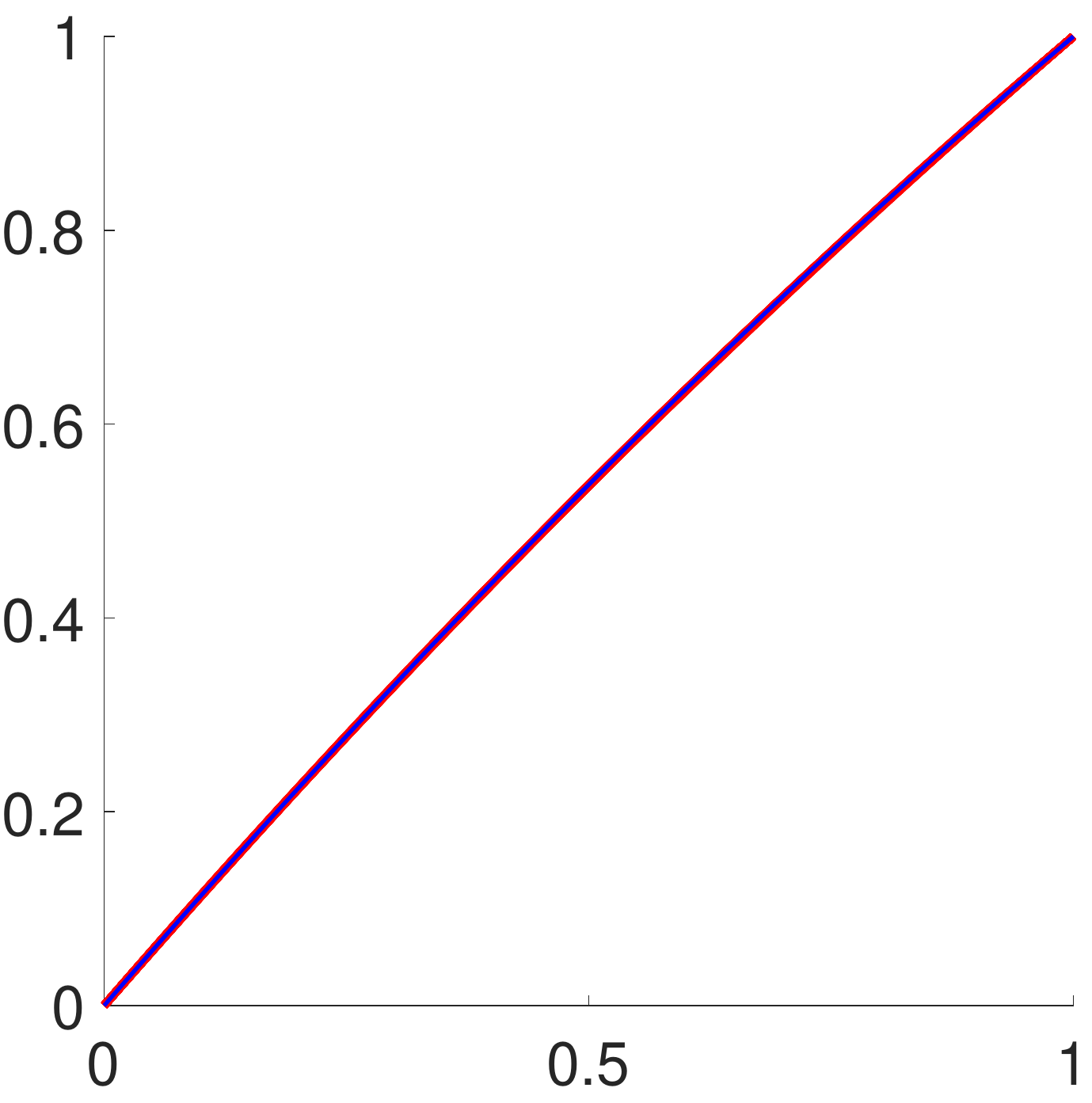}&\includegraphics[width=1in]{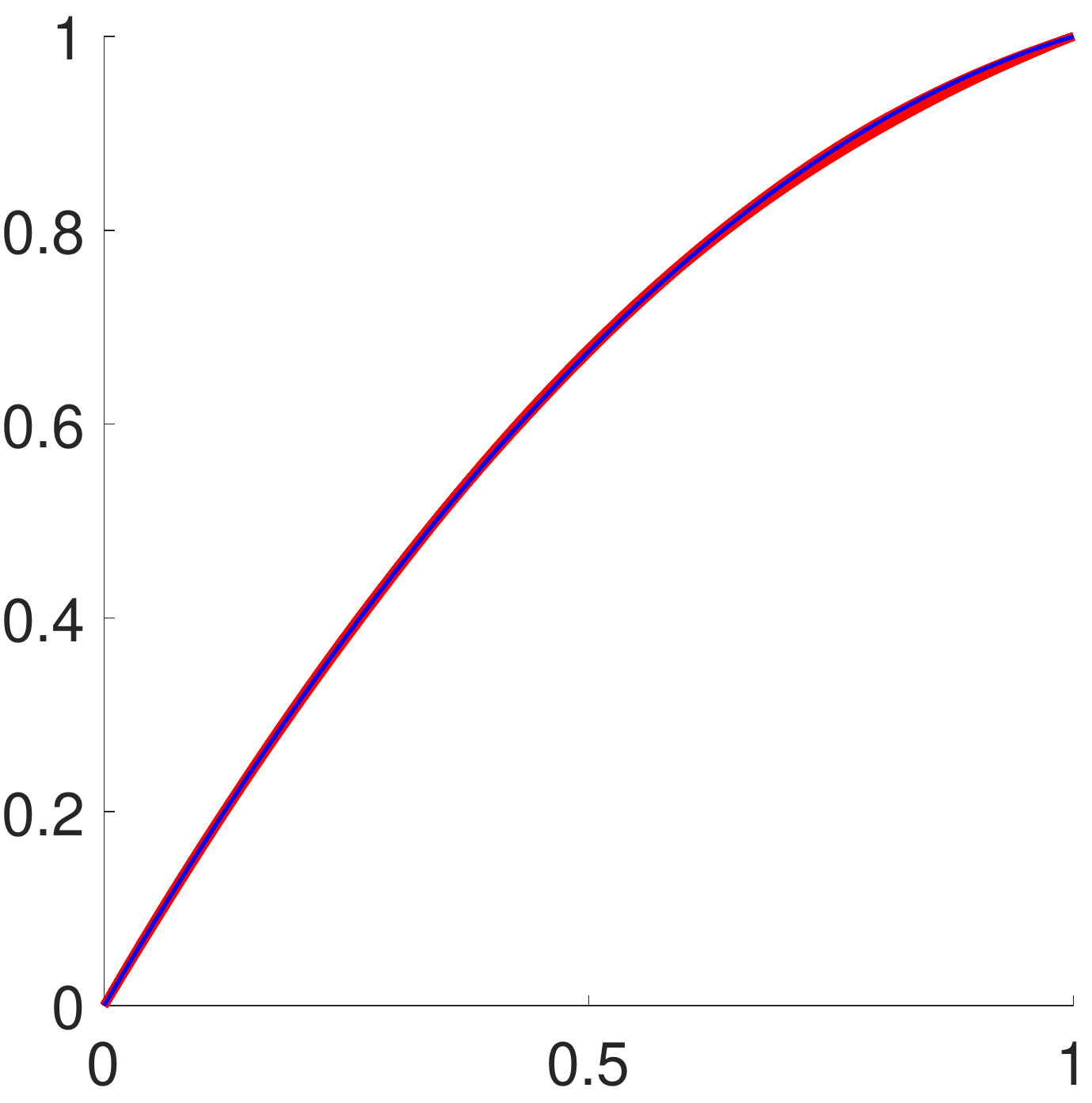}&\includegraphics[width=1in]{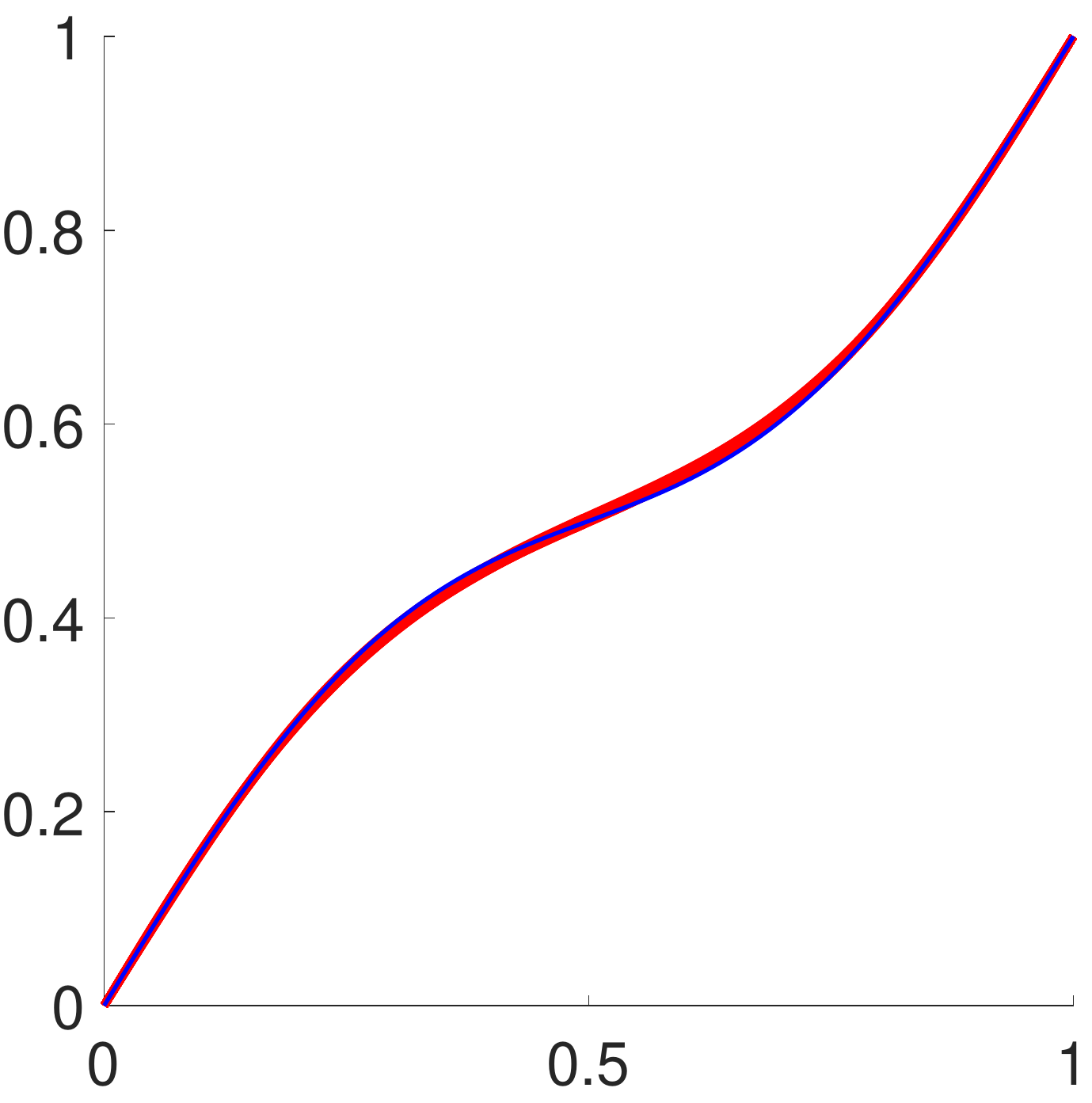}\\
\hline
\end{tabular}
\caption{(a) Simulated data with $f$ in blue, $f\circ\gamma_1$ in red, $f\circ\gamma_2$ in green, and $f\circ\gamma_3$ in black. (b)-(d) True warpings $\gamma_1$, $\gamma_2$ and $\gamma_3$ in blue, respectively, with 100 posterior means in red.}\label{fig:ex1sim}
\end{center}
\end{figure}

\begin{table}[!t]
\begin{center}
\caption{Simulation results for correct warping recovery for three different warping functions and sampling densities (SD) under quadratic decay of the prior standard deviations. (a) Average $d_{FR}(\gamma_{T},\bar{\gamma})$ with the standard deviations in parentheses (T=true). (b) $d_{FR}(\gamma_{T},\gamma_{DP})$. (c) Average $d_{FR}(\gamma_{T},\bar{\gamma}_{DIR})$ with the standard deviations in parentheses. (d) Average $DPD_{PM}$ (PM=proposed method). (e) $DPD_{DP}$. (f) Average $DPD_{DIR}$. Best results are in bold.} \label{tab:simu1}
\begin{tabular}{|c|c|ccc||ccc|}
\hline
Ex.&SD&(a)&(b)&(c)&(d)&(e)&(f)\\
\hline
&50&\textbf{0.0055 (0.0008)}&0.0359&0.0145 (0.0024)&\textbf{95.4}&85.4&93.6\\
$\gamma_1$&100&\textbf{0.0018 (0.0004)}&0.0366&0.0200 (0.0023)&\textbf{98.0}&88.3&89.8\\
&150&\textbf{0.0018 (0.0007)}&0.0384&0.0243 (0.0027)&\textbf{97.8}&88.8&87.4\\
\hline
&50&\textbf{0.0124 (0.0029)}&0.0312&0.0373 (0.0030)&\textbf{98.2}&96.5&94.6\\
$\gamma_2$&100&\textbf{0.0131 (0.0032)}&0.0211&0.0436 (0.0037)&\textbf{98.3}&98.1&93.4\\
&150&\textbf{0.0125 (0.0030)}&0.0182&0.0480 (0.0032)&\textbf{98.4}&98.3&92.2\\
\hline
&50&\textbf{0.0234 (0.0027)}&0.0284&0.0707 (0.0033)&\textbf{96.0}&95.9&88.1\\
$\gamma_3$&100&0.0258 (0.0029)&\textbf{0.0212}&0.0719 (0.0035)&94.8&\textbf{97.0}&87.5\\
&150&0.0259 (0.0028)&\textbf{0.0192}&0.0759 (0.0035)&95.0&\textbf{97.1}&86.7\\
\hline
\end{tabular}
\end{center}
\end{table}

In the first simulation study, we consider the effects of function sampling density and the order of decay of the standard deviation in the prior distribution. For this purpose, we simulated three different warping functions, $\gamma_1=t+0.15t(1-t)$, $\gamma_2=t+0.70t(1-t)$, $\gamma_3=t+0.1\sin(2\pi t)$, $t=[0,1]$, and applied them to a function with two modes denoted by $f$. We display the original function $f$ in Figure \ref{fig:ex1sim}(a) in blue and the same function under the three warpings, $f\circ\gamma_1$, $f\circ\gamma_2$ and $f\circ\gamma_3$, in red, green and black, respectively.

\begin{table}[!t]
\begin{center}
\caption{Simulation results for correct warping recovery for three different warping functions and sampling densities (SD) under linear and no decay of the prior standard deviations. (a) Average $d_{FR}(\gamma_{T},\bar{\gamma})$ with the standard deviations in parentheses (T=true). (b) Average $DPD_{PM}$ (PM=proposed method).} \label{tab:simu2}
\begin{tabular}{|c|c||c|c||c|c|}
\hline
&&\multicolumn{2}{c}{Linear Decay}&\multicolumn{2}{c|}{No Decay}\\
\hline
Ex.&SD&(a)&(b)&(a)&(b)\\
\hline
&50&0.0060 (0.0007)&94.9&0.0084 (0.0008)&92.9\\
$\gamma_1$&100&0.0028 (0.0006)&97.3&0.0090 (0.0008)&89.3\\
&150&0.0029 (0.0006)&96.6&0.0104 (0.0006)&83.7\\
\hline
&50&0.0176 (0.0033)&97.4&0.0389 (0.0035)&94.1\\
$\gamma_2$&100&0.0200 (0.0035)&97.4&0.0559 (0.0029)&91.4\\
&150&0.0202 (0.0035)&97.2&0.0674 (0.0028)&85.6\\
\hline
&50&0.0320 (0.0033)&94.6&0.0685 (0.0039)&88.4\\
$\gamma_3$&100&0.0347 (0.0031)&93.1&0.0875 (0.0032)&81.7\\
&150&0.0355 (0.0038)&93.1&0.1012 (0.0029)&75.4\\
\hline
\end{tabular}
\end{center}
\end{table}

We apply the proposed model to perform pairwise Bayesian alignment for each example using 100 replicates, and report the detailed results in Table \ref{tab:simu1} for quadratic decay of the prior standard deviations and sampling densities of 50, 100 and 150 points. For each example, we report the average Fisher--Rao distance between the true warping function and the estimated posterior mean $\bar{\gamma}$ in panel (a), the Fisher--Rao distance between the true warping function and the dynamic programming solution $\gamma_{DP}$ in panel (b), and the average Fisher--Rao distance between the true warping function and the estimated posterior mean when using a Dirichlet prior $\bar{\gamma}_{DIR}$ in panel (c). In all of the presented results, we set the parameters of the Dirichlet distribution to $\alpha_1=\dots=\alpha_{40}=1$ (i.e., uniform prior on warping functions specified in the same was as in \cite{Dryden:13}), and use importance sampling to sample from the posterior. The standard deviations of the distances are also provided in parentheses. We highlight the best performance for each example and sampling density in bold. In all examples, the proposed geometric Bayesian model outperforms a model with a Dirichlet prior on the warping functions. Furthermore, the performance of the proposed method is comparable to, and often better than, the commonly used dynamic programming algorithm.

In panels (d)-(f), we report the average percentage decrease in the distance between the two functions being registered, i.e., $DPD=\frac{\|q_1-q_2\|-\|q_1-(q_2,\gamma)\|}{\|q_1-q_2\|}$. Again, the proposed model performs very well according to this metric. It is important to note that the gains in performance are small when the sampling density is increased from 100 to 150 points. Thus, for fairly smooth functions, as is the case in this simulation and the applications presented in the subsequent section, we will sample the functions with 100 points for computational efficiency. The replicate posterior means for the proposed method are displayed in red in Figure \ref{fig:ex1sim}(b)-(d) with the true warping in blue. It is clear from this figure that there is little variation across replicates and that we are able to recover the true warping very well.

Table \ref{tab:simu2} reports the same set of results for linear and no decay in the prior standard deviations for the proposed method across the three sampling densities. Linear decay performs comparably to quadratic decay, while no decay does not perform well as expected. Throughout the rest of the paper we utilize quadratic decay as indicated by these simulation results.

\subsection{Simulation 2}

\begin{figure}[!t]
\begin{center}
\begin{tabular}{|c|c|}
\hline
(a)&(b)\\
\hline
\includegraphics[width=1in]{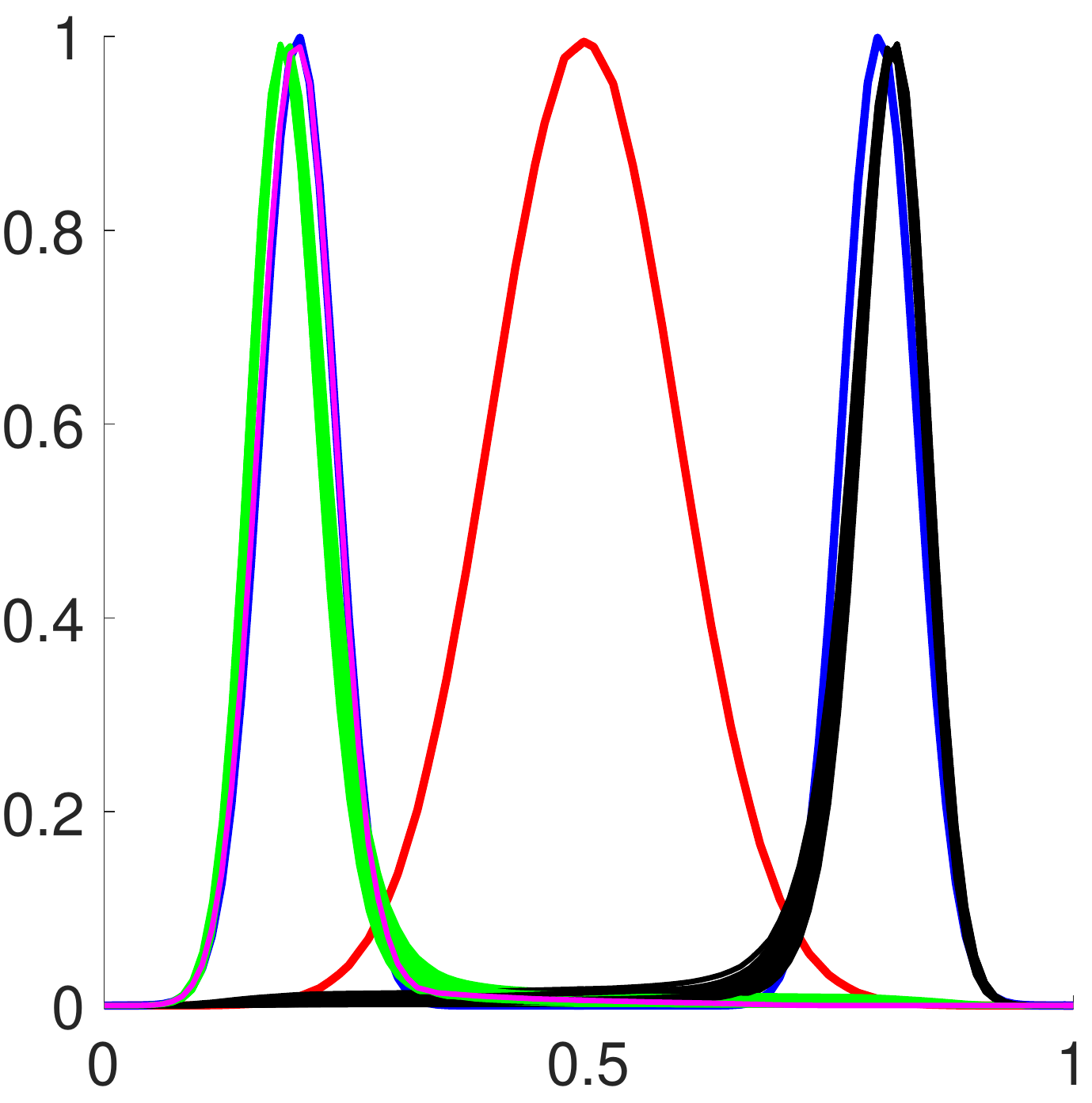}&\includegraphics[width=1in]{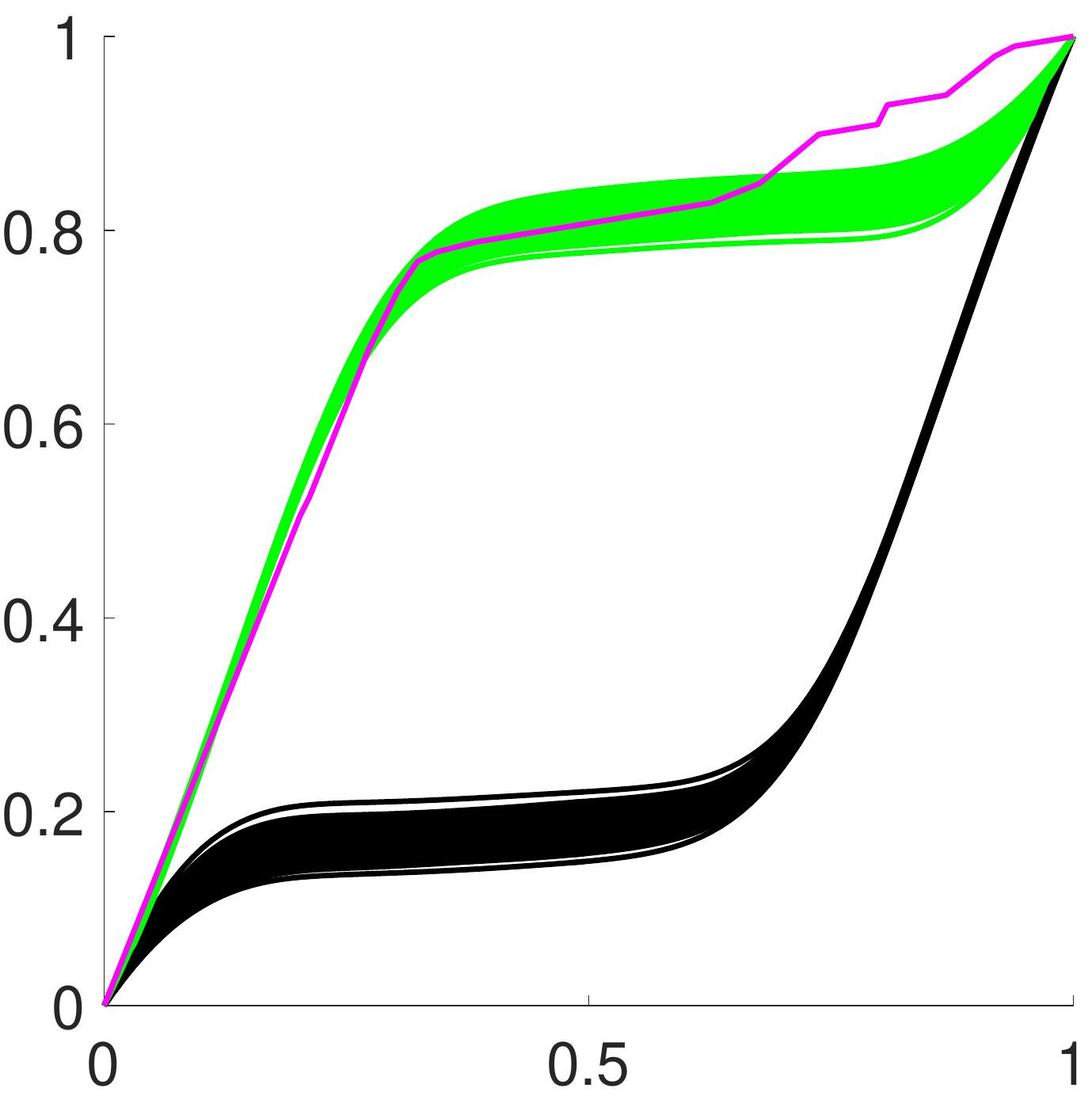}\\
\hline
\end{tabular}
\caption{(a) Simulated data with $f_1$ in blue, $f_2$ in red, $f_2\circ\bar{\gamma}_1$ in green, $f_2\circ\bar{\gamma}_2$ in black and $f_2\circ\gamma_{DP}$ in magenta. (b) Estimated average warping functions in cluster 1, ($\bar{\gamma}_1$) in green, in cluster 2 ($\bar{\gamma}_2$) in black, and using dynamic programming ($\gamma_{DP}$) in magenta.}\label{fig:ex2sim}
\end{center}
\end{figure}

\begin{table}[!t]
\begin{center}
\caption{Clusterwise summaries of the posterior distribution. (a) Average cluster size. (b)-(d) Average $\ltwo$ distance between $q_1$ and $q_2$ after warping using the mean, median and MAP of each cluster, respectively. The standard deviations are given in parentheses.} \label{tab:simu3}
\begin{tabular}{|c|c|c|c|c|}
\hline
&(a)&(b)&(c)&(d)\\
\hline
Cluster 1&98.95 (6.48)&1.5016 (0.0143)&1.3769 (0.0150)&1.3020 (0.0474)\\
Cluster 2&101.05 (6.48)&1.5071 (0.0141)&1.3784 (0.0163)&1.3028 (0.0475)\\
\hline
\end{tabular}
\end{center}
\end{table}

In the second simulation, we explore the performance of the proposed alignment model when two modes are present in the posterior distribution. The two functions to be aligned, $f_1$ and $f_2$, are shown in Figure \ref{fig:ex2sim}(a) in blue and red, respectively. In the same panel, we show the alignment results, across 100 replicates, using the mean of each posterior cluster in green and black. For comparison, we also display the dynamic programming result in magenta. Note that in this simulation we have treated the number of clusters as known ($k=2$) and applied the $k$-means clustering algorithm as described in Section \ref{sec:warpfun}. In panel (b), we display the two clusters of warping functions representing the two posterior modes (again in green and black) as well as the dynamic programming result (in magenta). The clusterwise posterior mean warping functions are much smoother than the dynamic programming solution and achieve essentially the same level of alignment between the two functions.

In Table \ref{tab:simu3}, we provide a few summaries for each posterior cluster. In particular, we report the average cluster size, and the average distance between the two functions based on clusterwise posterior mean, median and MAP alignment. We expect the clusters to be balanced as the peaks in the bimodal function are approximately equidistant from the peak in the unimodal function. This should also be reflected in the post-alignment, clusterwise distances between the two functions. The original distance between them is 2.6668, and the distance after dynamic programming alignment is 1.4221. The reported clusterwise distances are comparable to the dynamic programming solution when using mean warping, and better when using median and MAP warping. This shows that in addition to being able to discover multiple plausible alignments as modes of the posterior distribution, we are able to better explore the full space of warping functions than the deterministic dynamic programming algorithm.


\section{Applications}
\label{sec:expres}

Next, we consider pairwise alignment of functions using the proposed Bayesian model for various types of real data. We start with three types of biomedical signals: gait pressure functions, PQRST complexes extracted from an ECG and respiration functions. For a detailed description of these datasets please see \cite{biosignals}. We proceed to show examples on growth velocity functions for boys and girls obtained from the Berkeley Growth Dataset (BGD) \cite{tuddenham1954physicalGrowth}. Finally, we show two examples on signature (tangential) acceleration functions from a subset of the data described in \cite{Yeung04svc2004:first}. In each example, we first determine whether multiple modes exist in the posterior distribution of warping functions. If this is the case, we cluster the posterior samples using $k$-means clustering, where $k$ is selected based on the average silhouette measure. Finally, we show the clusterwise alignment results and assess registration uncertainty in each cluster.

\begin{figure}[!t]
\begin{center}
\begin{tabular}{|c|c|c|}
\hline
(a)&(b)&(c)\\
\hline
\includegraphics[width=1.2in]{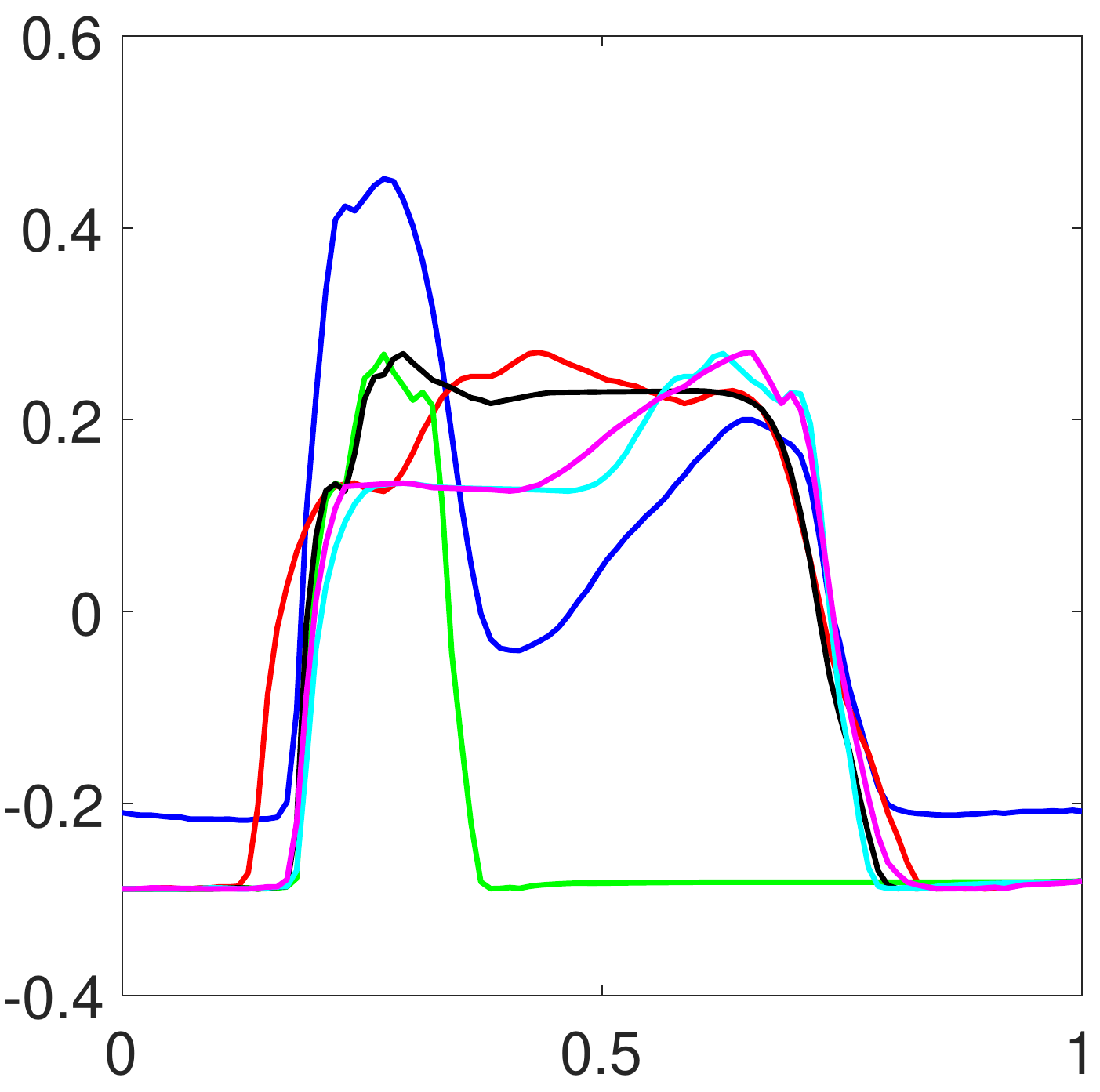}&\includegraphics[width=1.2in]{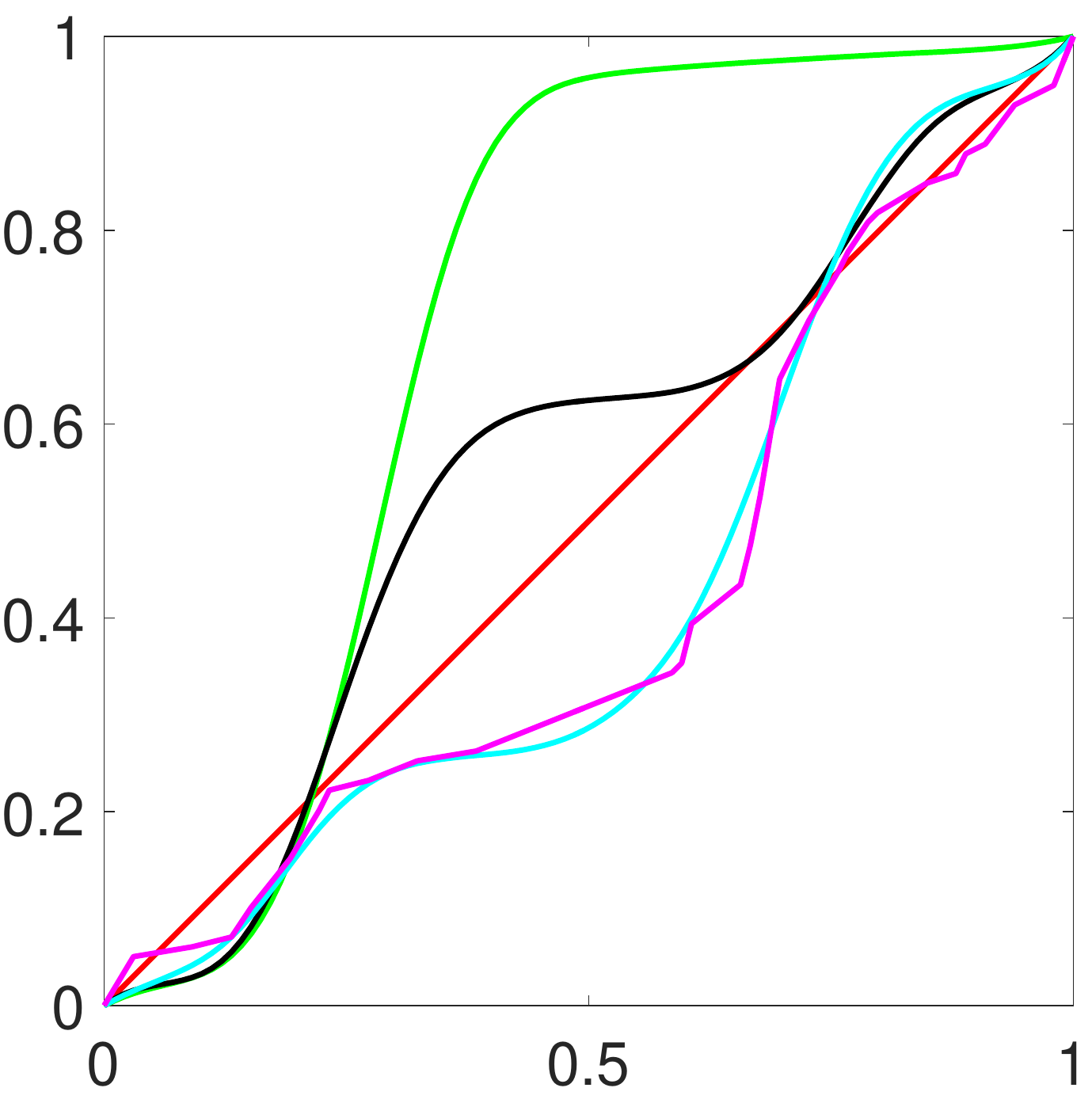}&\includegraphics[width=1.2in]{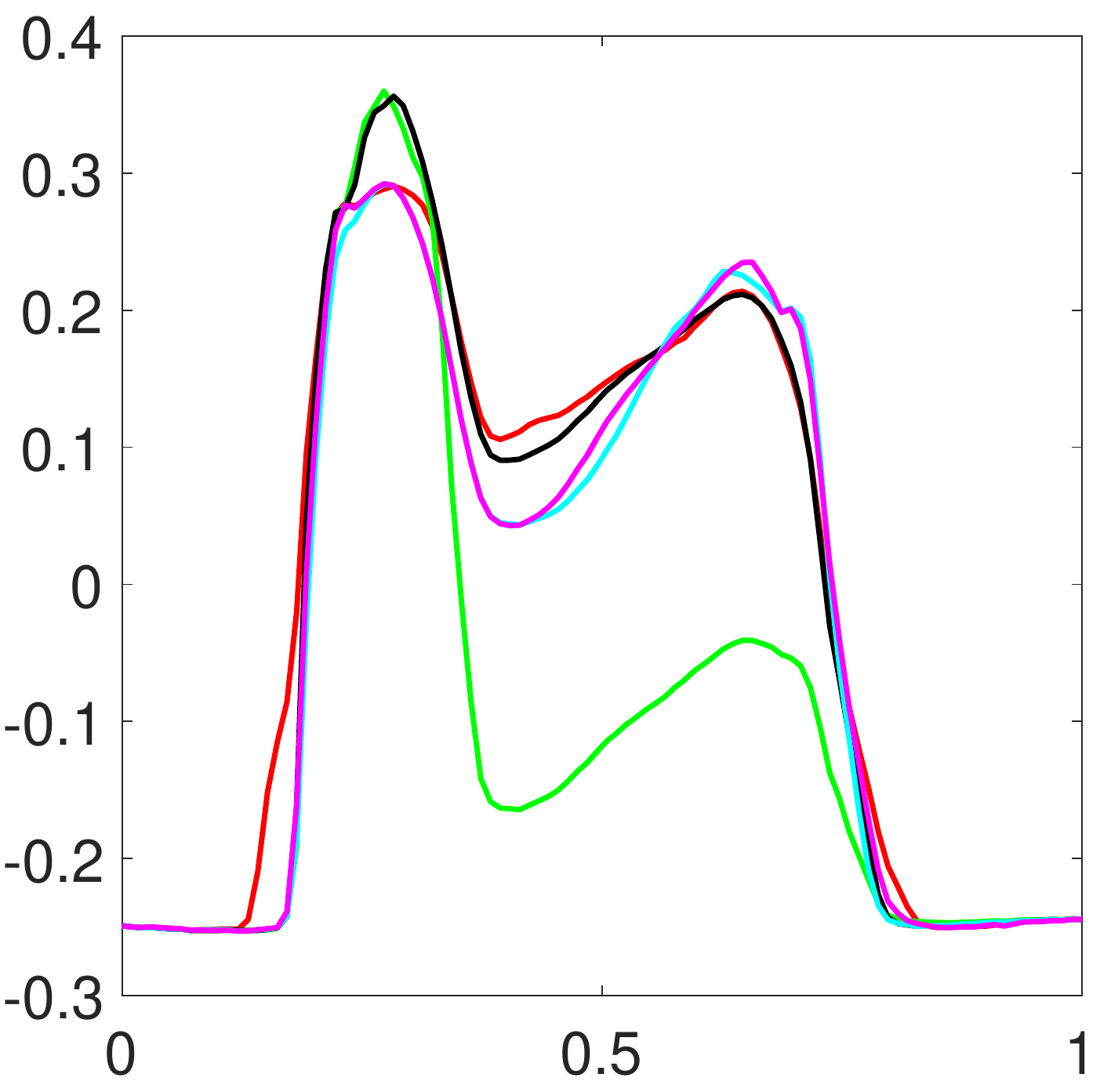}\\
\hline
(d)&(e)&(f)\\
\hline
\includegraphics[width=1.2in]{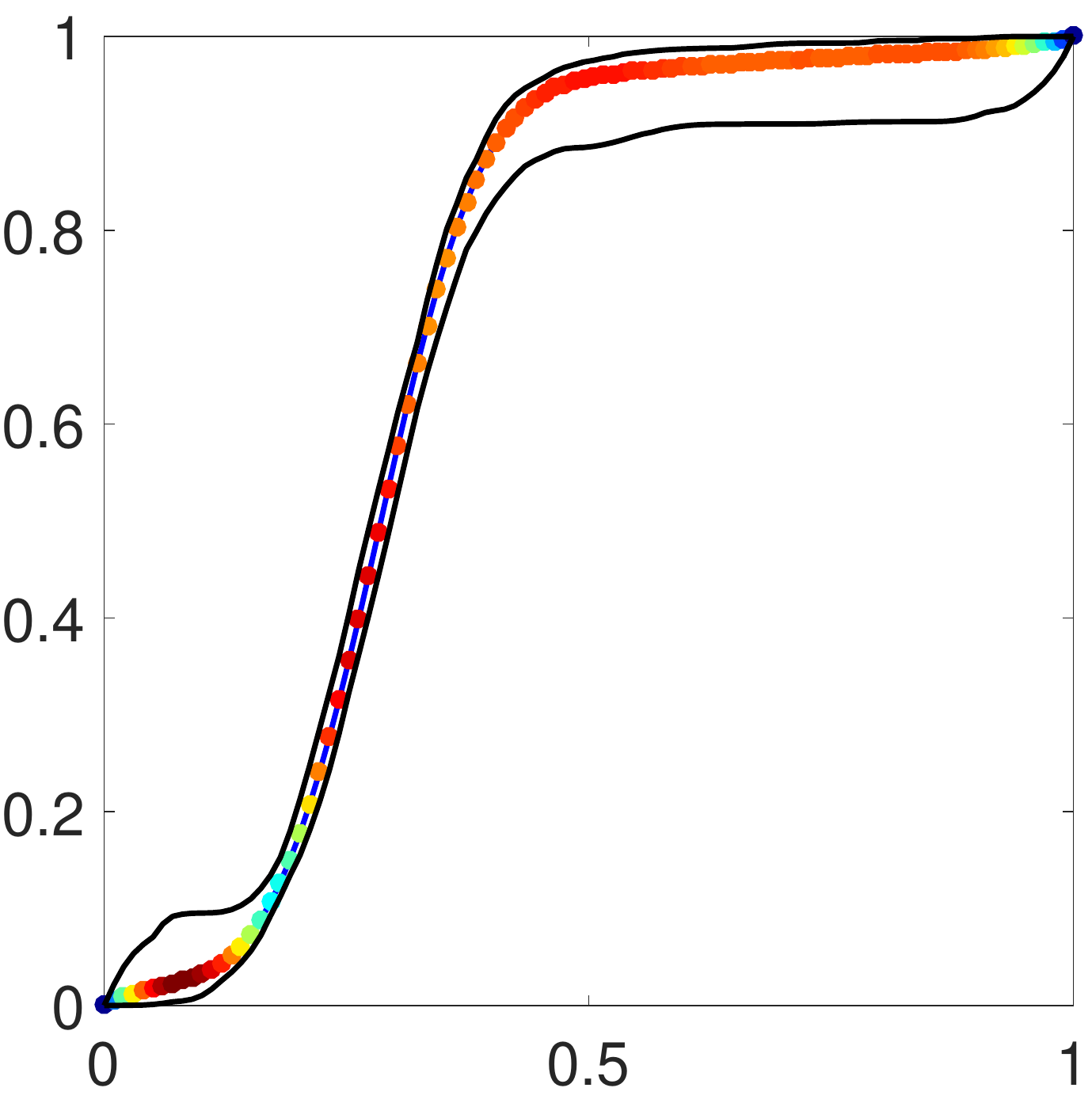}&\includegraphics[width=1.2in]{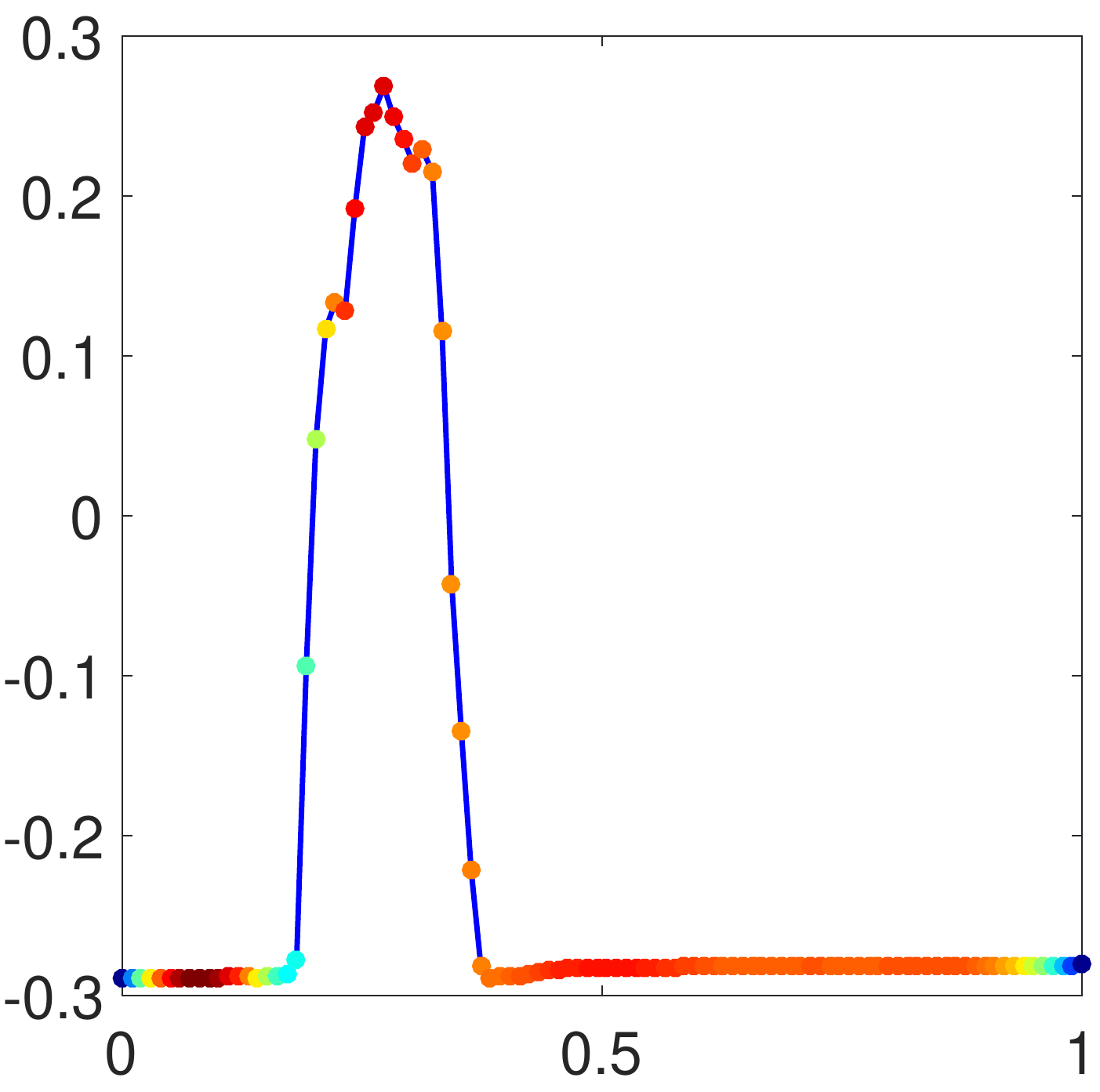}&\includegraphics[width=1.2in]{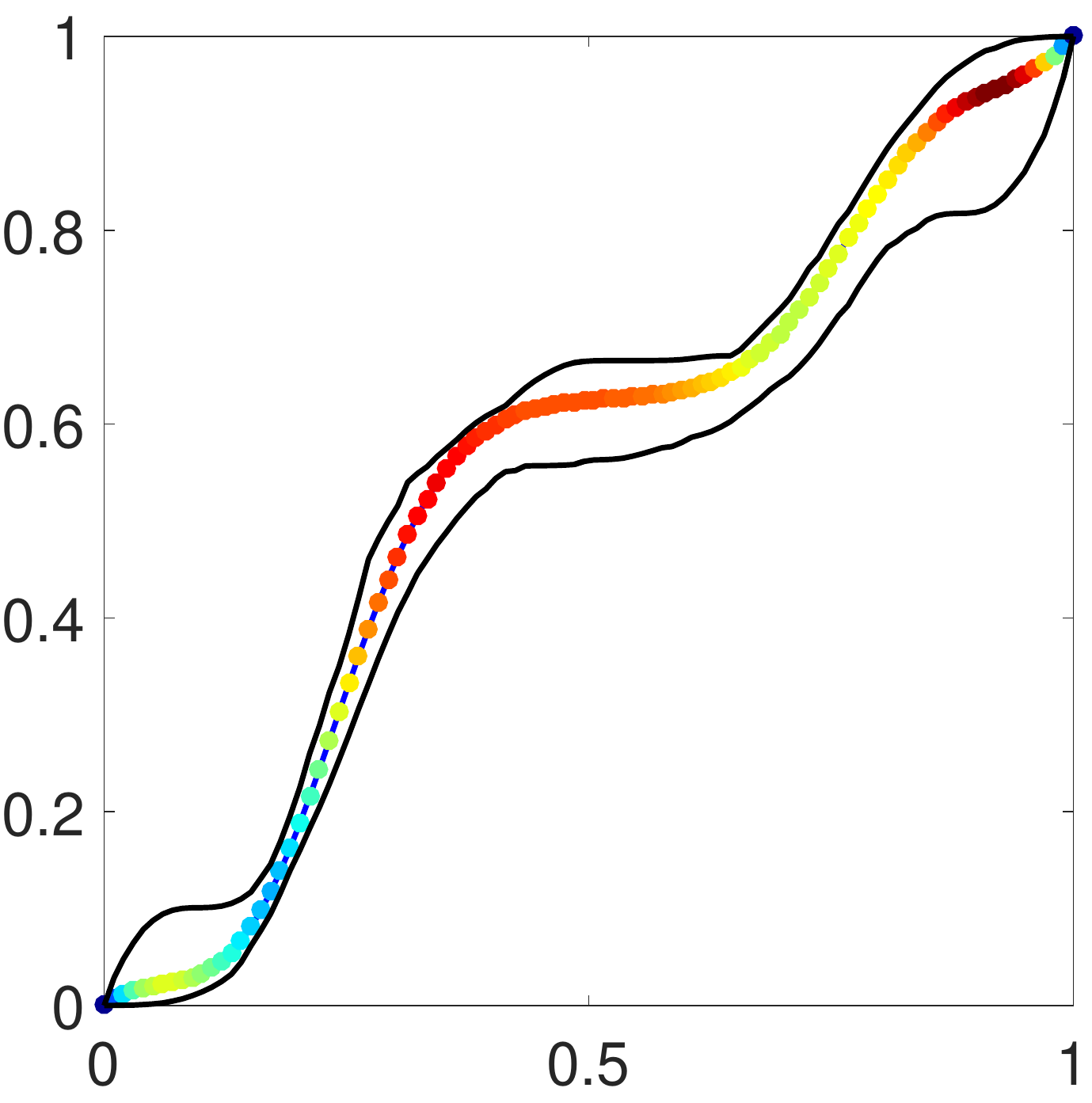}\\
\hline
(g)&(h)&(i)\\
\hline
\includegraphics[width=1.2in]{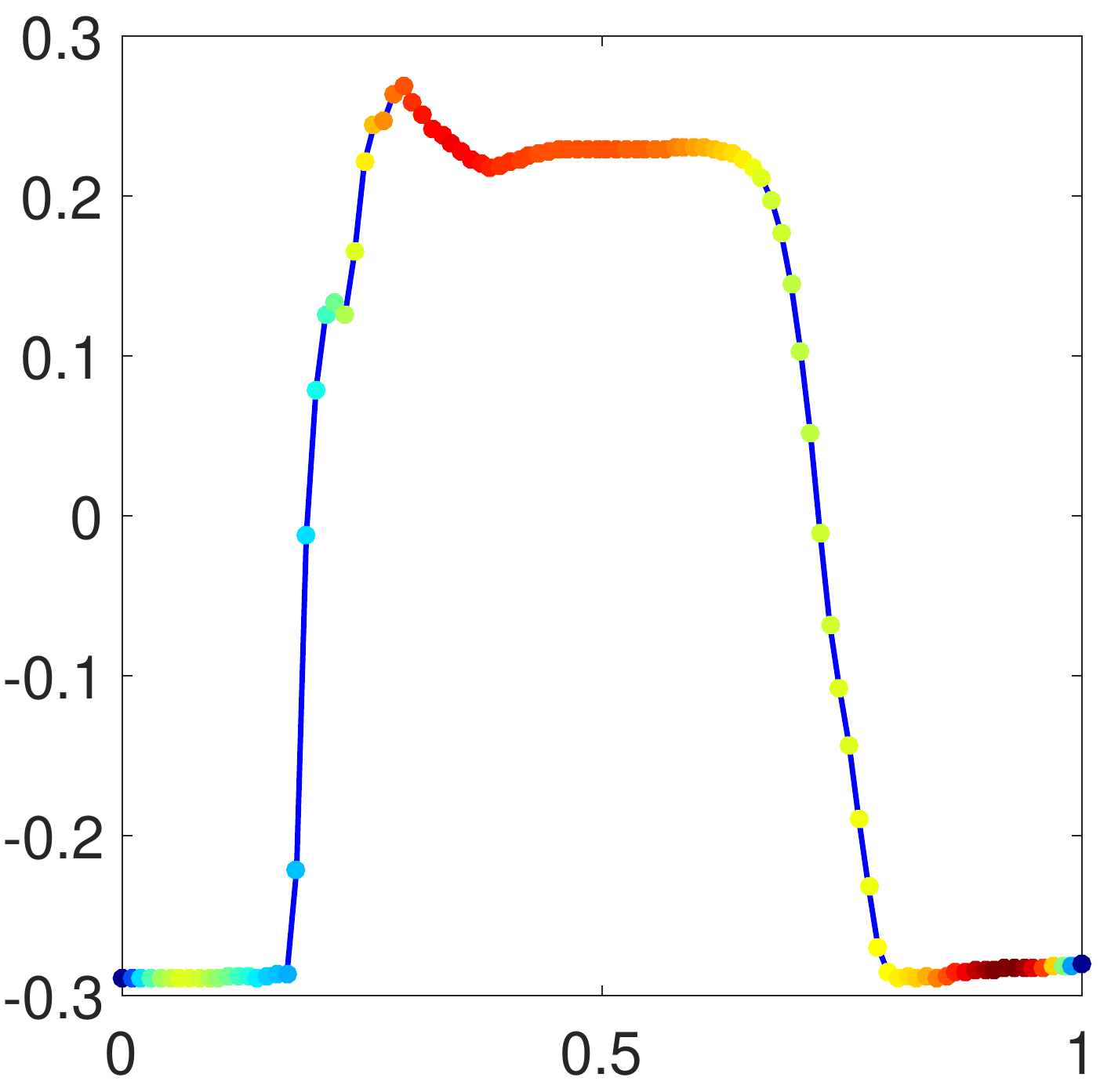}&\includegraphics[width=1.2in]{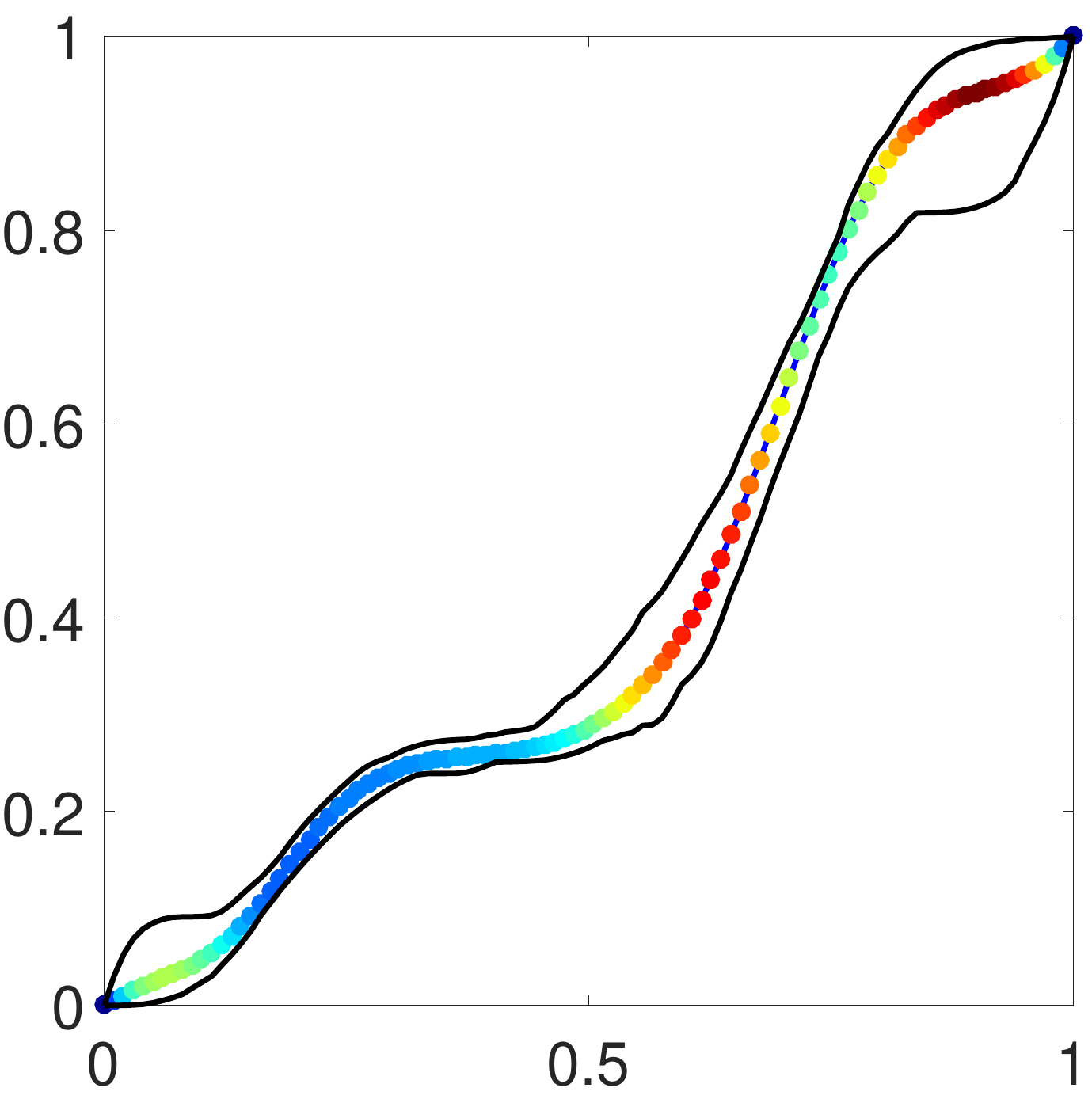}&\includegraphics[width=1.2in]{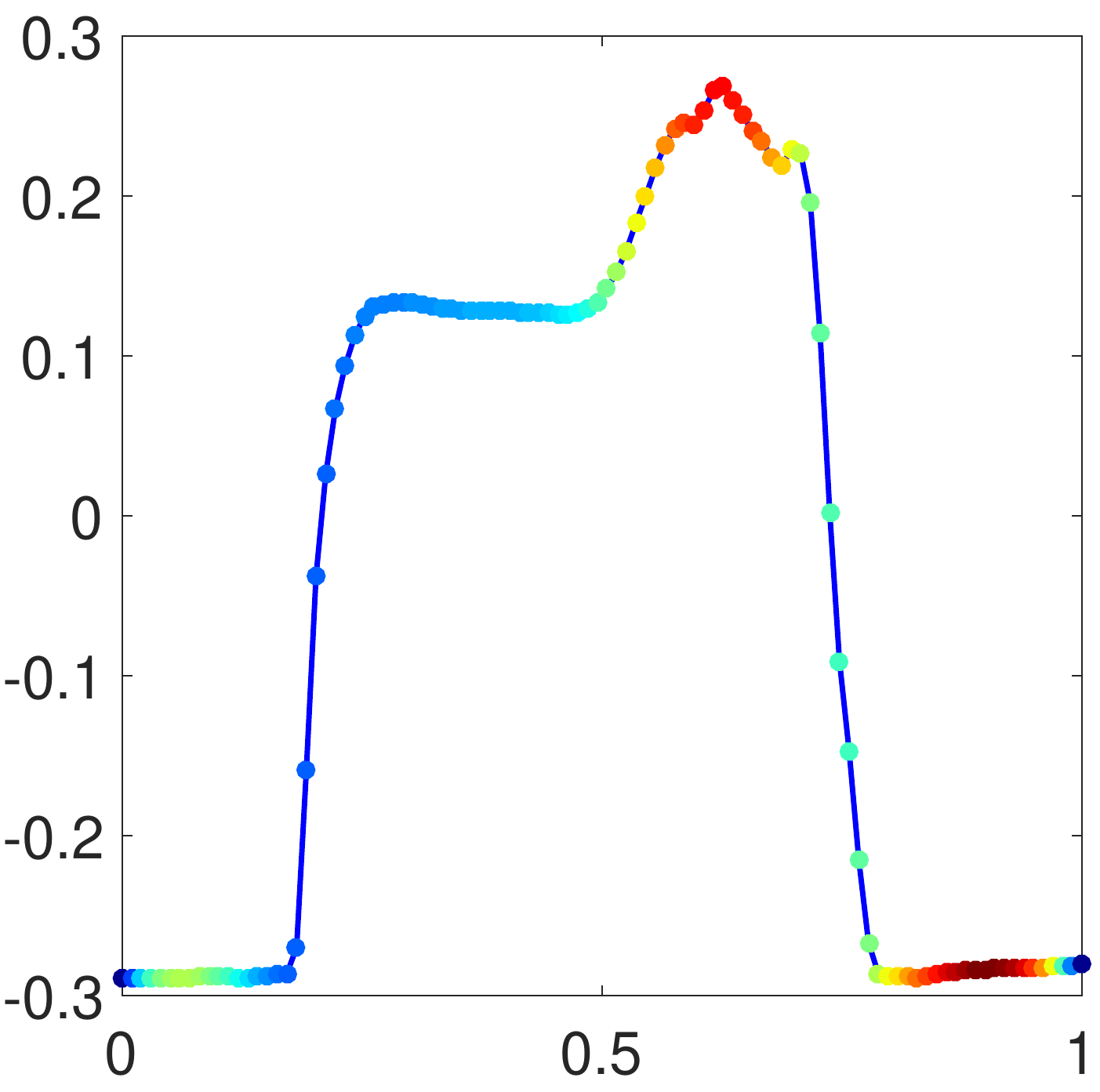}\\
\hline
\end{tabular}
\caption{Pairwise alignment of two gait pressure functions. (a) Original functions $f_1$ and $f_2$ in blue and red, respectively; $f_2\circ\gamma_{DP}$ in magenta, $f_2\circ\bar{\gamma}_1$ in green (cluster 1), $f_2\circ\bar{\gamma}_2$ in black (cluster 2) and $f_2\circ\bar{\gamma}_3$ in cyan (cluster 3). (b) $\gamma_{DP}$ in magenta, $\gamma_{id}$ in red, and $\bar{\gamma}_1$, $\bar{\gamma}_2$ and $\bar{\gamma}_3$ in green, black and cyan, respectively. (c) Pointwise average of $f_1$ and $f_2$ for each alignment result (colored in the same way as (a) and (b)). (d) Pointwise standard deviation (hot colors correspond to higher values) plotted on $\bar{\gamma}_1$, and the $95\%$ credible interval in black. (e) Pointwise standard deviation (hot colors correspond to higher values) plotted on $f_2\circ\bar{\gamma}_1$. (f)-(i) Same as (d) and (e) but for clusters 2 and 3, respectively.} \label{fig:ex1gait}
\end{center}
\end{figure}

\begin{figure}[!t]
\begin{center}
\begin{tabular}{|c|c|c|}
\hline
(a)&(b)&(c)\\
\hline
\includegraphics[width=1.2in]{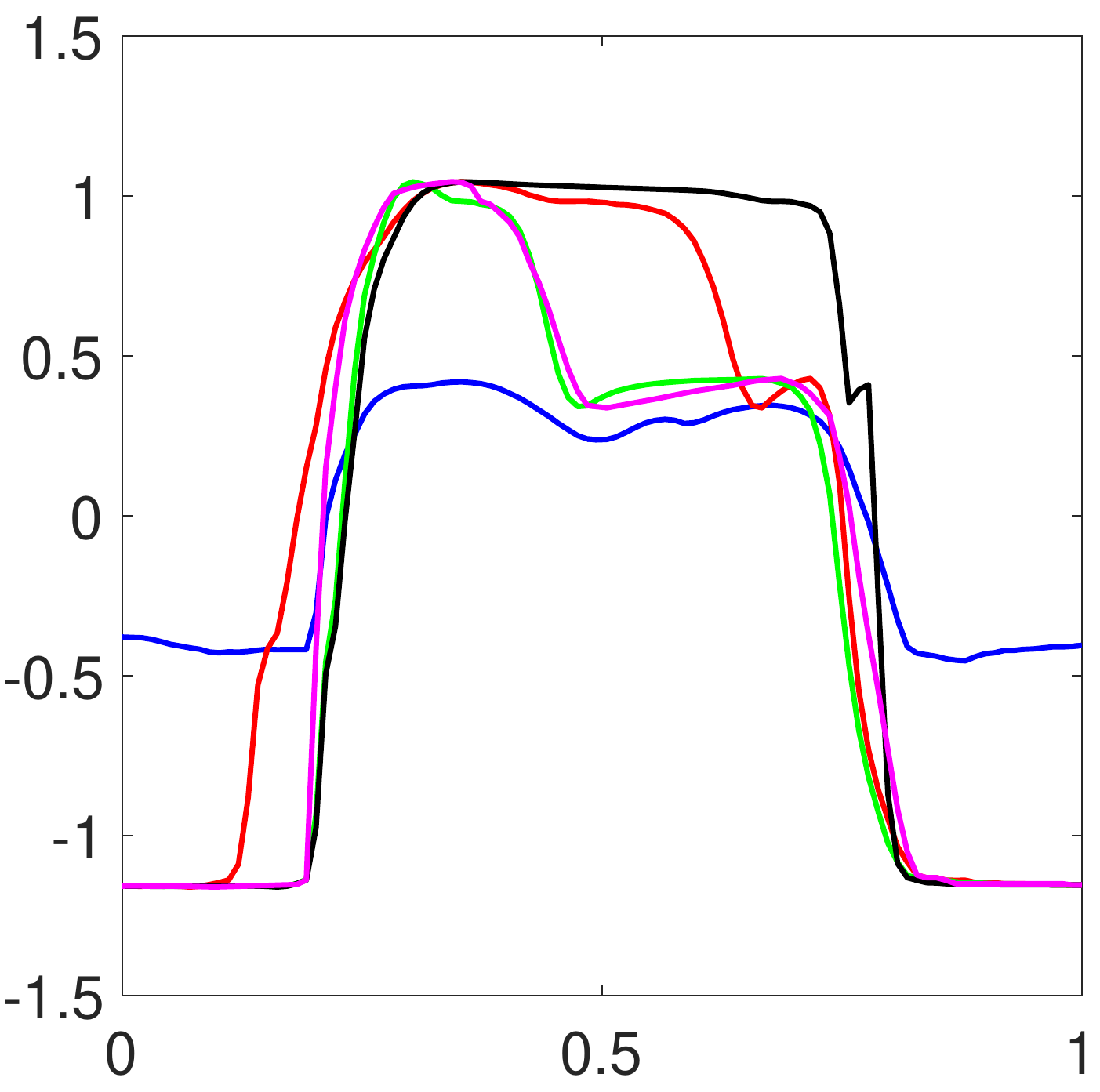}&\includegraphics[width=1.2in]{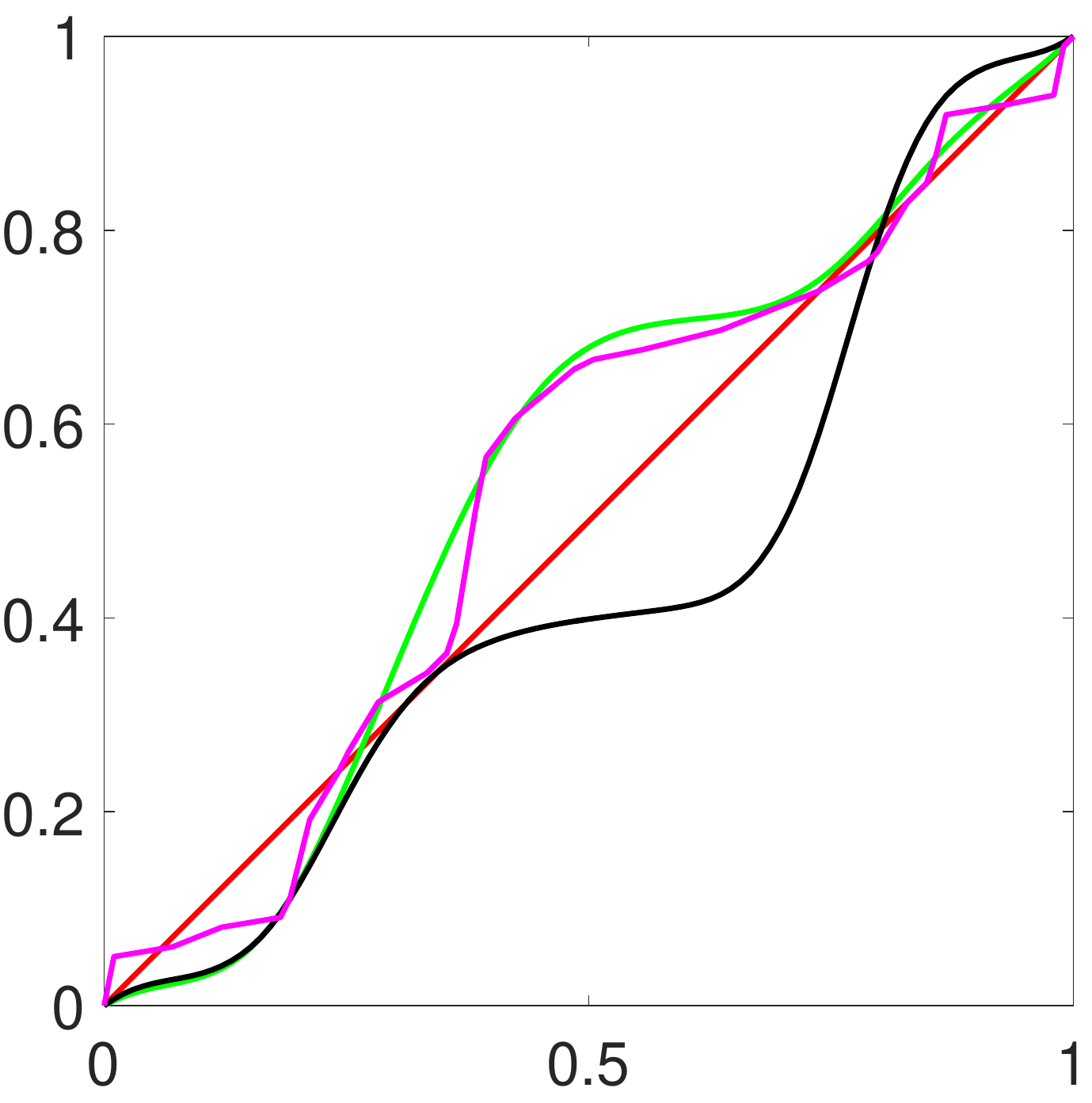}&\includegraphics[width=1.2in]{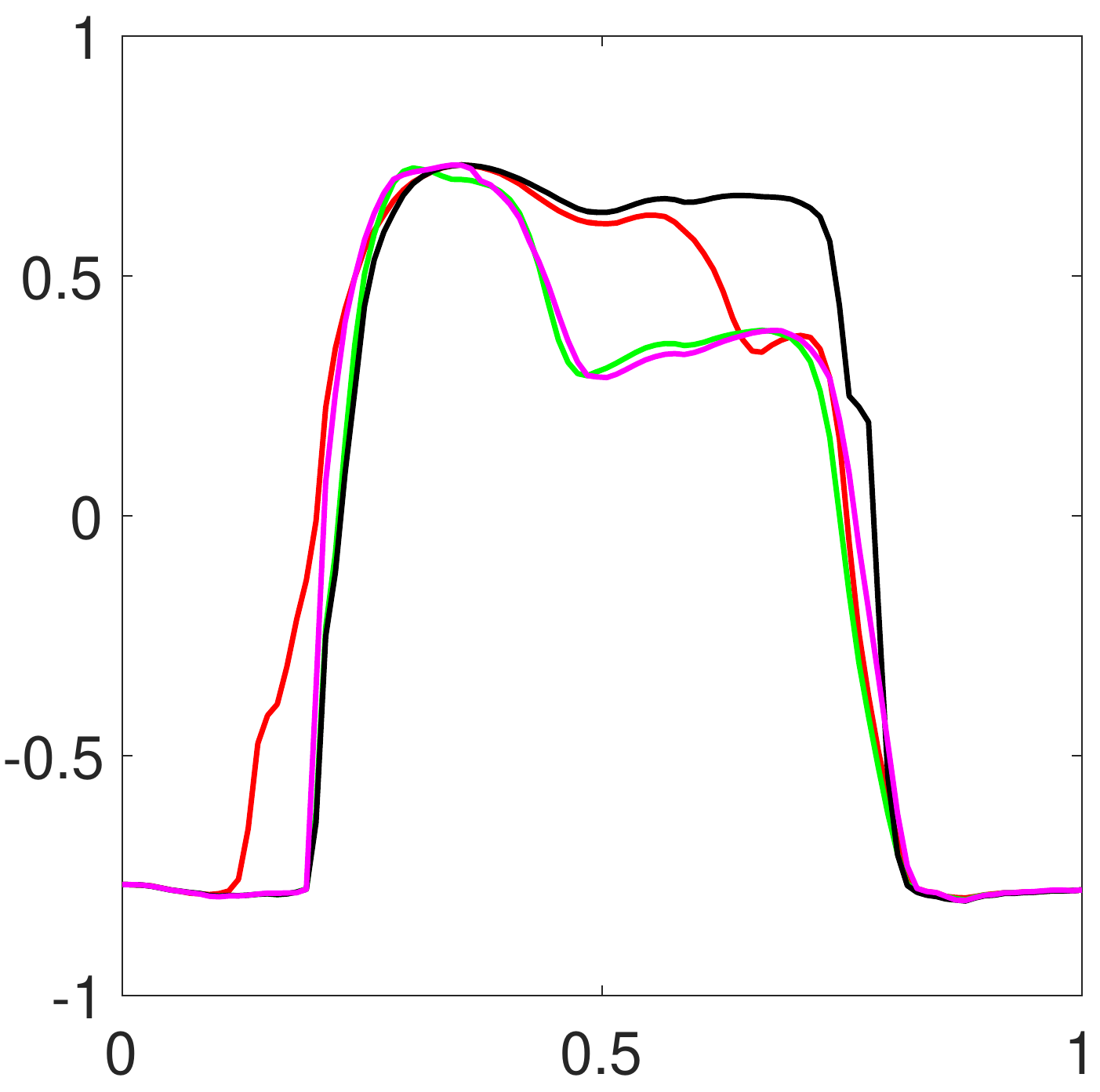}\\
\end{tabular}
\begin{tabular}{|c|c|c|c|}
\hline
(d)&(e)&(f)&(g)\\
\hline
\includegraphics[width=.9in]{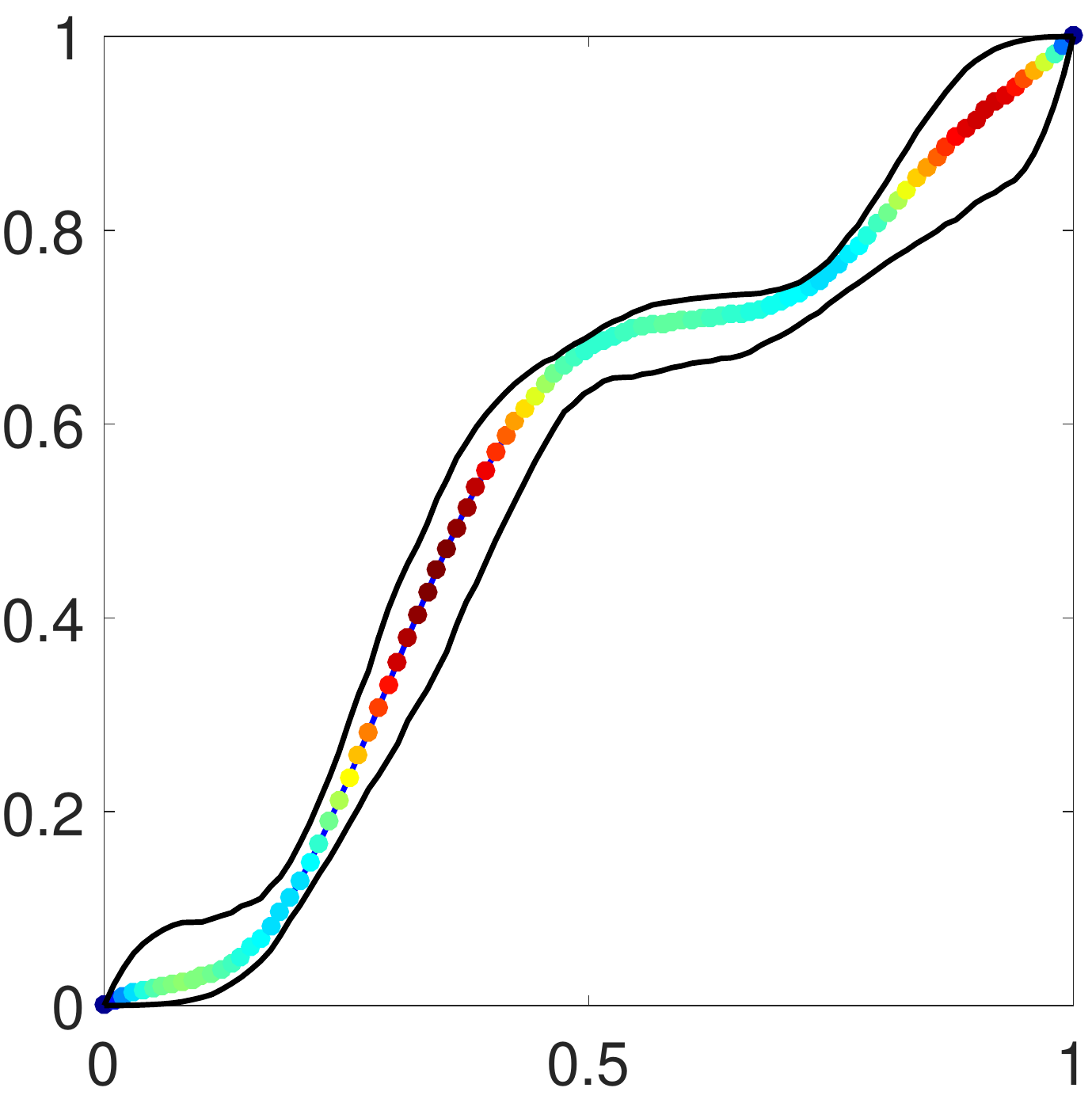}&\includegraphics[width=.9in]{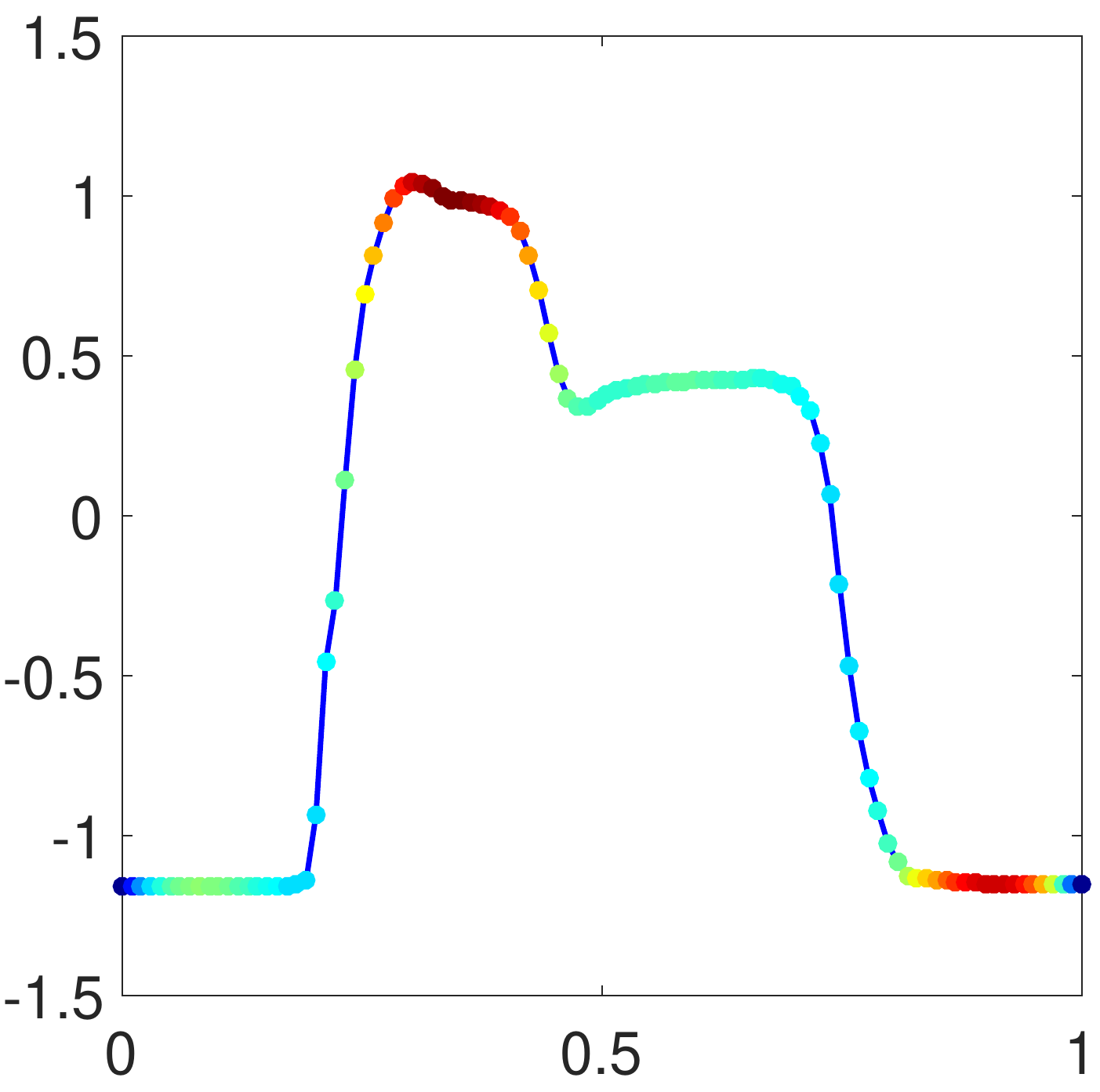}&\includegraphics[width=.9in]{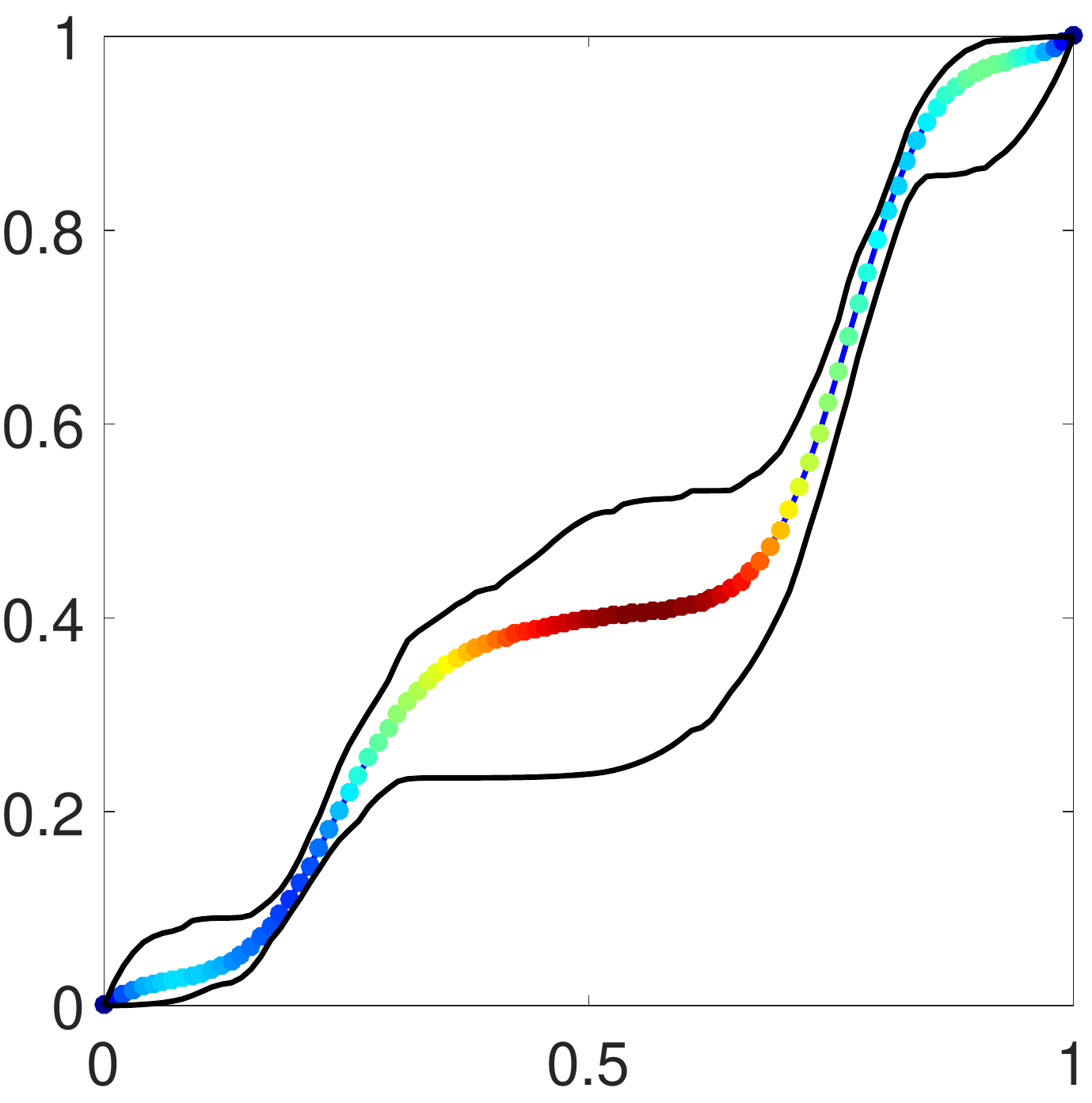}&\includegraphics[width=.9in]{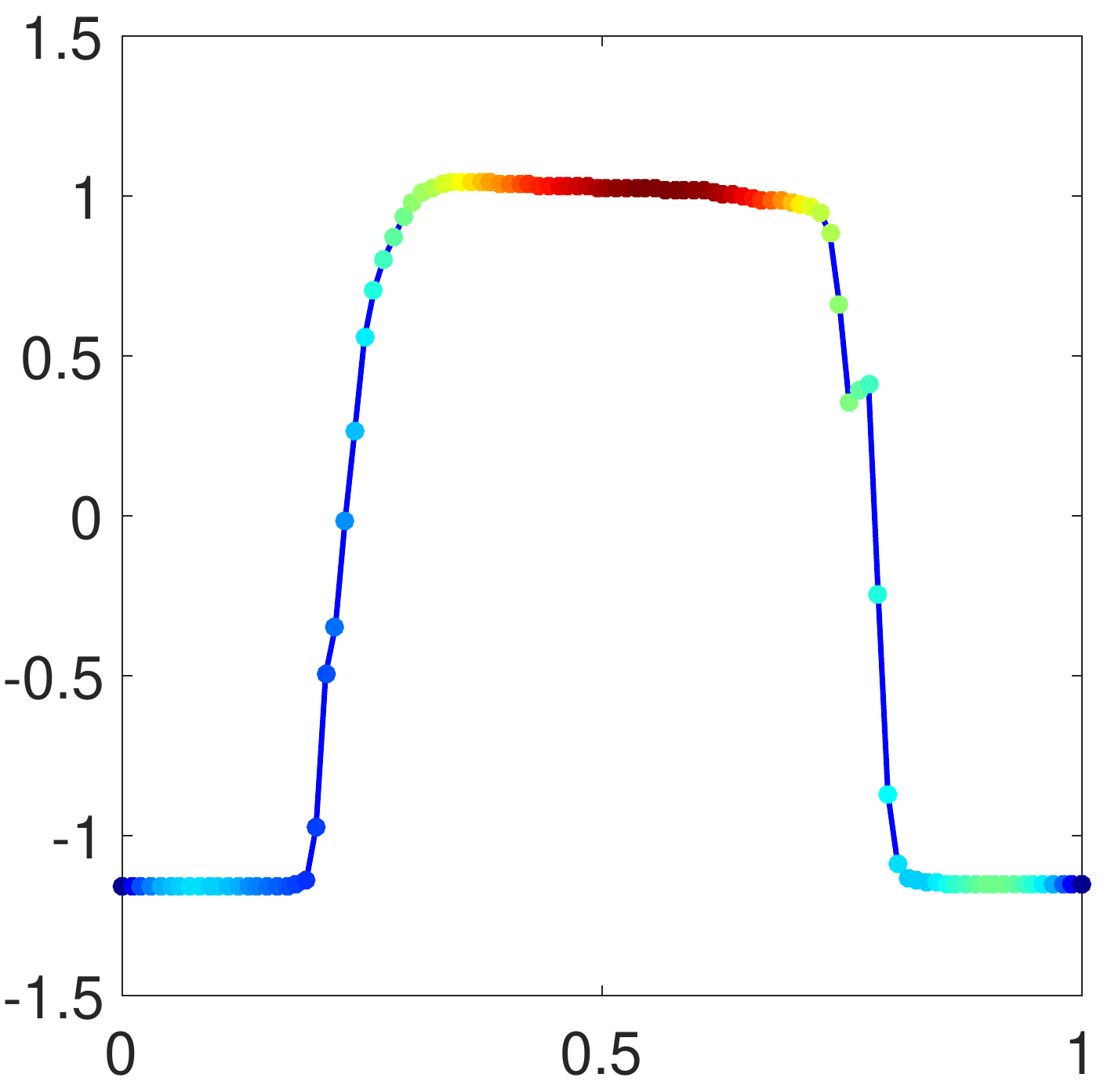}\\
\hline
\end{tabular}
\caption{Pairwise alignment of two gait pressure functions. (a) Original functions $f_1$ and $f_2$ in blue and red, respectively; $f_2\circ\gamma_{DP}$ in magenta, $f_2\circ\bar{\gamma}_1$ in green (cluster 1) and $f_2\circ\bar{\gamma}_2$ in black (cluster 2). (b) $\gamma_{DP}$ in magenta, $\gamma_{id}$ in red, and $\bar{\gamma}_1$ and $\bar{\gamma}_2$ in green and black, respectively. (c) Pointwise average of $f_1$ and $f_2$ for each alignment result (colored in the same way as (a) and (b)). (d) Pointwise standard deviation (hot colors correspond to higher values) plotted on $\bar{\gamma}_1$, and the $95\%$ credible interval in black. (e) Pointwise standard deviation (hot colors correspond to higher values) plotted on $f_2\circ\bar{\gamma}_1$. (f)-(g) Same as (d) and (e) but for cluster 2.} \label{fig:ex2gait}
\end{center}
\end{figure}

\begin{figure}[!t]
\begin{center}
\begin{tabular}{|c|c|c|}
\hline
(a)&(b)&(c)\\
\hline
\includegraphics[width=1.2in]{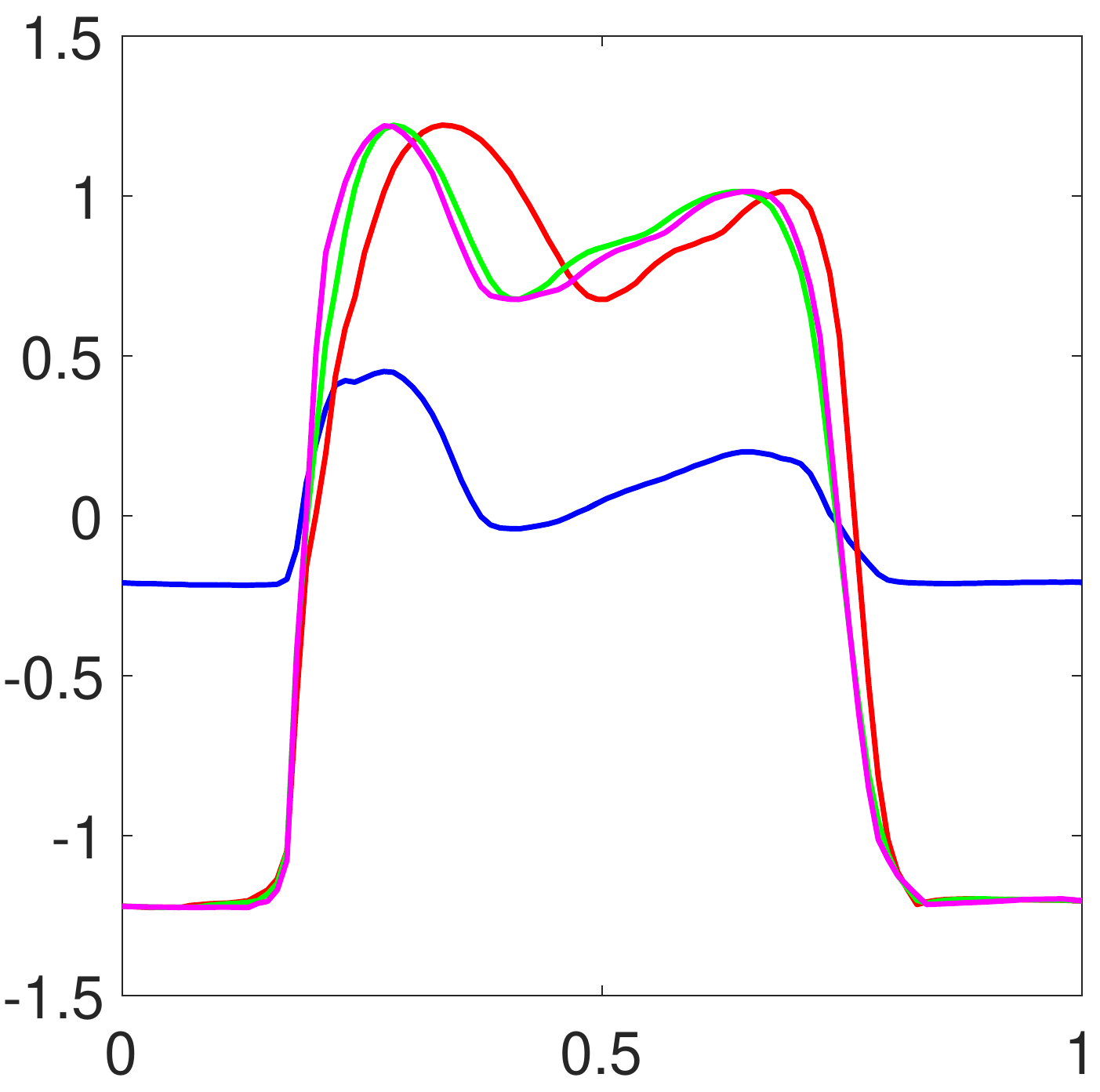}&\includegraphics[width=1.2in]{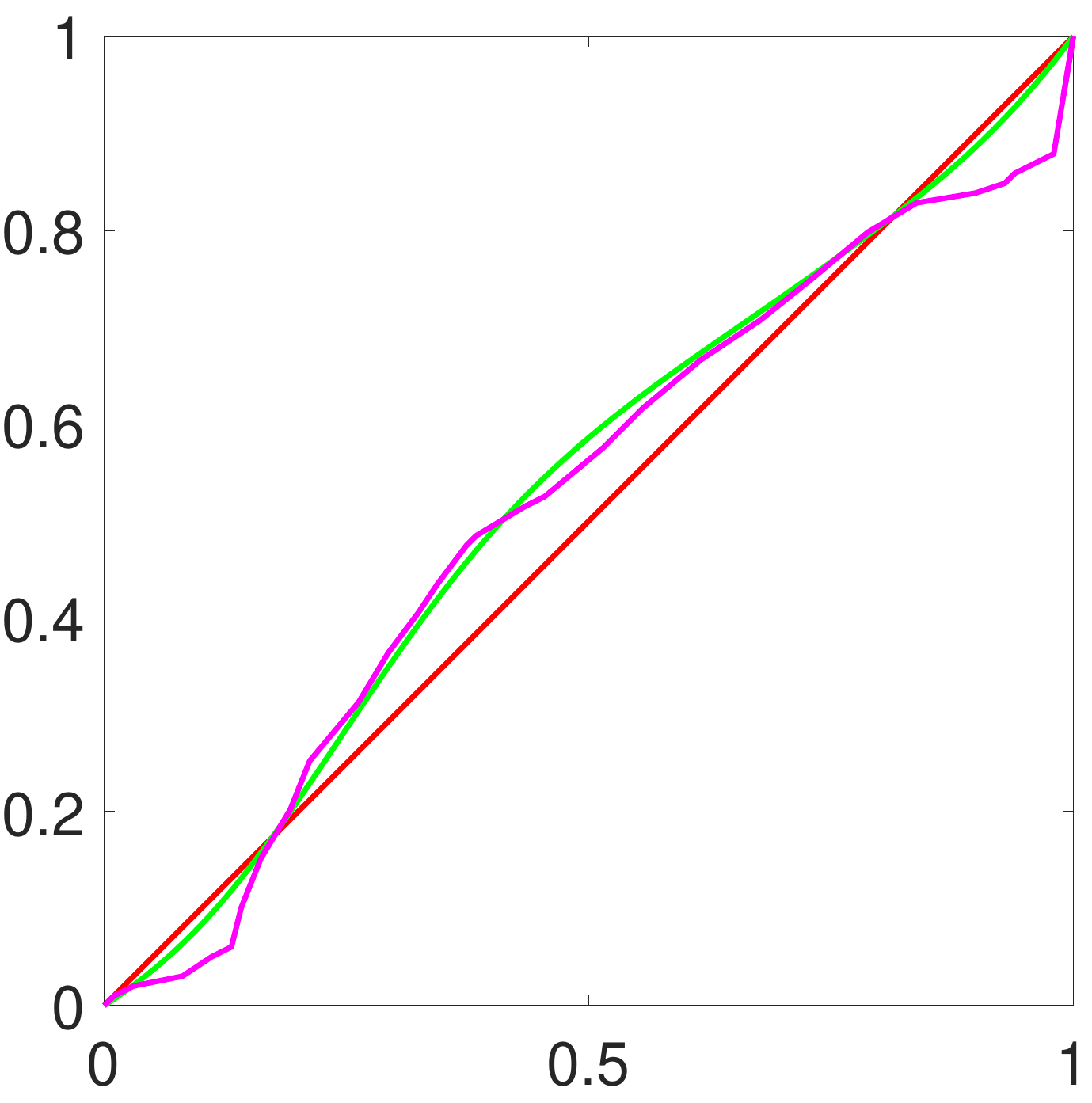}&\includegraphics[width=1.2in]{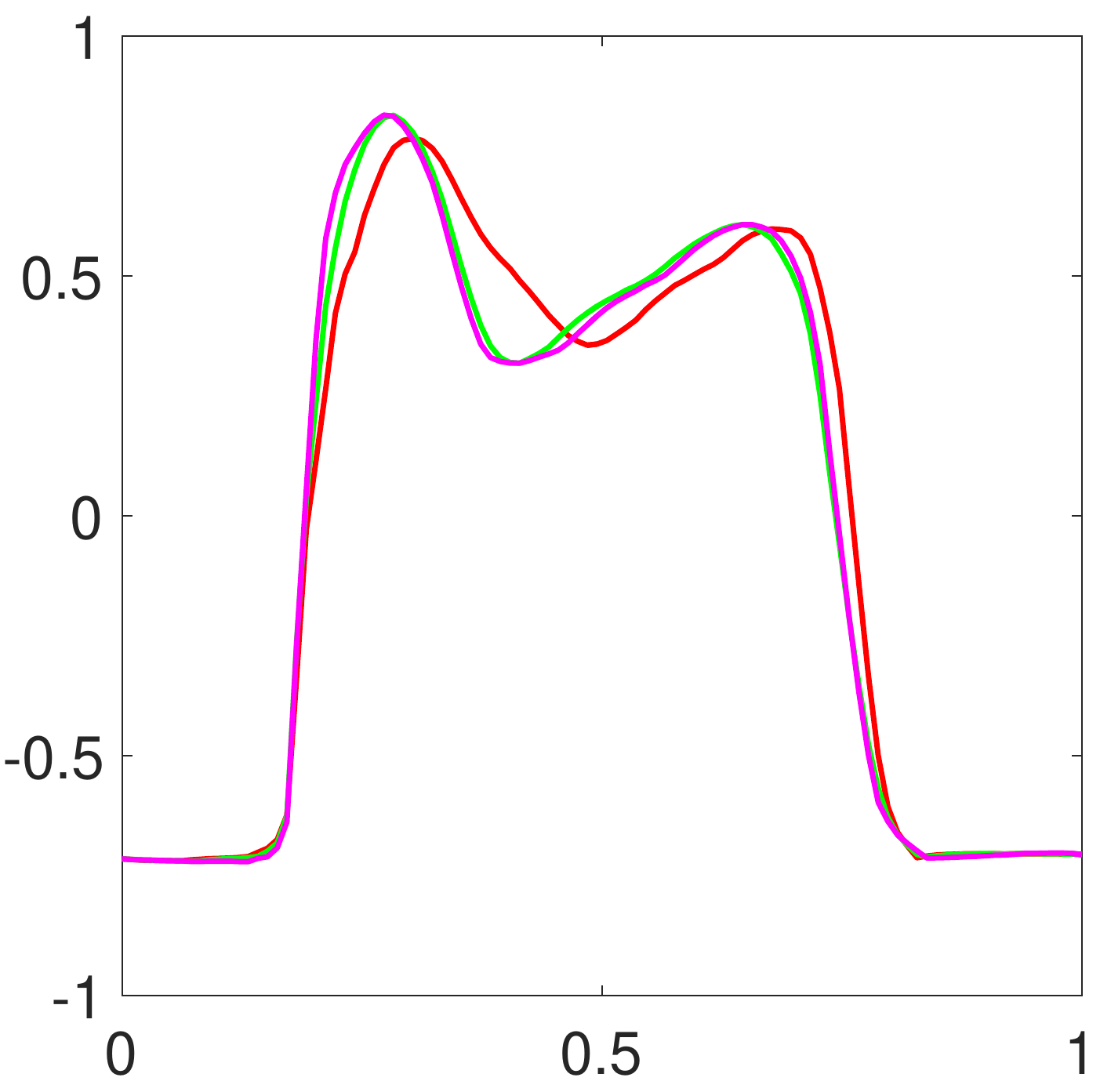}\\
\hline
\end{tabular}
\begin{tabular}{|c|c|}
(d)&(e)\\
\hline
\includegraphics[width=1.2in]{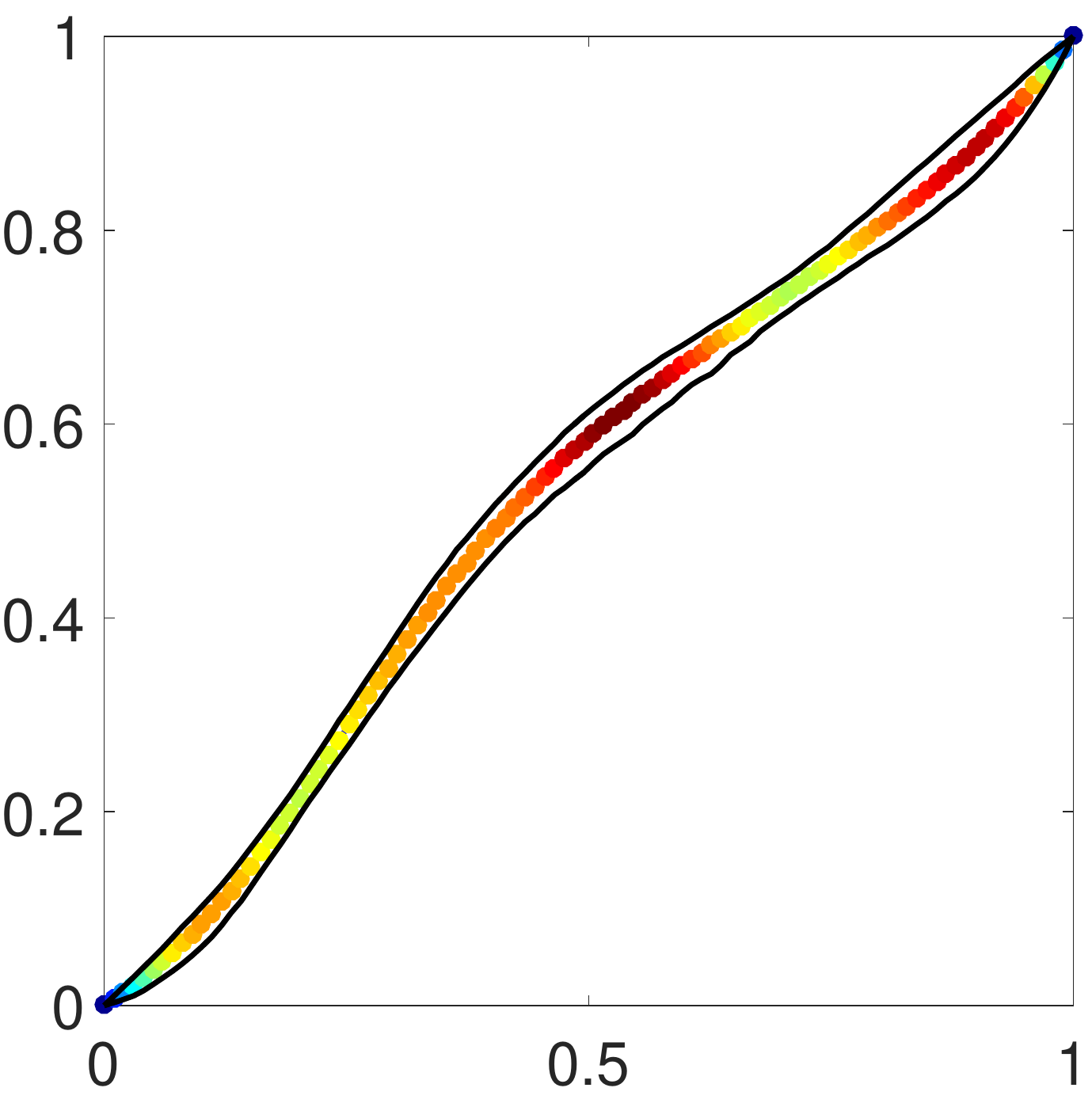}&\includegraphics[width=1.2in]{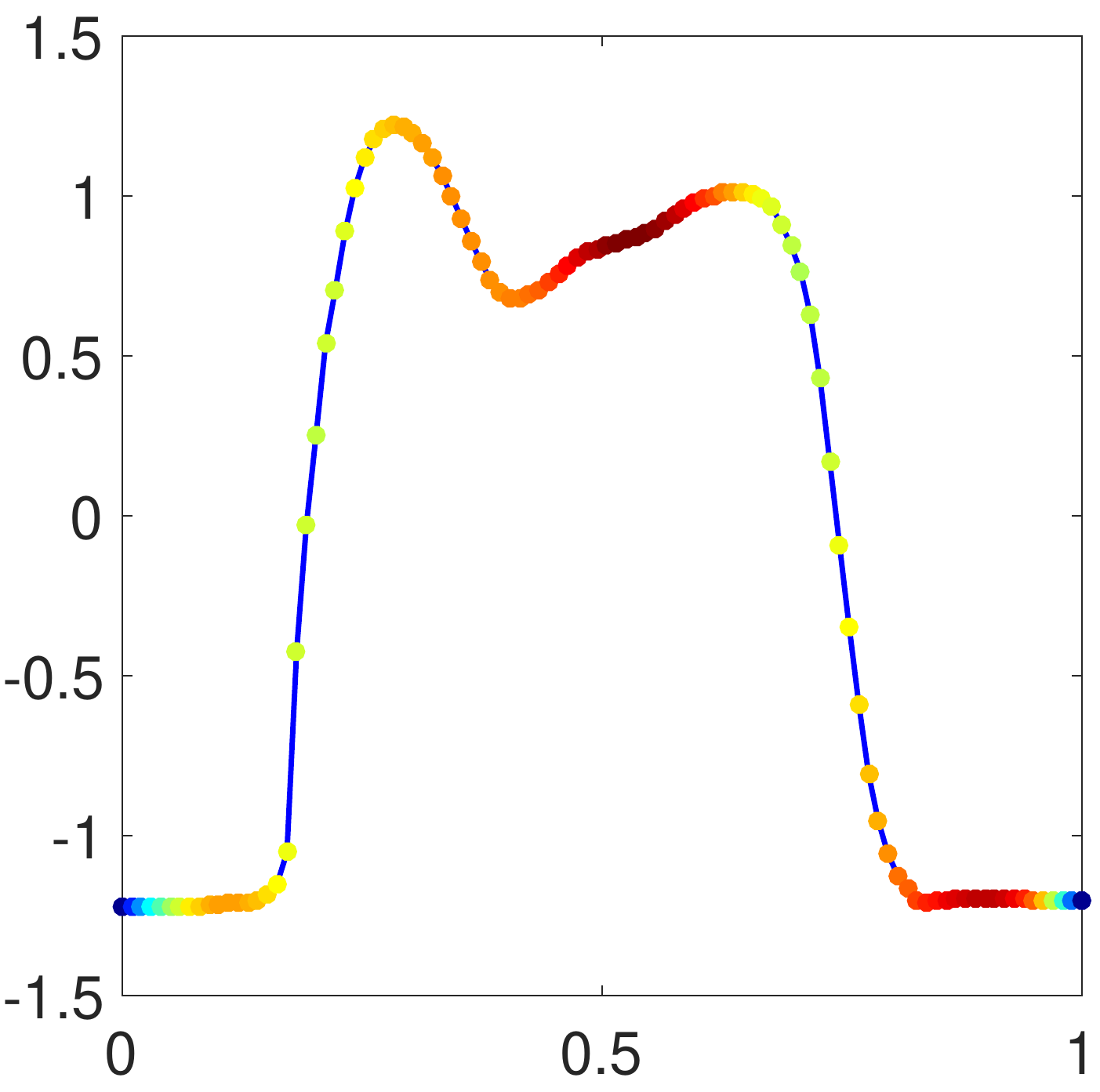}\\
\hline
\end{tabular}
\caption{Pairwise alignment of two gait pressure functions. (a) Original functions $f_1$ and $f_2$ in blue and red, respectively; $f_2\circ\gamma_{DP}$ in magenta and $f_2\circ\bar{\gamma}$ in green. (b) $\gamma_{DP}$ in magenta, $\gamma_{id}$ in red, and $\bar{\gamma}$ in green. (c) Pointwise average of $f_1$ and $f_2$ for each alignment result (colored in the same way as (a) and (b)). (d) Pointwise standard deviation (hot colors correspond to higher values) plotted on $\bar{\gamma}$, and the $95\%$ credible interval in black. (e) Pointwise standard deviation (hot colors correspond to higher values) plotted on $f_2\circ\bar{\gamma}$.} \label{fig:ex3gait}
\end{center}
\end{figure}

\subsection{Biomedical Signals}

We describe several alignment examples for biomedical signals. In all of the presented datasets, the functions must first be properly registered to align important features across the functional observations. At times, due to significant structural differences, registration ambiguities result in multiple plausible alignments, which cannot be detected using optimization-based registration algorithms. This is especially seen in the gait pressure functional data, which we consider in the first set of examples.

\textbf{Gait pressure functions:} We begin with three examples of pairwise alignment of gait pressure functions. In the first example, shown in Figure \ref{fig:ex1gait}, we discover three modes in the posterior distribution. Panel (a) displays the registration results using the mean warping function in each cluster. The functions to be registered are plotted in blue and red. The deterministic dynamic programming solution is given in magenta, and the three clusterwise alignments are shown in green (cluster 1), black (cluster 2) and cyan (cluster 3). The corresponding avereage warping functions are shown in panel (b) with the identity element in red. Panel (c) displays the pointwise average of the two functions for each registration result. First, the cluster 3 alignment is nearly identical to the dynamic programming solution. The main benefit of the proposed Bayesian approach is the discovery of two other plausible alignments. The cluster 1 alignment emphasizes the first mode in the gait functions (green average in panel (c)) while the cluster 2 registration weights both modes equally (black average in panel (c)). The cluster 3 alignment as well as the dynamic programming solution emphasize both modes as well as the large dip toward the midpoint of the gait cycle. Panels (d)-(i) show the uncertainty in each cluster using two displays: (1) pointwise standard deviation as a color (blue to red=low to high) on the mean warping as well as the pointwise $95\%$ credible interval in black, and (2) pointwise standard deviation as a color on the warped version of the second function. We usually observe lower standard deviation along the pronounced features such as the steep increase and decrease in pressure at the beginning and end of the gait cycle. On the other hand, the standard deviation is inflated in flat regions where many types of warping provide a satisfactory solution.

The second example is displayed in Figure \ref{fig:ex2gait}. In this case, we find two modes in the posterior distribution and display the same set of results as for the first example. The results are similar as in the previous case where different modes of the pressure functions are emphasized in each cluster. Again, the cluster 2 mean is very similar to the dynamic programming solution. Importantly, the result based on the proposed Bayesian model is always much smoother while achieving very similar alignment. Finally, in Figure \ref{fig:ex3gait}, we display an example where the posterior distribution of warping functions is unimodal. In this case, the two functions to be aligned have two very clear gait pressure modes, and thus, there is little uncertainty in the registration.

\begin{figure}[!t]
\begin{center}
\begin{tabular}{|c|c|c|}
\hline
(a)&(b)&(c)\\
\hline
\includegraphics[width=1.2in]{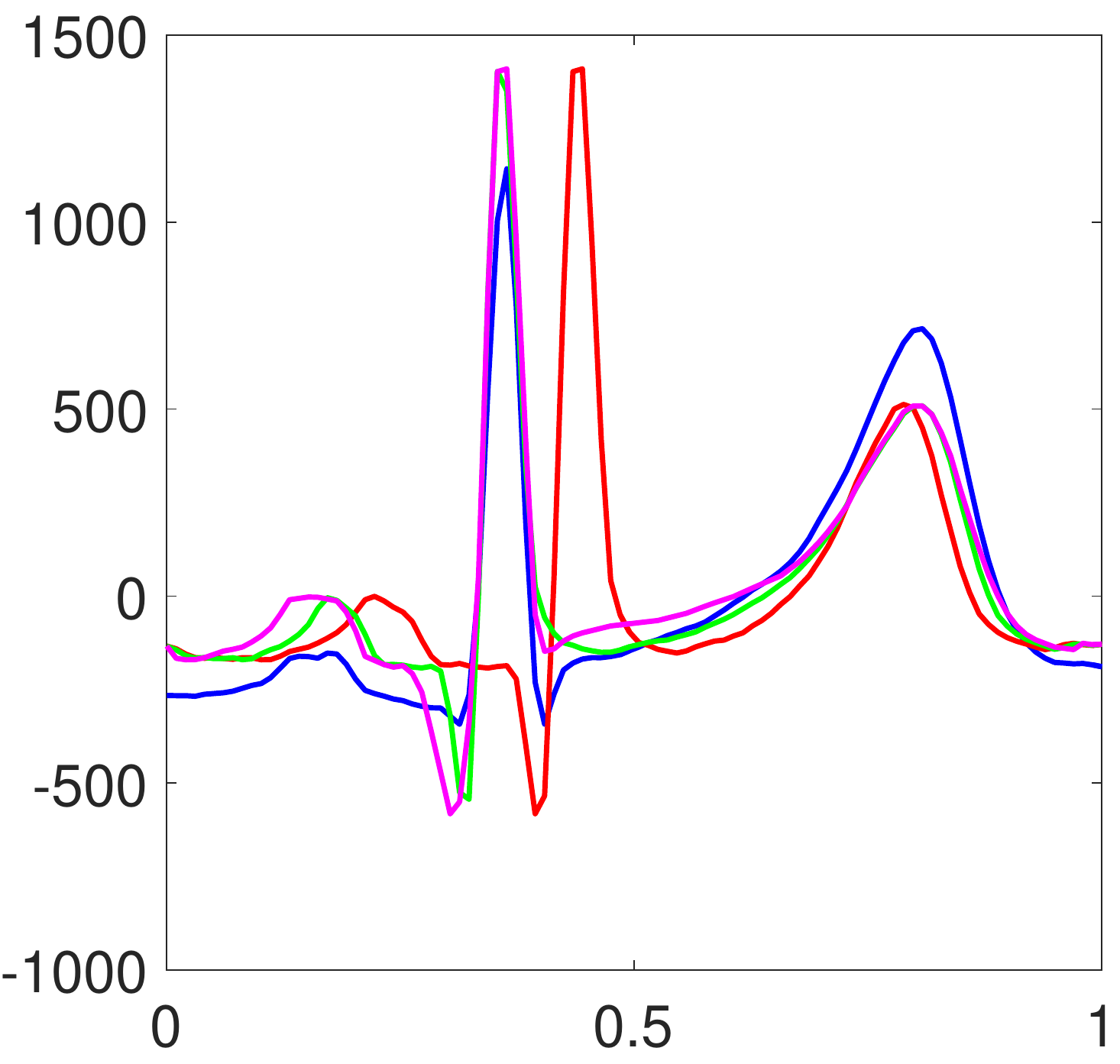}&\includegraphics[width=1.2in]{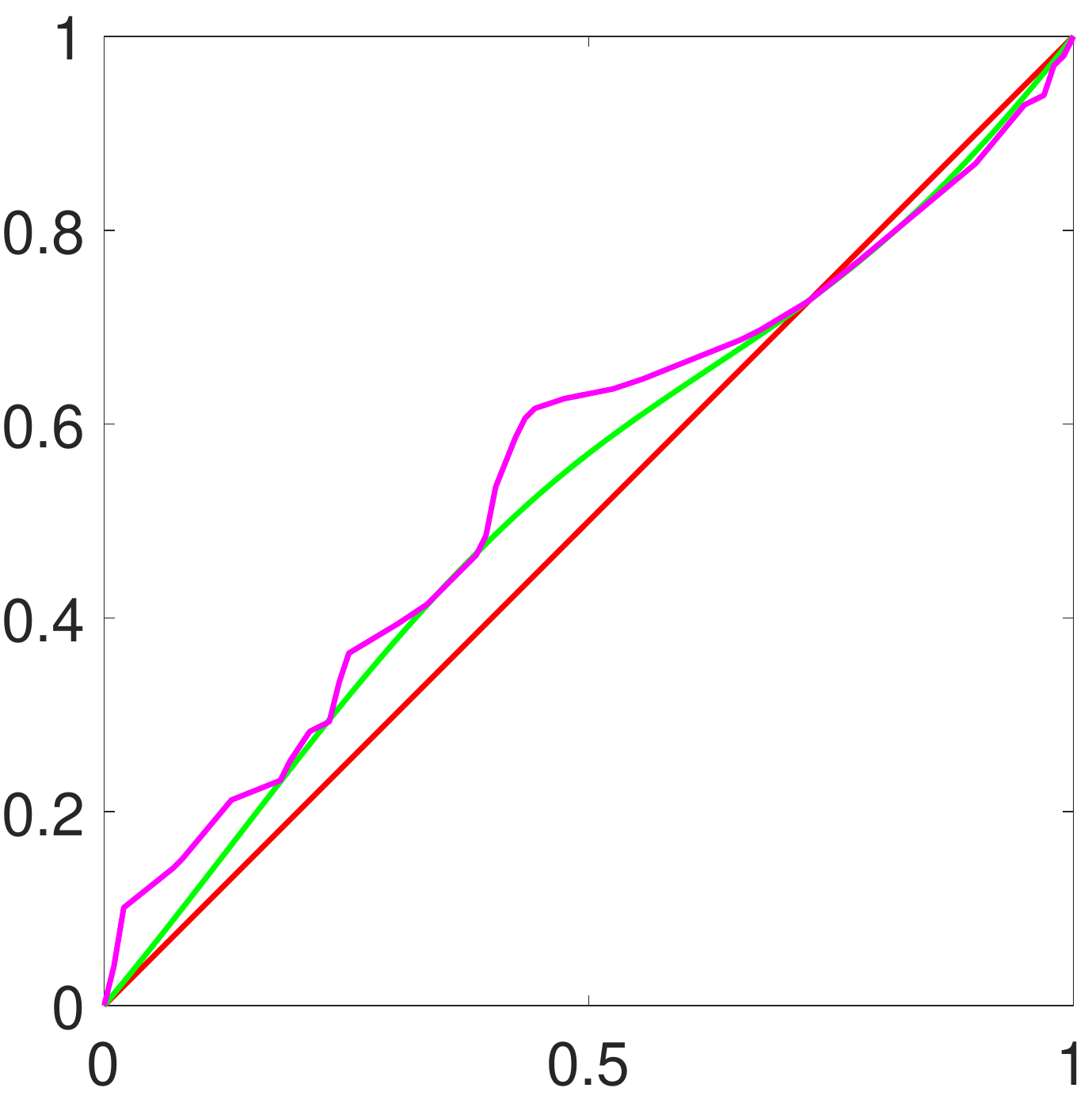}&\includegraphics[width=1.2in]{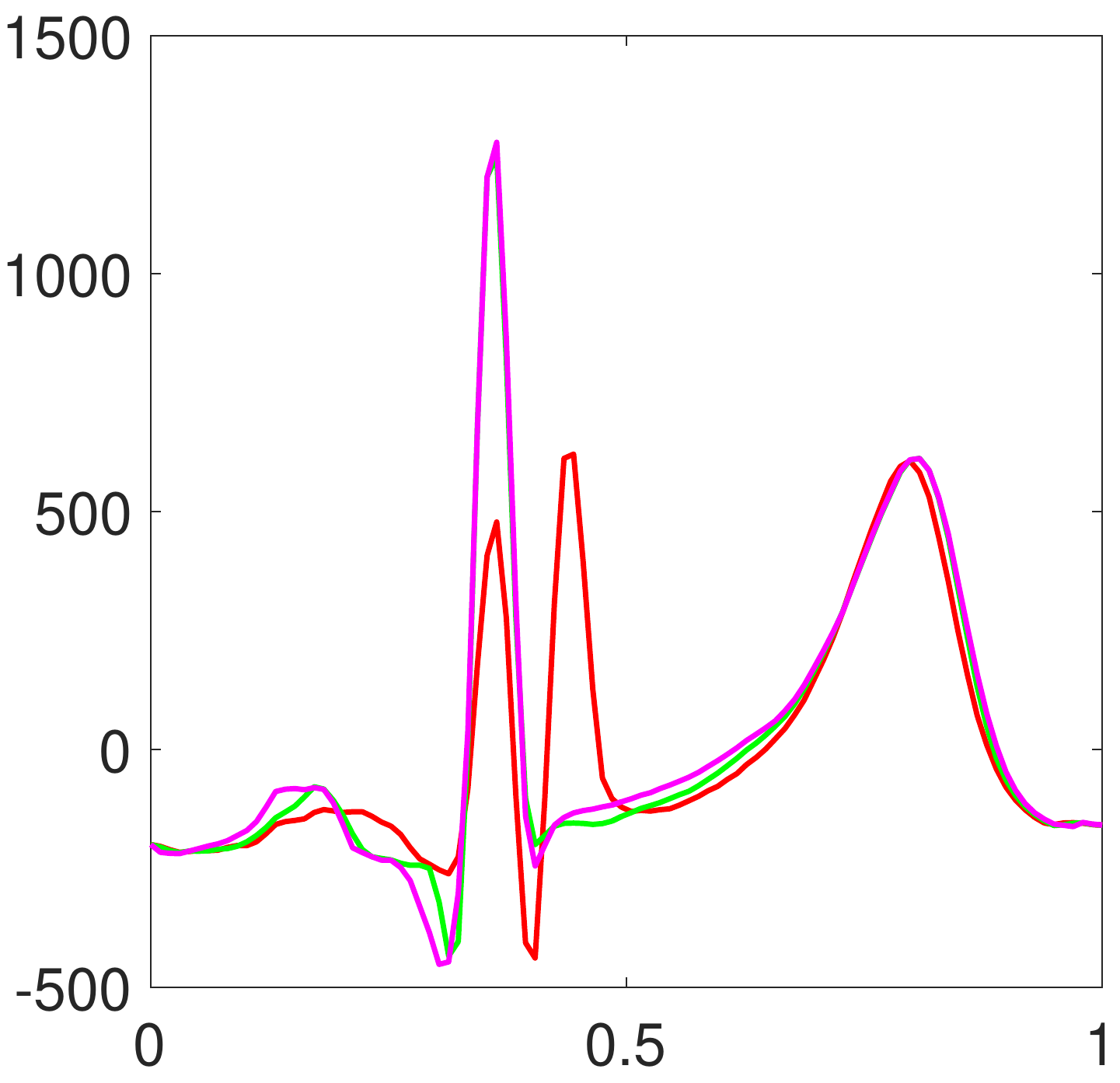}\\
\hline
\end{tabular}
\begin{tabular}{|c|c|}
(d)&(e)\\
\hline
\includegraphics[width=1.2in]{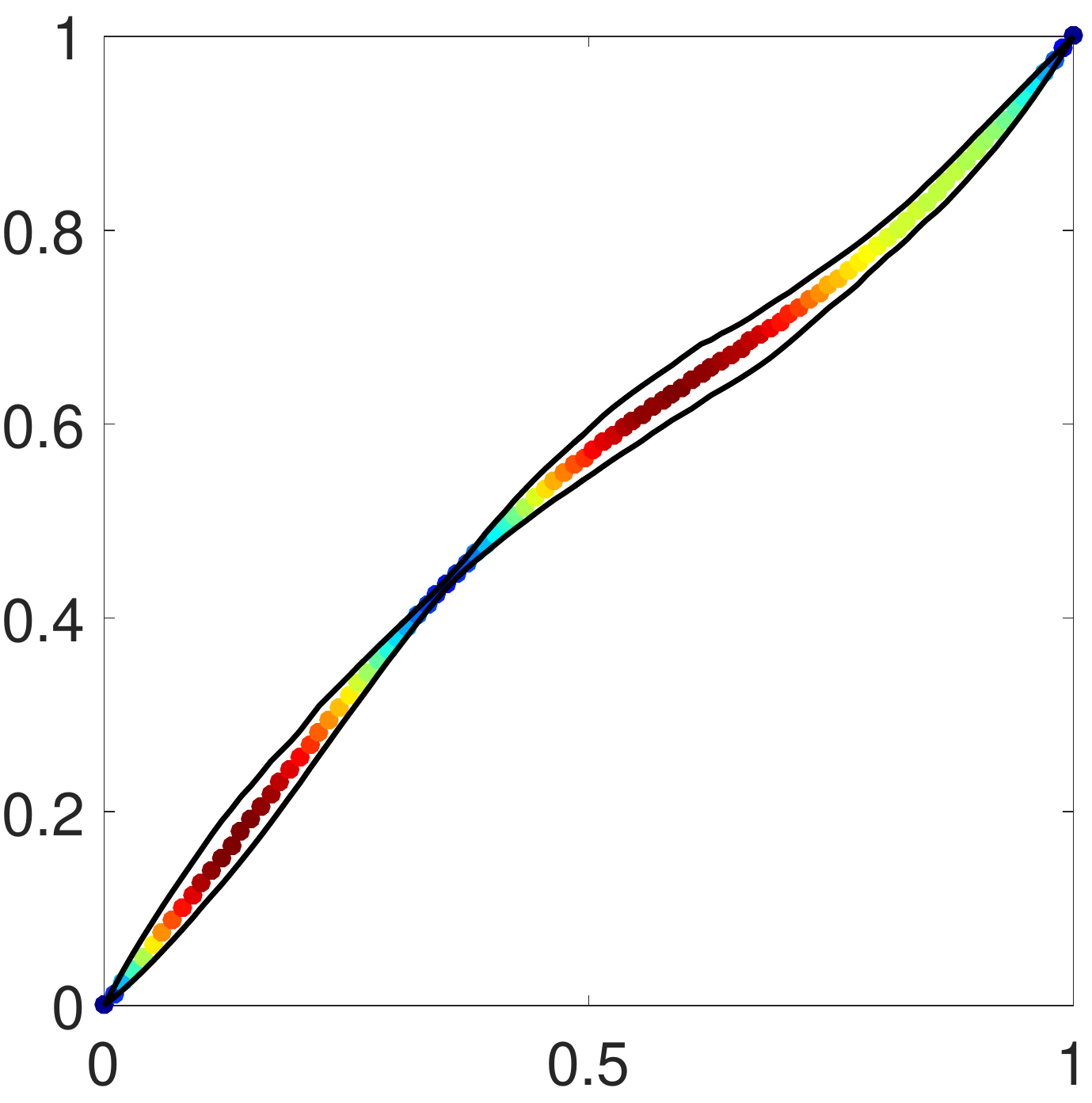}&\includegraphics[width=1.2in]{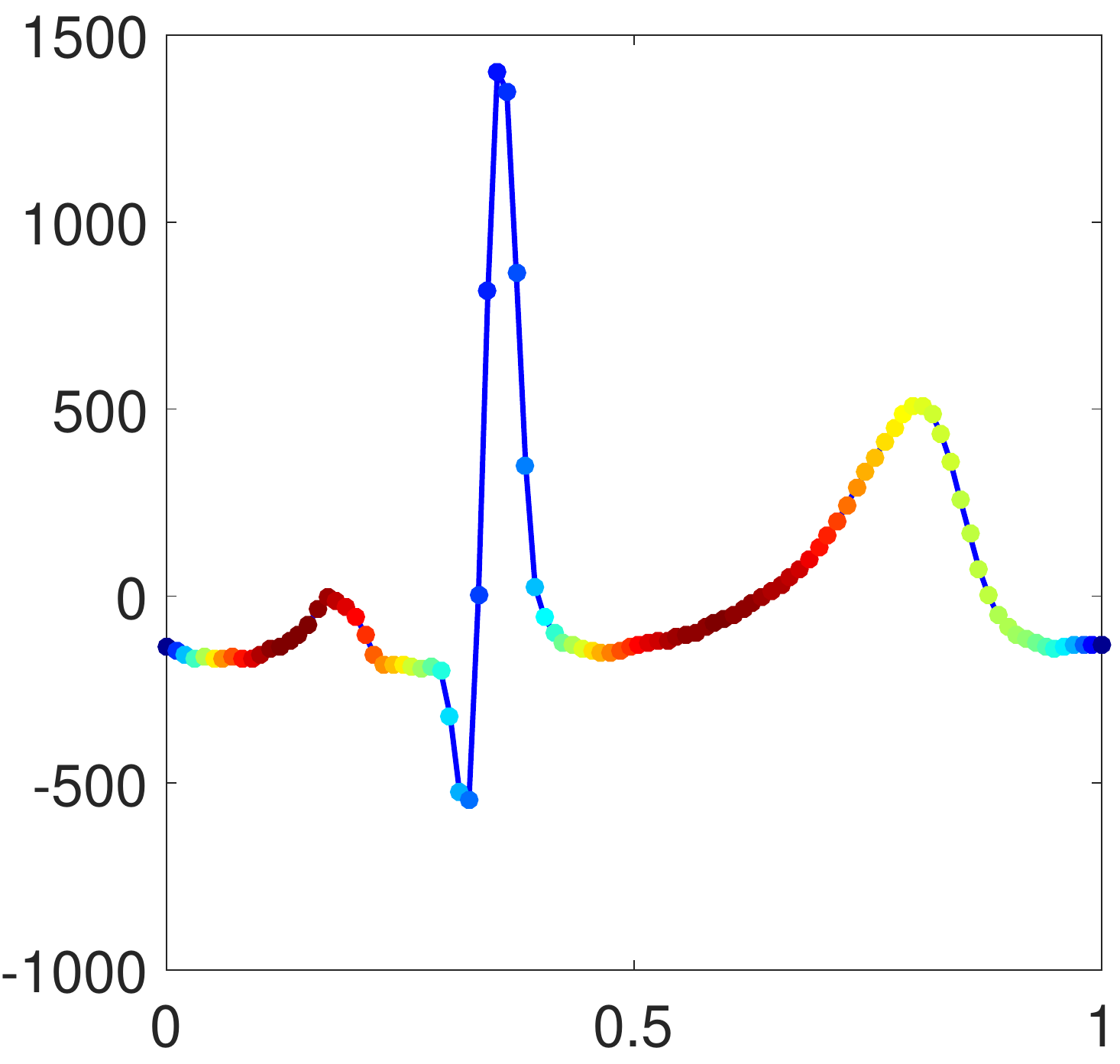}\\
\hline
\end{tabular}
\caption{Pairwise alignment of two PQRST complexes. (a) Original functions $f_1$ and $f_2$ in blue and red, respectively; $f_2\circ\gamma_{DP}$ in magenta and $f_2\circ\bar{\gamma}$ in green. (b) $\gamma_{DP}$ in magenta, $\gamma_{id}$ in red, and $\bar{\gamma}$ in green. (c) Pointwise average of $f_1$ and $f_2$ for each alignment result (colored in the same way as (a) and (b)). (d) Pointwise standard deviation (hot colors correspond to higher values) plotted on $\bar{\gamma}$, and the $95\%$ credible interval in black. (e) Pointwise standard deviation (hot colors correspond to higher values) plotted on $f_2\circ\bar{\gamma}$.} \label{fig:ex1ECG}
\end{center}
\end{figure}

\begin{figure}[!t]
\begin{center}
\begin{tabular}{|c|c|c|}
\hline
(a)&(b)&(c)\\
\hline
\includegraphics[width=1.2in]{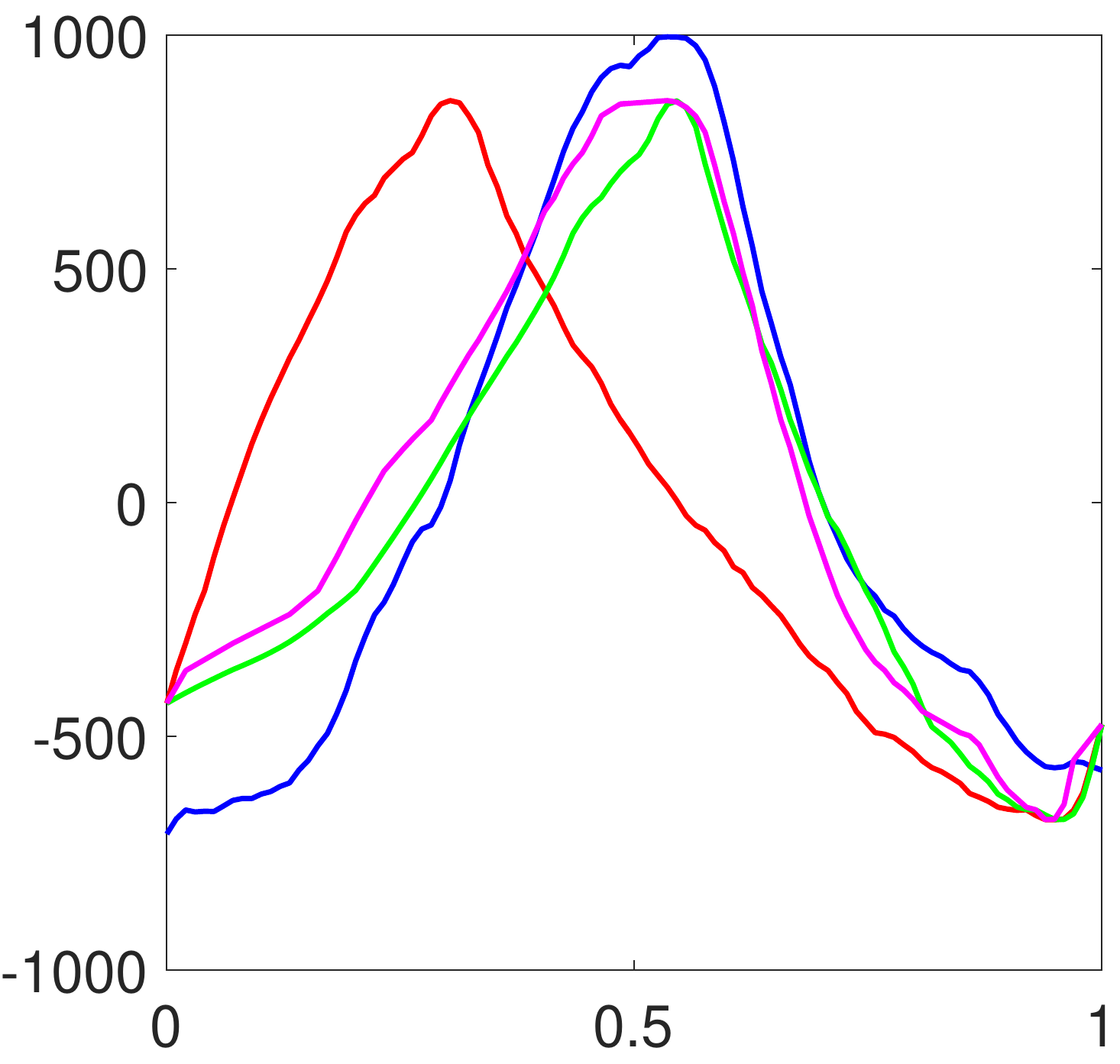}&\includegraphics[width=1.2in]{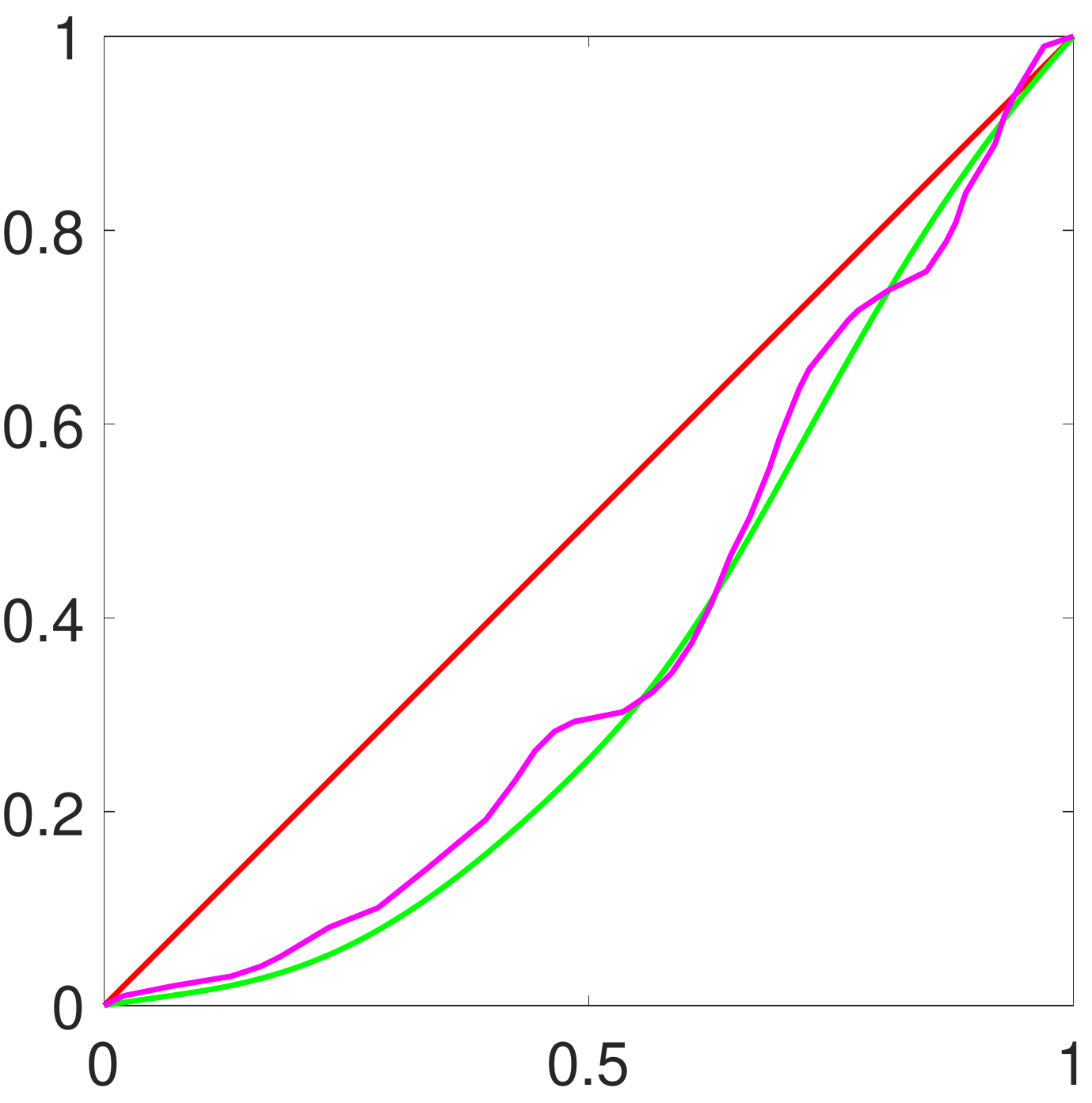}&\includegraphics[width=1.2in]{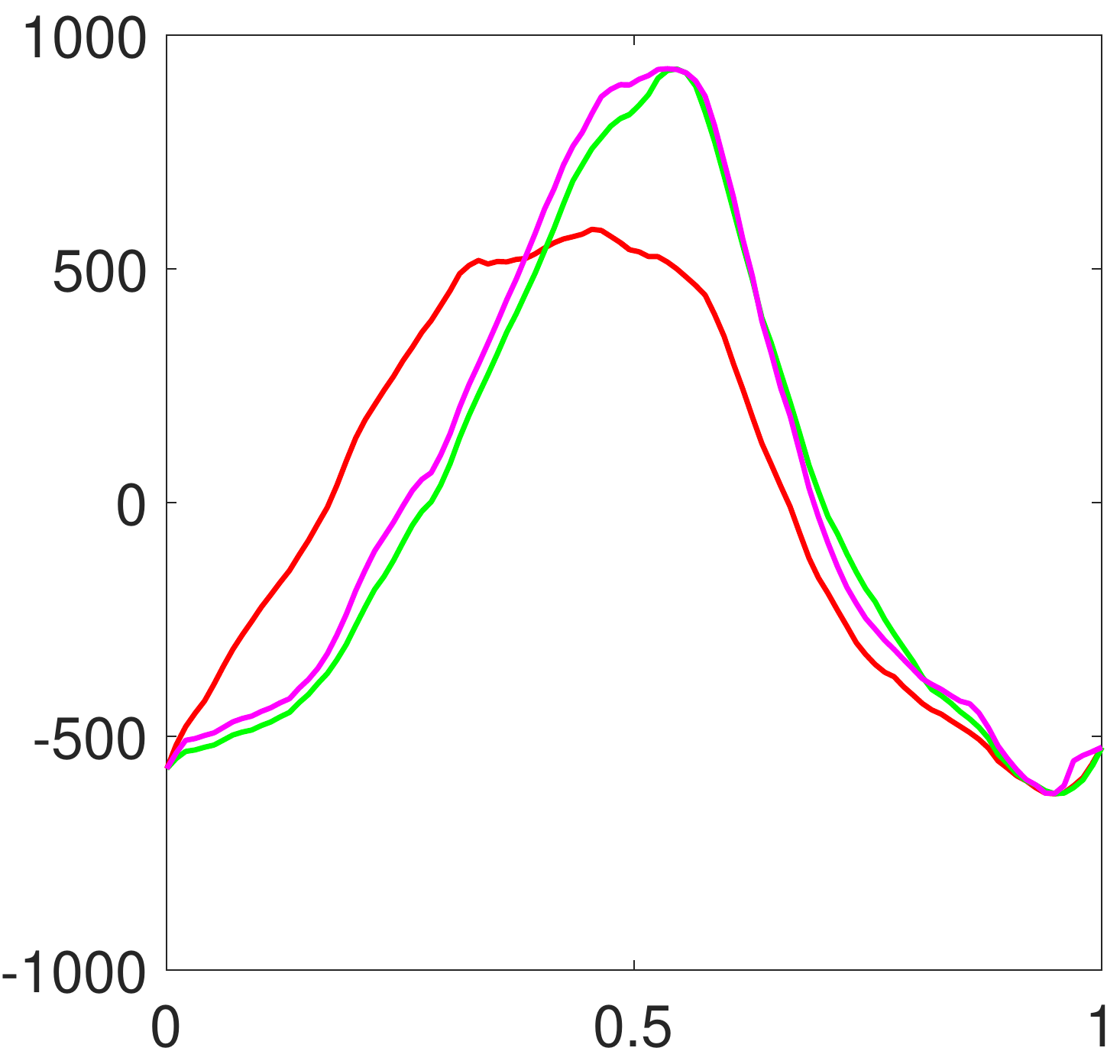}\\
\hline
\end{tabular}
\begin{tabular}{|c|c|}
(d)&(e)\\
\hline
\includegraphics[width=1.2in]{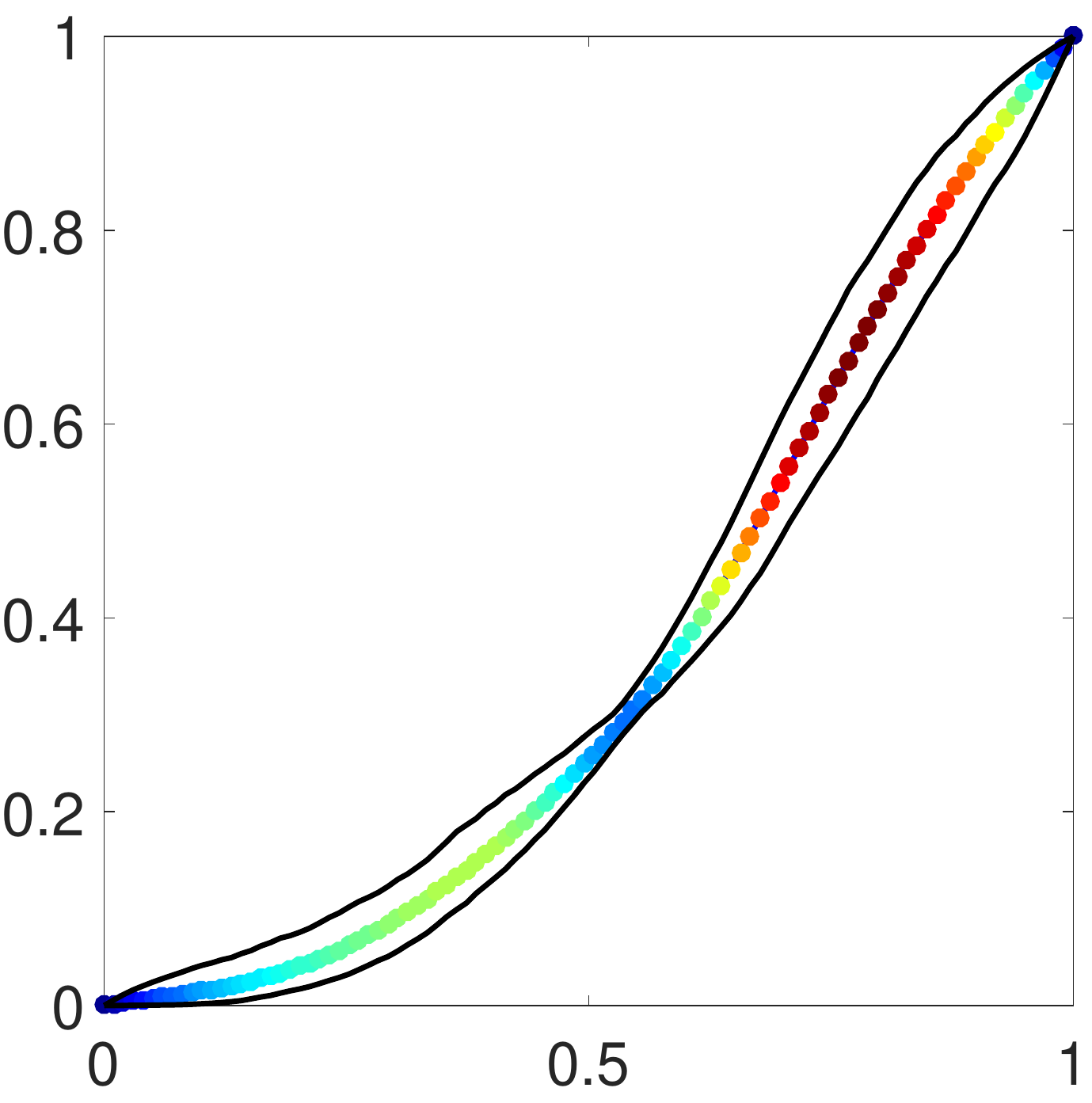}&\includegraphics[width=1.2in]{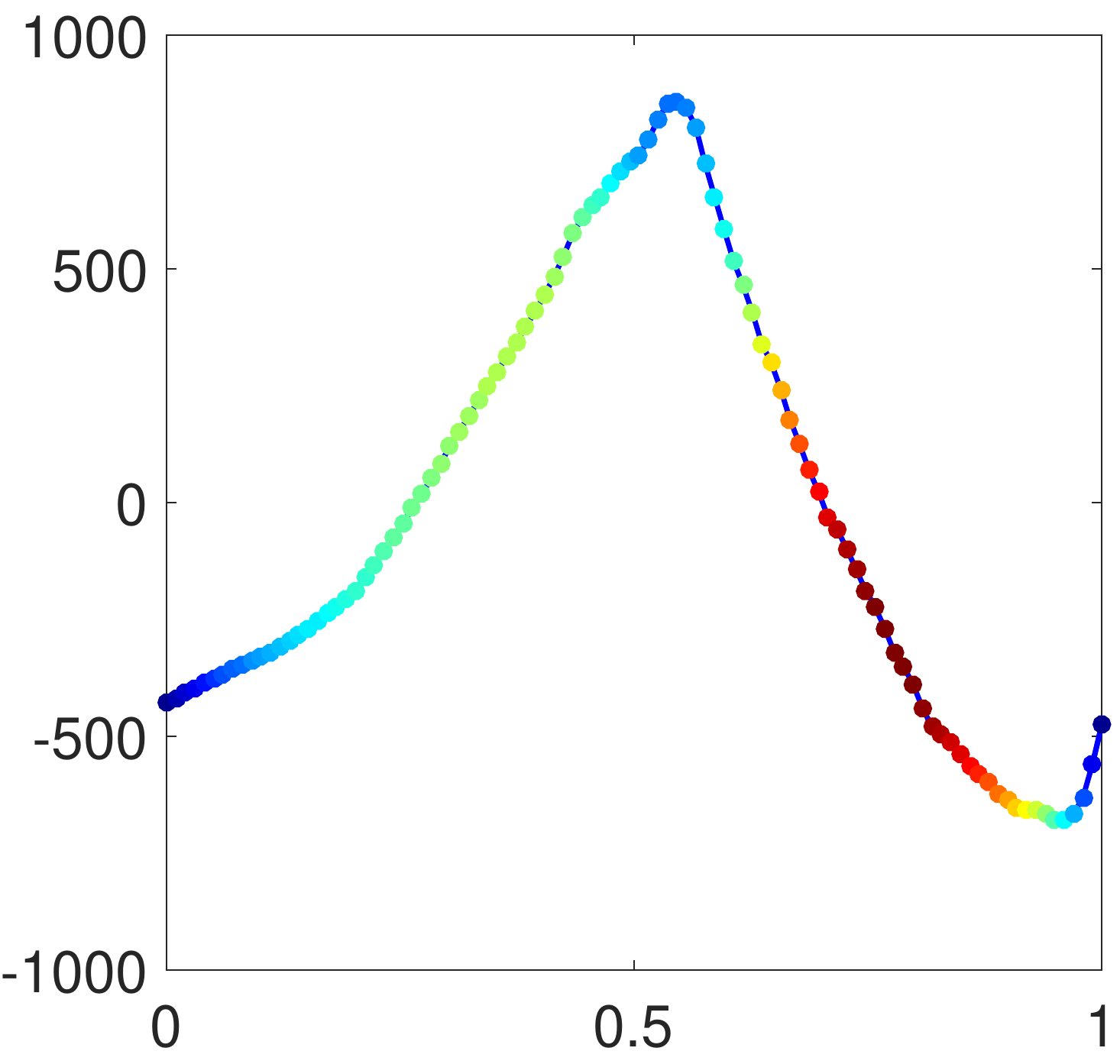}\\
\hline
\end{tabular}
\caption{Pairwise alignment of two respiration functions. (a) Original functions $f_1$ and $f_2$ in blue and red, respectively; $f_2\circ\gamma_{DP}$ in magenta and $f_2\circ\bar{\gamma}$ in green. (b) $\gamma_{DP}$ in magenta, $\gamma_{id}$ in red, and $\bar{\gamma}$ in green. (c) Pointwise average of $f_1$ and $f_2$ for each alignment result (colored in the same way as (a) and (b)). (d) Pointwise standard deviation (hot colors correspond to higher values) plotted on $\bar{\gamma}$, and the $95\%$ credible interval in black. (e) Pointwise standard deviation (hot colors correspond to higher values) plotted on $f_2\circ\bar{\gamma}$.} \label{fig:ex1resp}
\end{center}
\end{figure}

\textbf{PQRST complexes:} The PQRST complex in ECG refers to the first peak (P wave), the sharp second peak (QRS complex), and the third peak (T wave). These functions have very pronounced features, and thus, most of the pairwise alignment results on this data yield a unimodal posterior distribution. We display one example of such an alignment in Figure \ref{fig:ex1ECG}. The posterior mean warping is very similar to the dynamic programming solution, albeit smoother. Also, there is very little registration uncertainty around the QRS complex. Alignment uncertainty is also low at the T wave, which is much more pronounced than the P wave in this example. The red (no warping) pointwise average of the two PQRST complexes displayed in panel (c) is clearly not a valid PQRST complex. As a result, warping in this application is necessary to obtain reasonable functional summaries.

\textbf{Respiration data:} Each function in this dataset represents lung volume during a breathing cycle. Respiration cycle alignment is important for understanding breathing variation as well as radiotherapy in lung cancer \cite{biosignals}. In this application, the posterior distribution of warping functions is also almost always unimodal due to the very simple structure of each breathing cycle. Figure \ref{fig:ex1resp} displays one example of pairwise alignment of two such respiration functions. Again, the produced posterior mean alignment is very good, with little uncertainty in the area of the peak of the breathing cycle.

\subsection{Berkeley Growth Velocity Data}

\begin{figure}[!t]
\begin{center}
\begin{tabular}{|c|c|c|}
\hline
(a)&(b)&(c)\\
\hline
\includegraphics[width=1.2in]{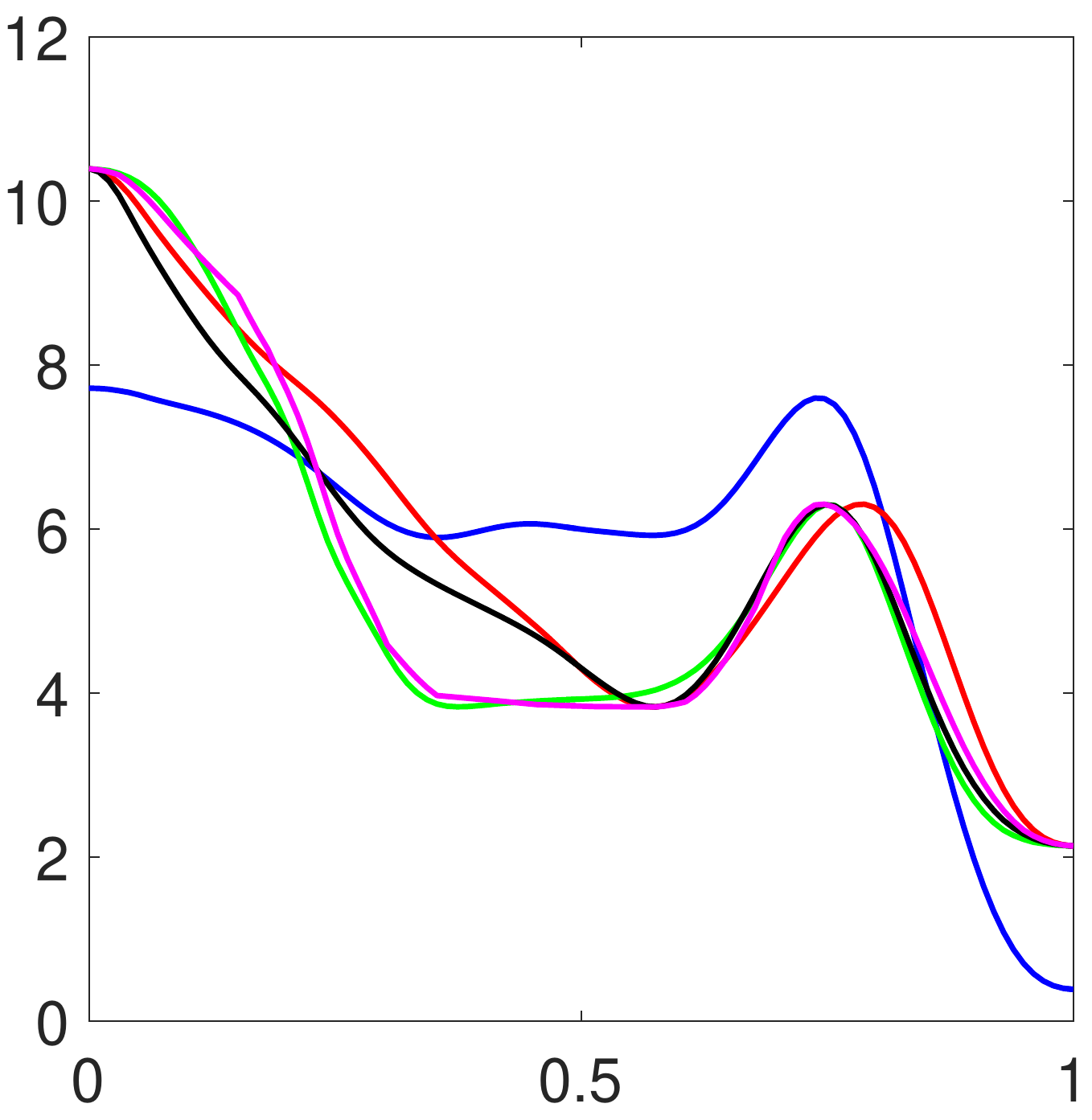}&\includegraphics[width=1.2in]{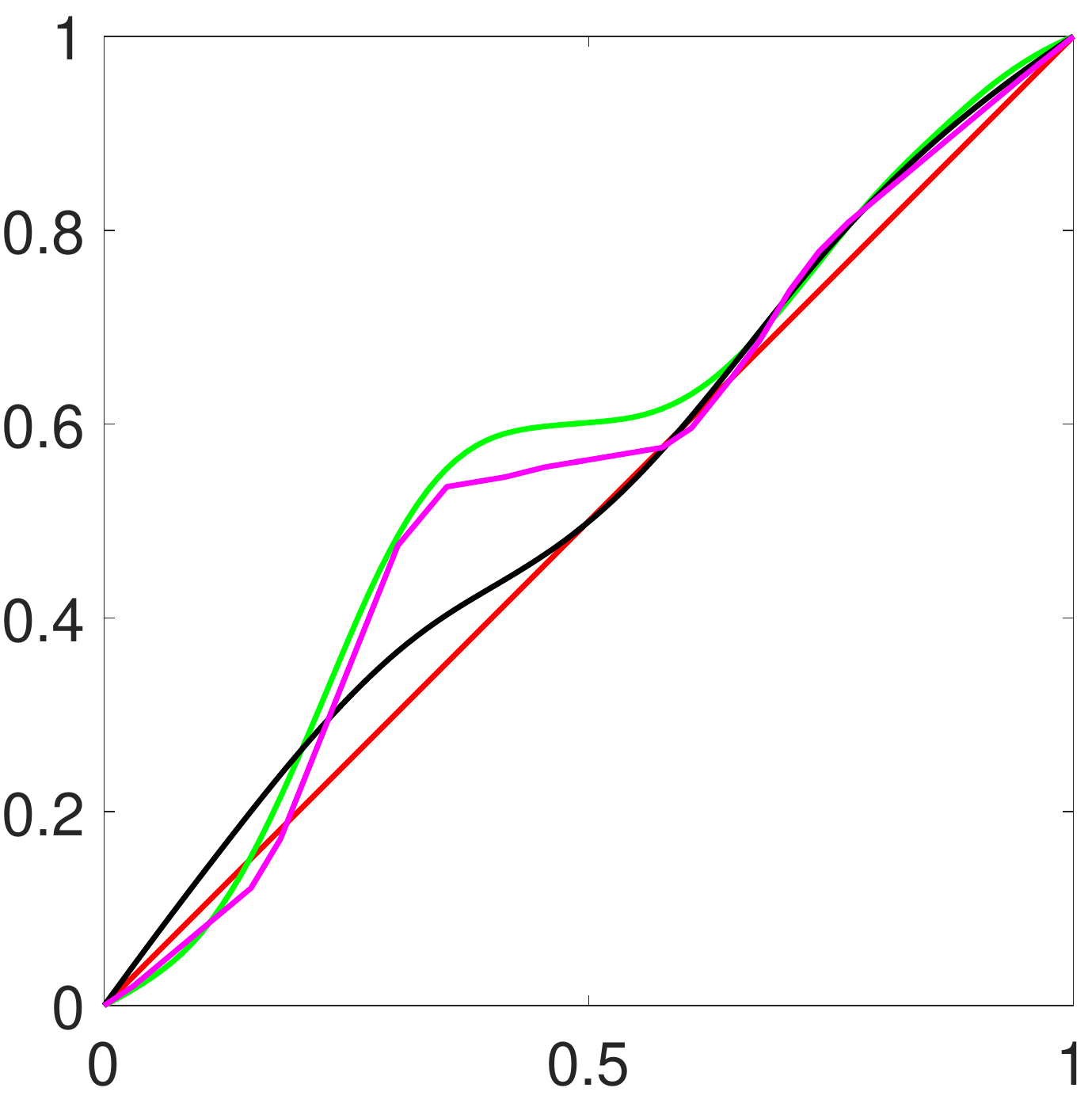}&\includegraphics[width=1.2in]{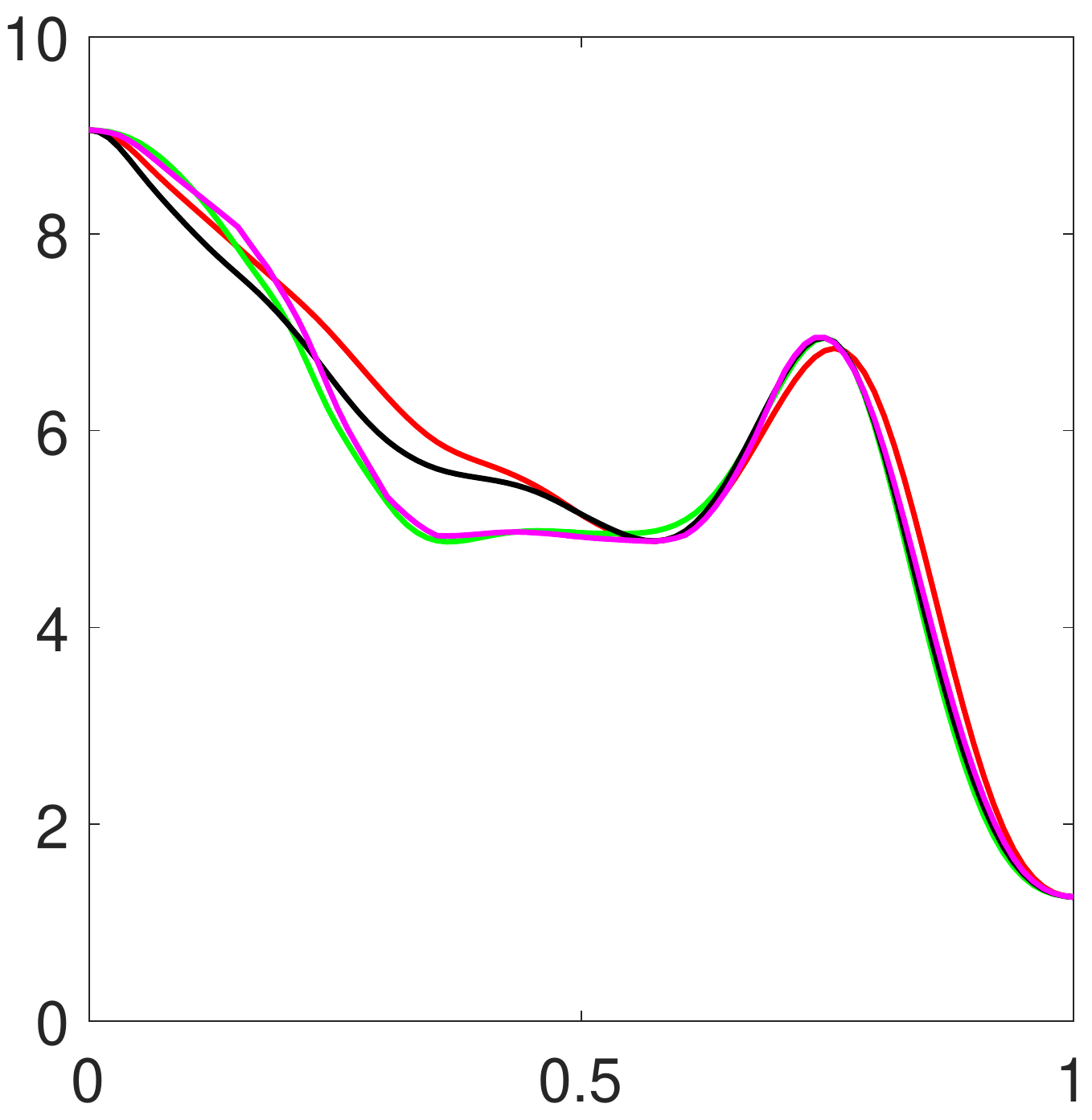}\\
\end{tabular}
\begin{tabular}{|c|c|c|c|}
\hline
(d)&(e)&(f)&(g)\\
\hline
\includegraphics[width=.9in]{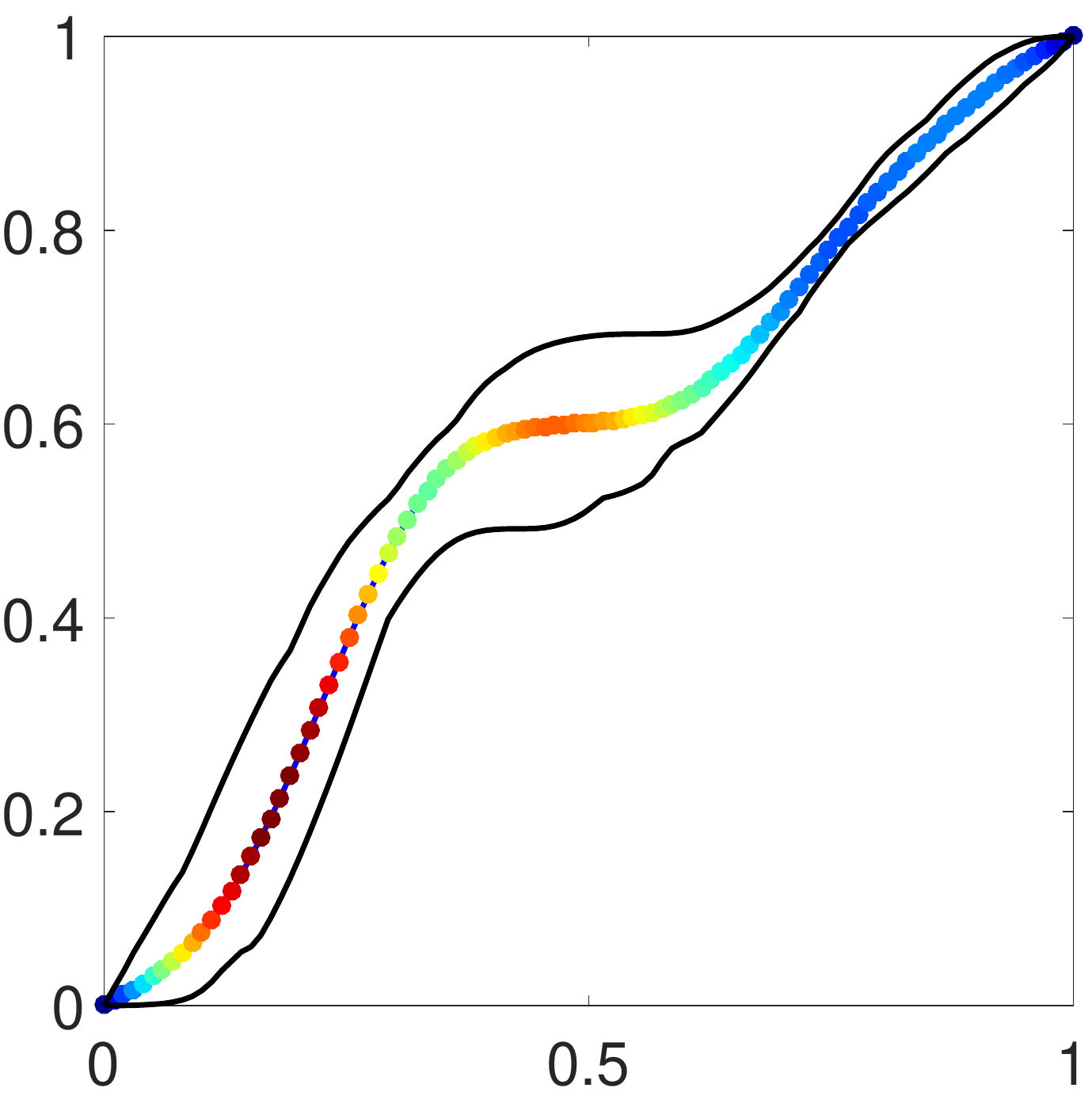}&\includegraphics[width=.9in]{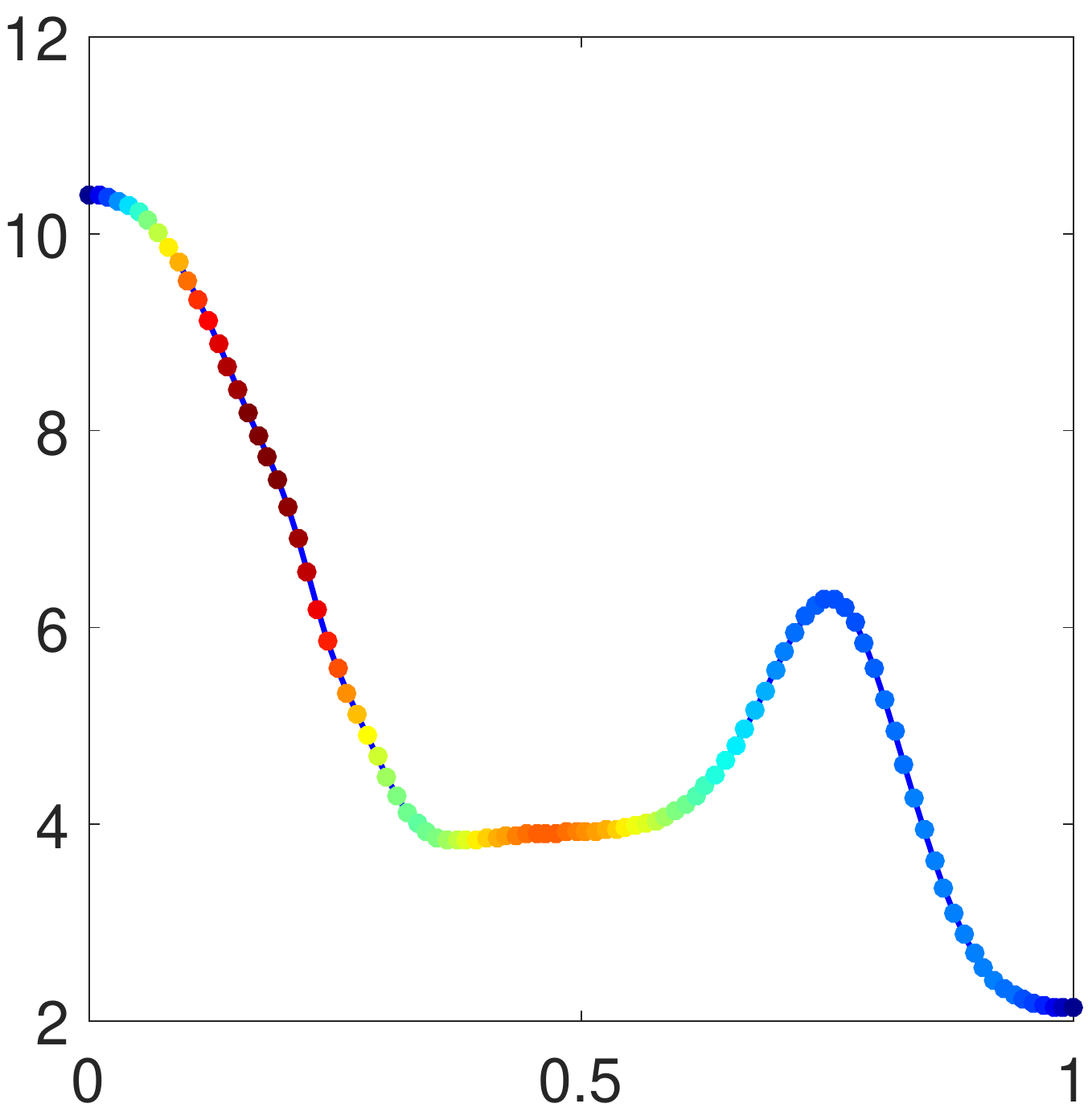}&\includegraphics[width=.9in]{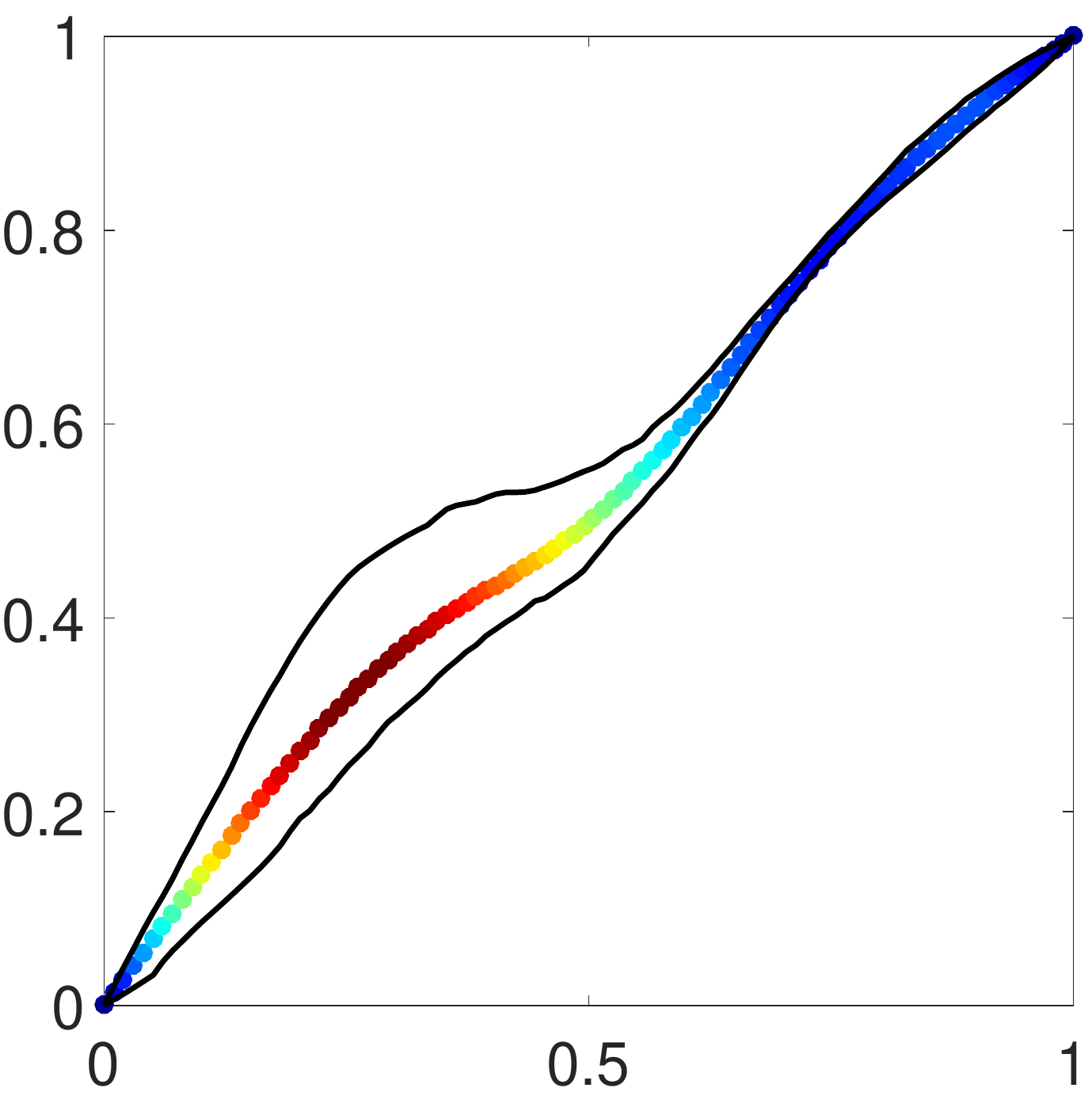}&\includegraphics[width=.9in]{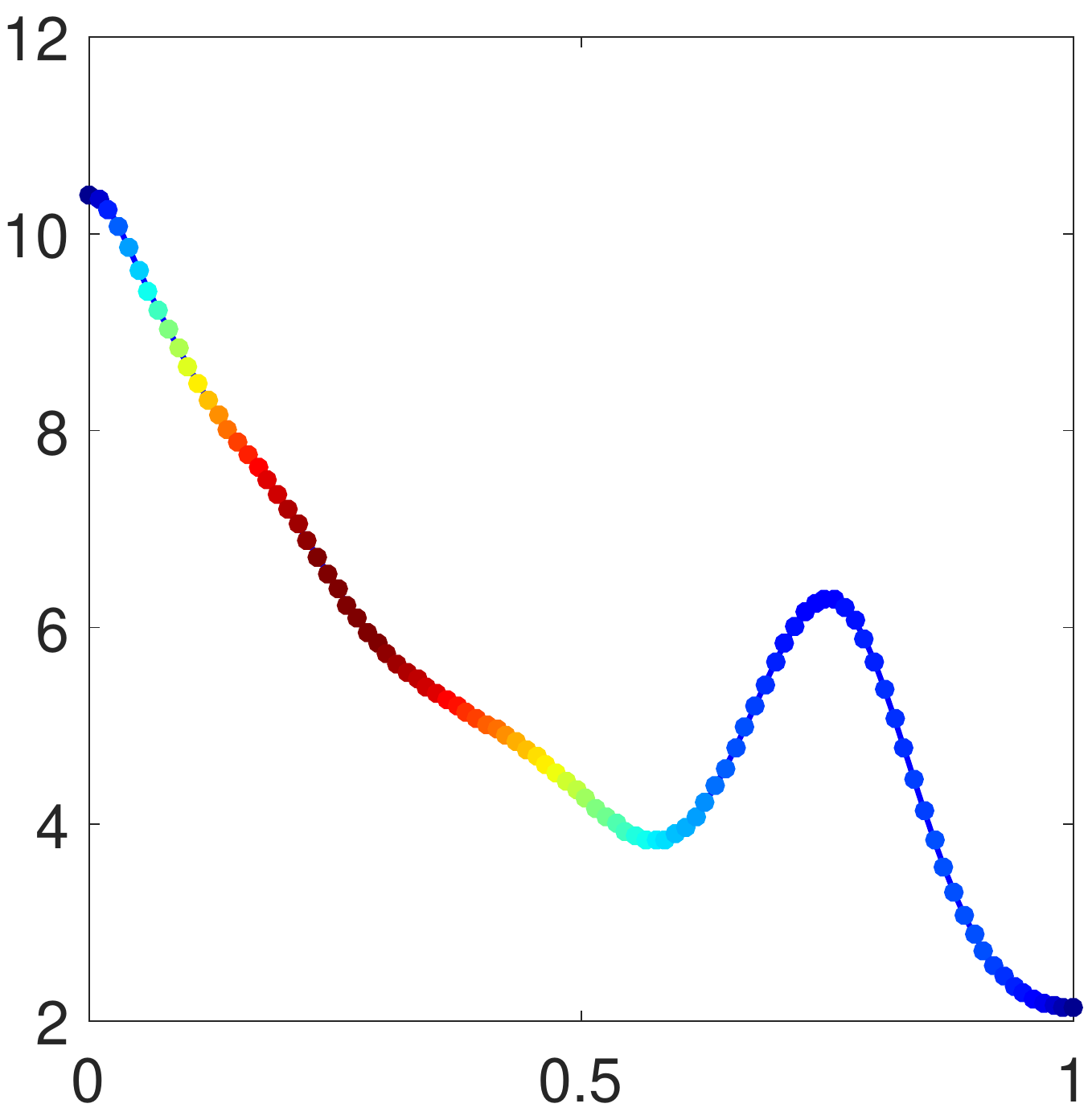}\\
\hline
\end{tabular}
\begin{tabular}{|c|c|c|}
\hline
(a)&(b)&(c)\\
\hline
\includegraphics[width=1.2in]{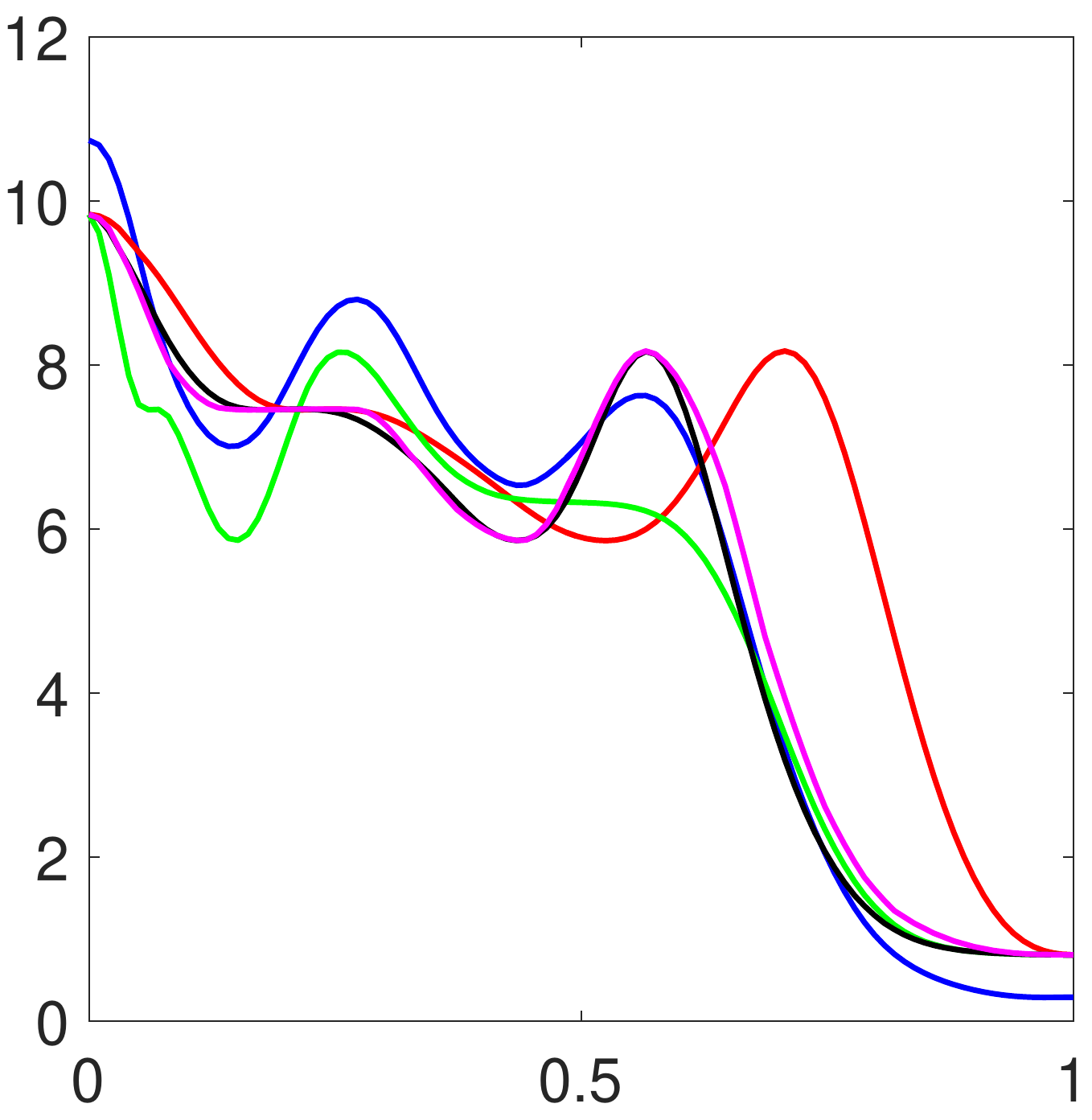}&\includegraphics[width=1.2in]{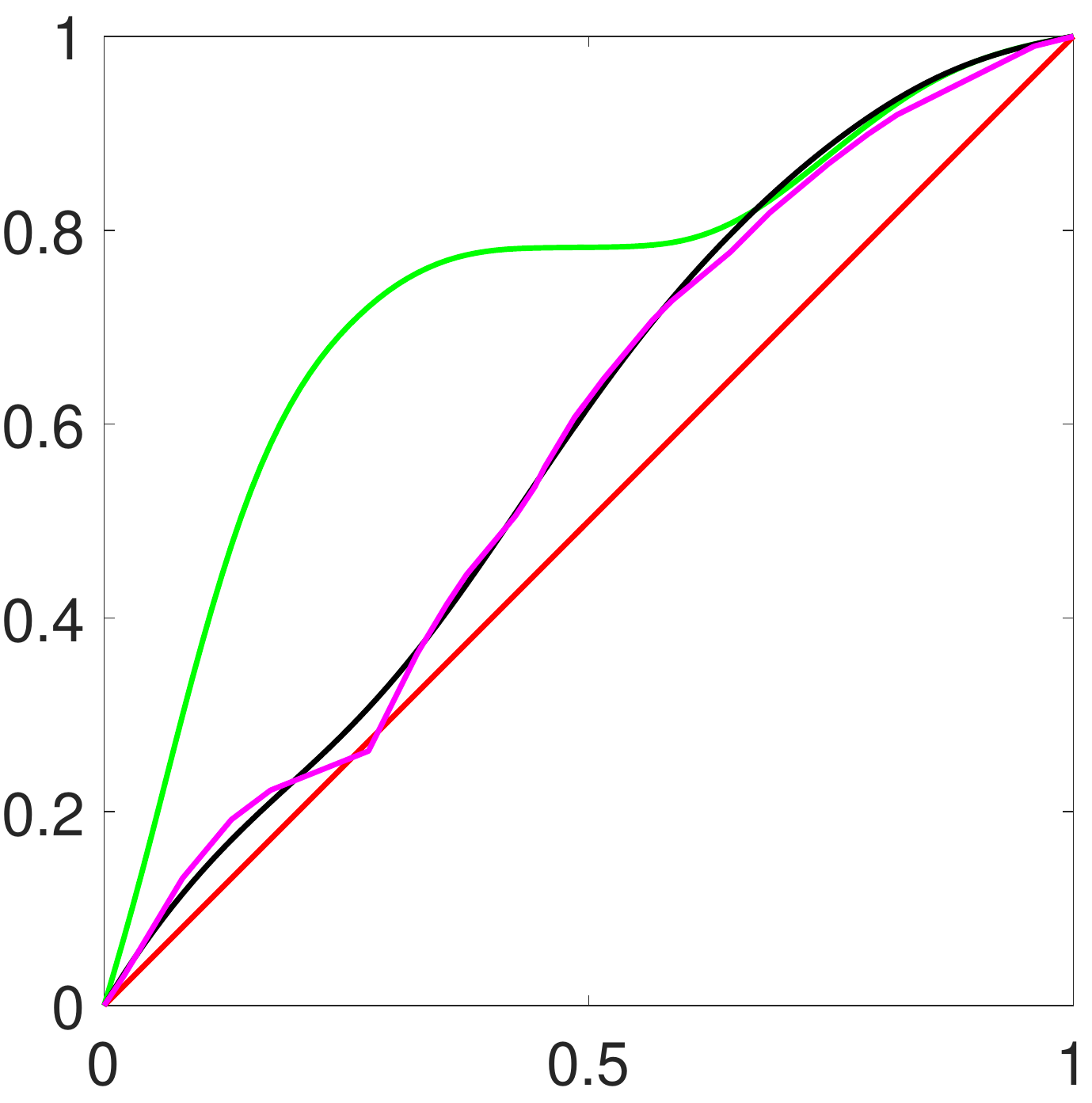}&\includegraphics[width=1.2in]{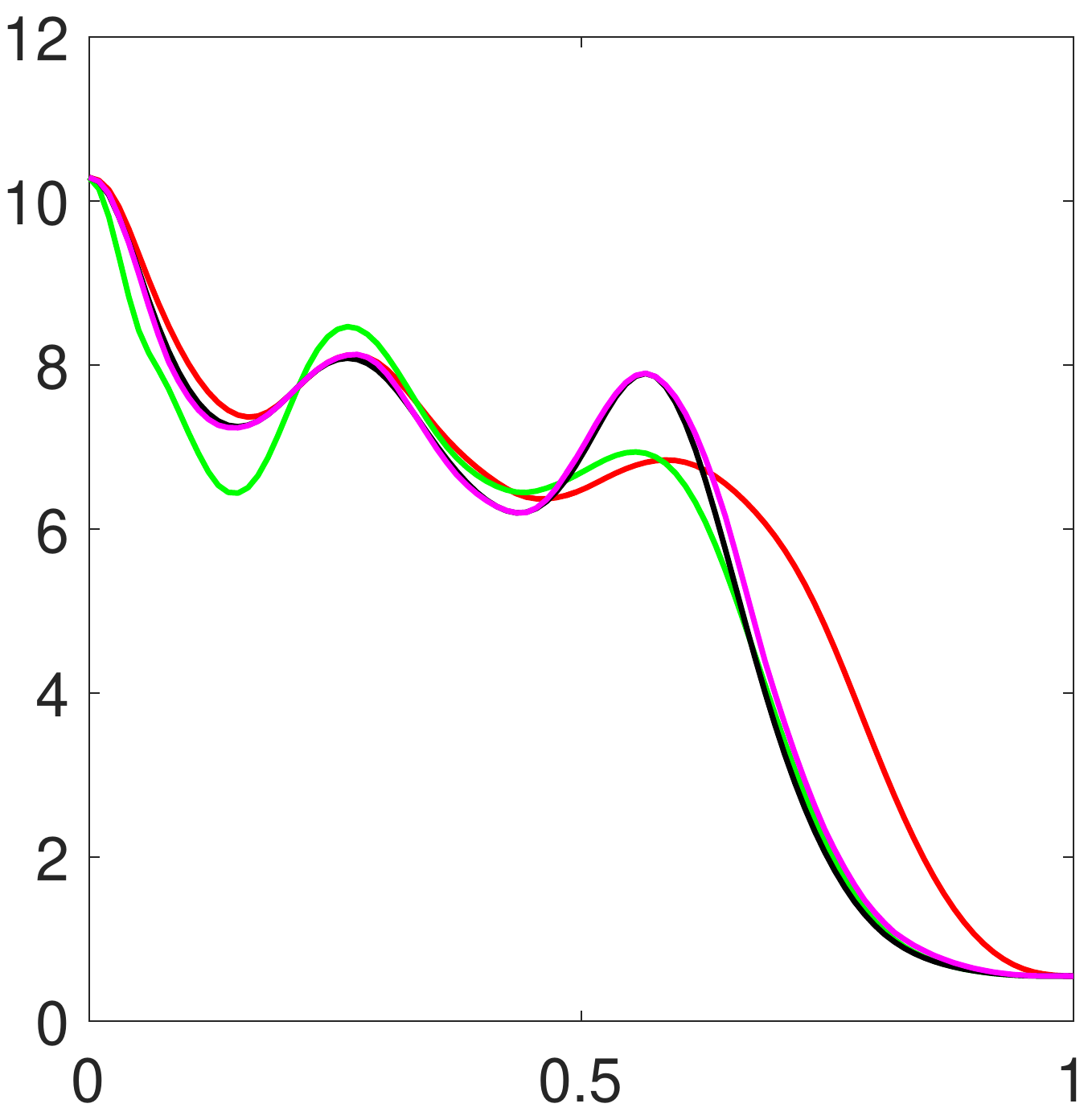}\\
\end{tabular}
\begin{tabular}{|c|c|c|c|}
\hline
(d)&(e)&(f)&(g)\\
\hline
\includegraphics[width=.9in]{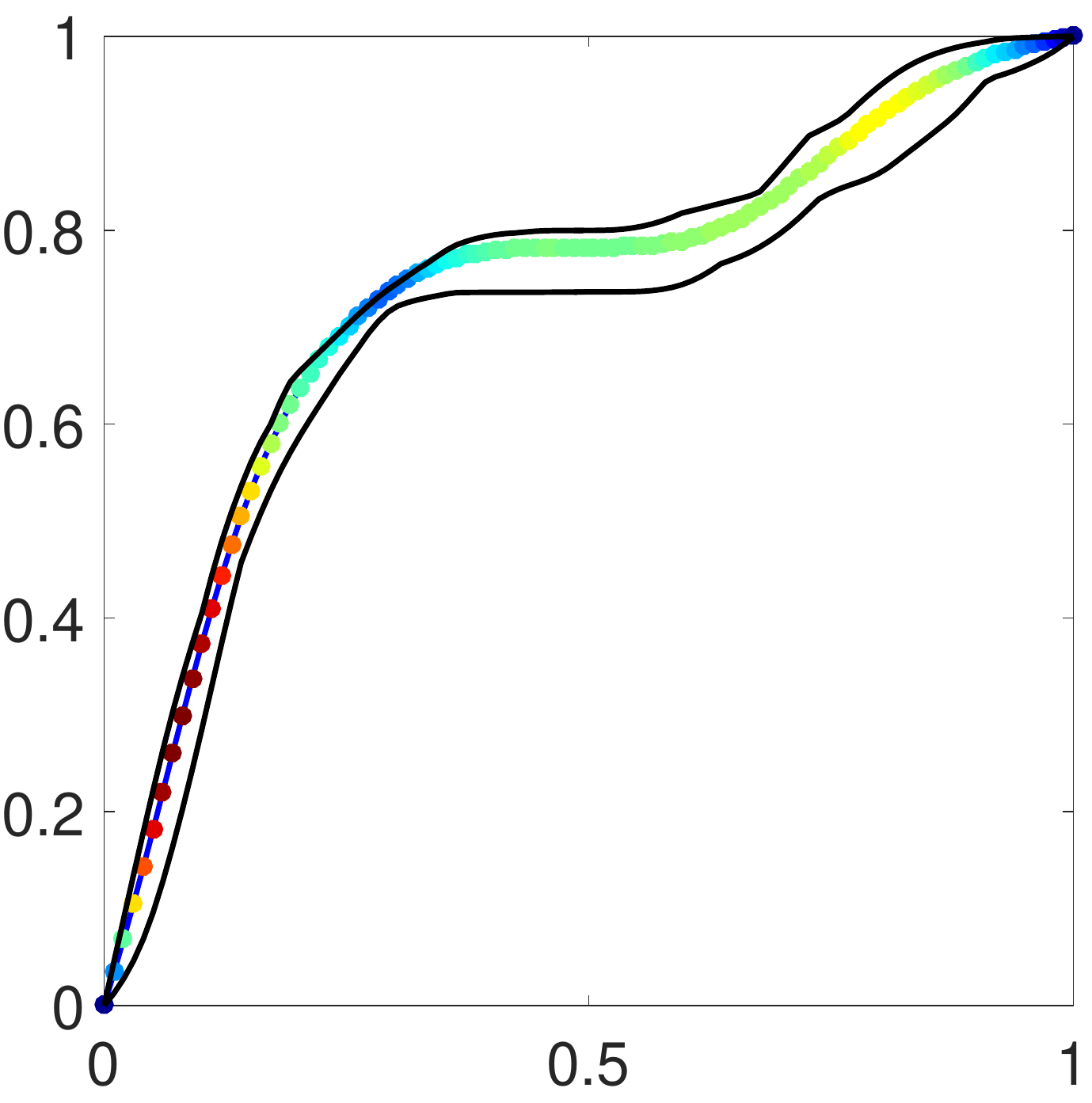}&\includegraphics[width=.9in]{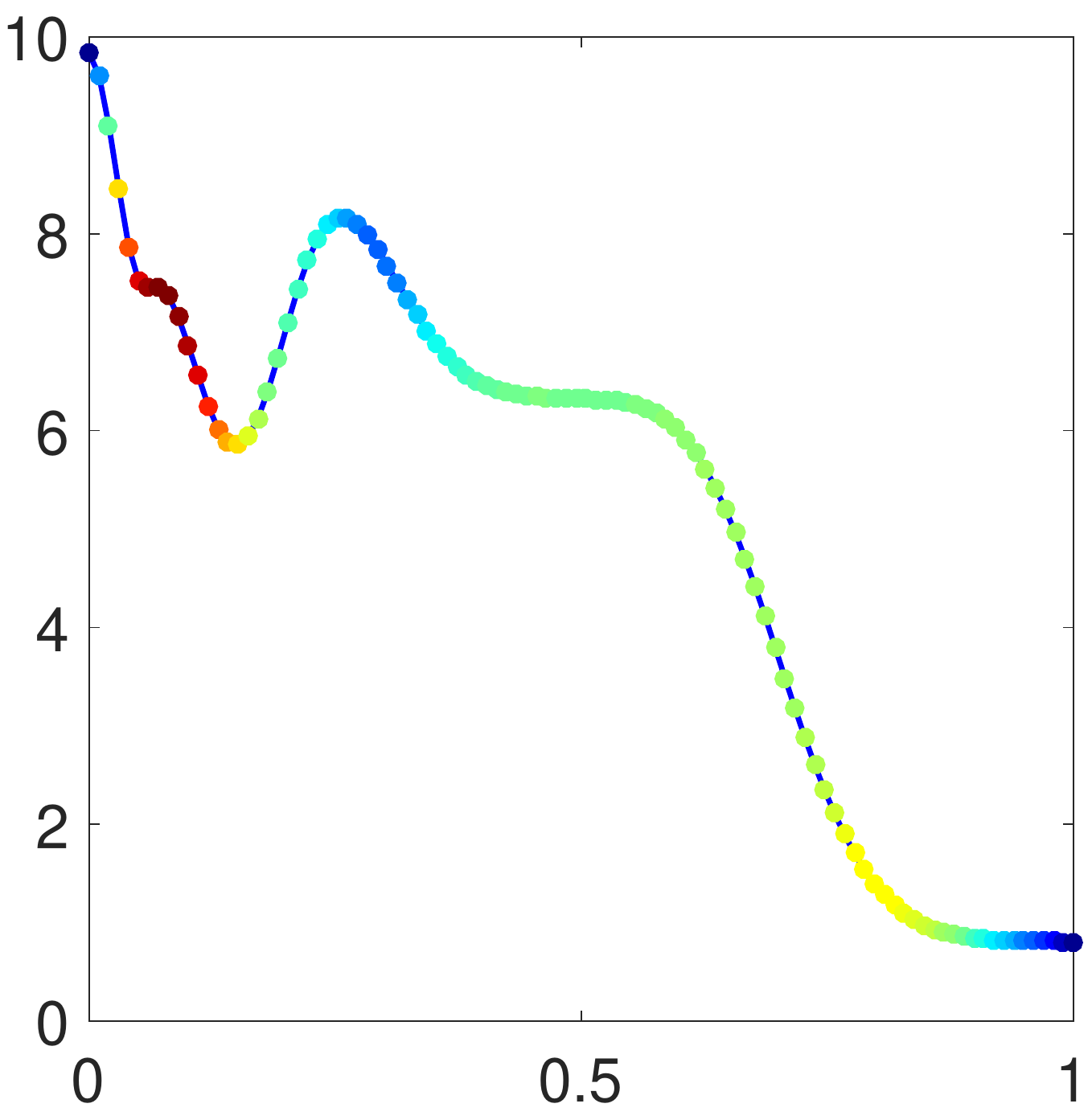}&\includegraphics[width=.9in]{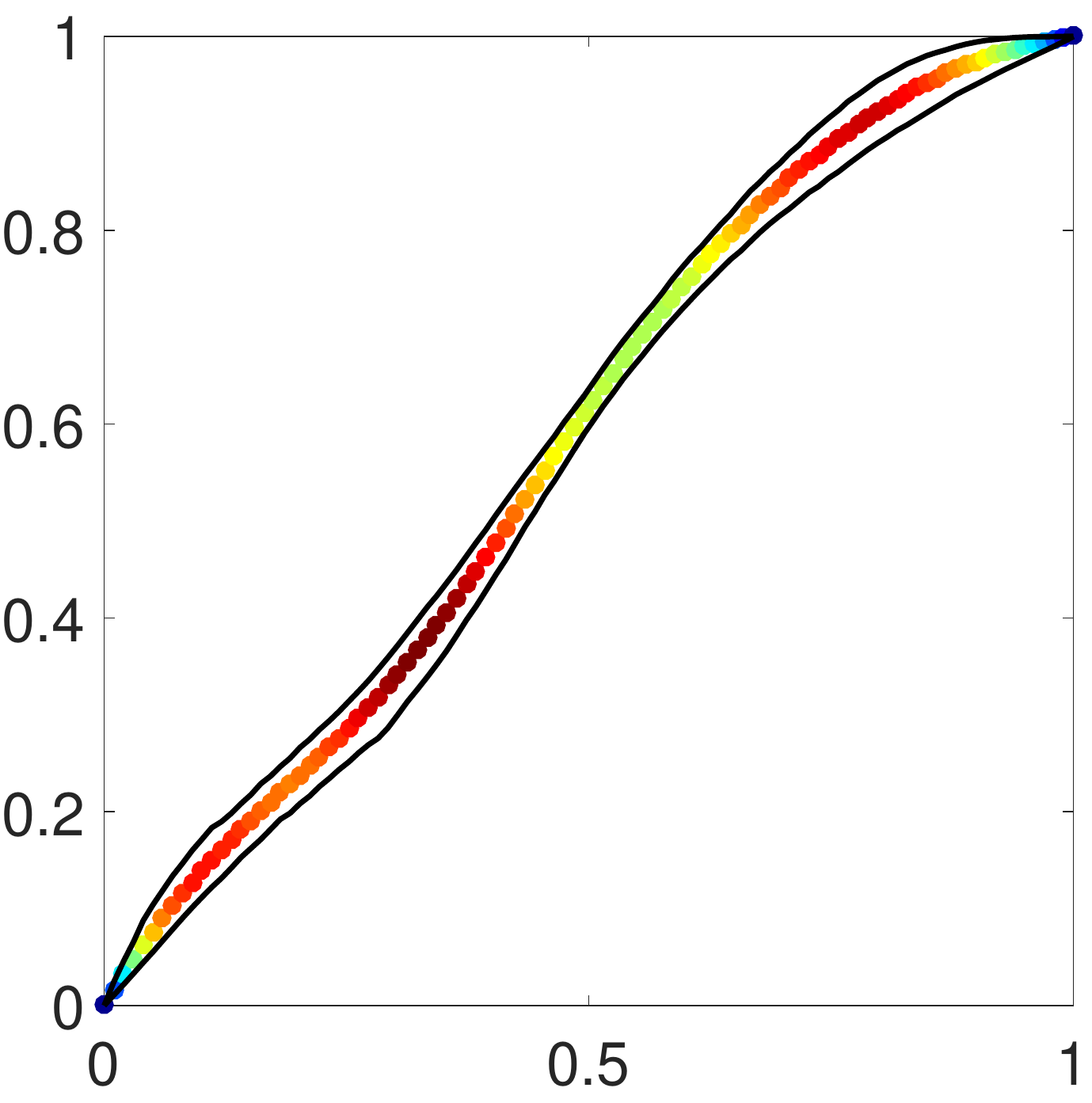}&\includegraphics[width=.9in]{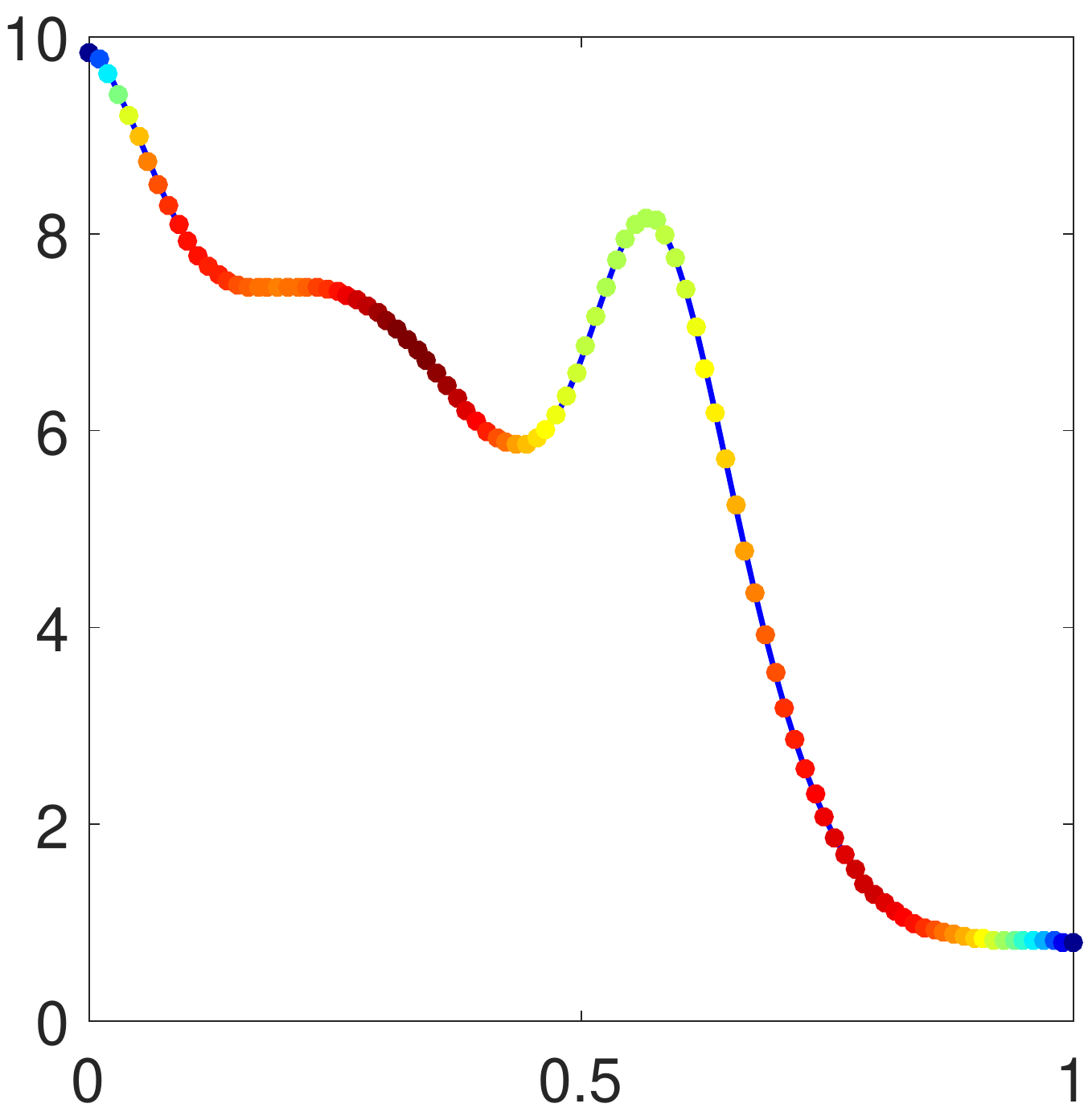}\\
\hline
\end{tabular}
\caption{Top: Pairwise alignment of two growth velocity functions for boys. Bottom: Pairwise alignment of two growth velocity functions for girls. (a) Original functions $f_1$ and $f_2$ in blue and red, respectively; $f_2\circ\gamma_{DP}$ in magenta, $f_2\circ\bar{\gamma}_1$ in green (cluster 1) and $f_2\circ\bar{\gamma}_2$ in black (cluster 2). (b) $\gamma_{DP}$ in magenta, $\gamma_{id}$ in red, and $\bar{\gamma}_1$ and $\bar{\gamma}_2$ in green and black, respectively. (c) Pointwise average of $f_1$ and $f_2$ for each alignment result (colored in the same way as (a) and (b)). (d) Pointwise standard deviation (hot colors correspond to higher values) plotted on $\bar{\gamma}_1$, and the $95\%$ credible interval in black. (e) Pointwise standard deviation (hot colors correspond to higher values) plotted on $f_2\circ\bar{\gamma}_1$. (f)-(g) Same as (d) and (e) but for cluster 2.} \label{fig:ex1boy}
\end{center}
\end{figure}

A major goal in studying the growth velocity functions of children is to characterize the number and timing of growth spurts in boys and girls. The BGD has been studied for these purposes before \cite{doi:10.1080/03014469500004092}. In the current paper, we emphasize that there may be multiple plausible time warpings that align growth spurts across children. In the first example, presented in the top part of Figure \ref{fig:ex1boy}, we examine two growth velocity functions for boys. The resulting posterior distribution on the space of warping functions is bimodal. The mean warping in both clusters nicely aligns the large growth spurt. But, the average growth velocity patterns, as seen in panel (c), are quite different depending on which alignment is used. The cluster 1 alignment (green) results in a long constant velocity growth period in the average, while cluster 2 (black) results in a decreasing velocity (at an approximately constant rate) during the same period. This presents two very different growth mechanisms, which are useful for characterizing growth functions. The second example, shown in the bottom portion of Figure \ref{fig:ex1boy}, considers alignment of two growth velocity curves for girls. Again, we discover two modes in the posterior distribution. As seen in panel (c), the mean warping in cluster 1 (green) emphasizes the first growth spurt and is followed by a smaller second spurt. On the other hand, the mean alignment in cluster 2 results in an average growth pattern where the two growth spurts are approximately of the same size.

\begin{figure}[!t]
\begin{center}
\begin{tabular}{|c|c|c|}
\hline
(a)&(b)&(c)\\
\hline
\includegraphics[width=1.2in]{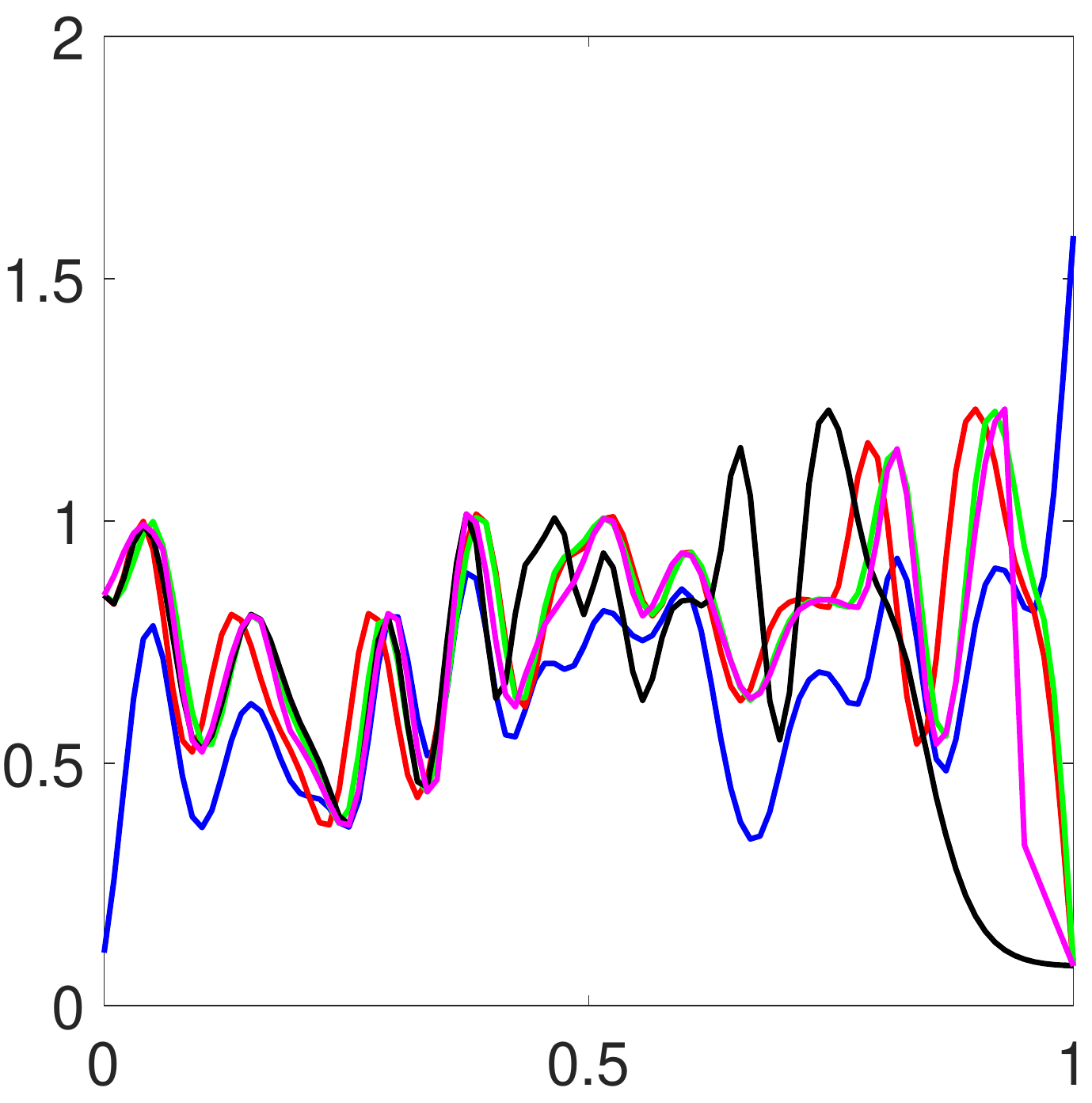}&\includegraphics[width=1.2in]{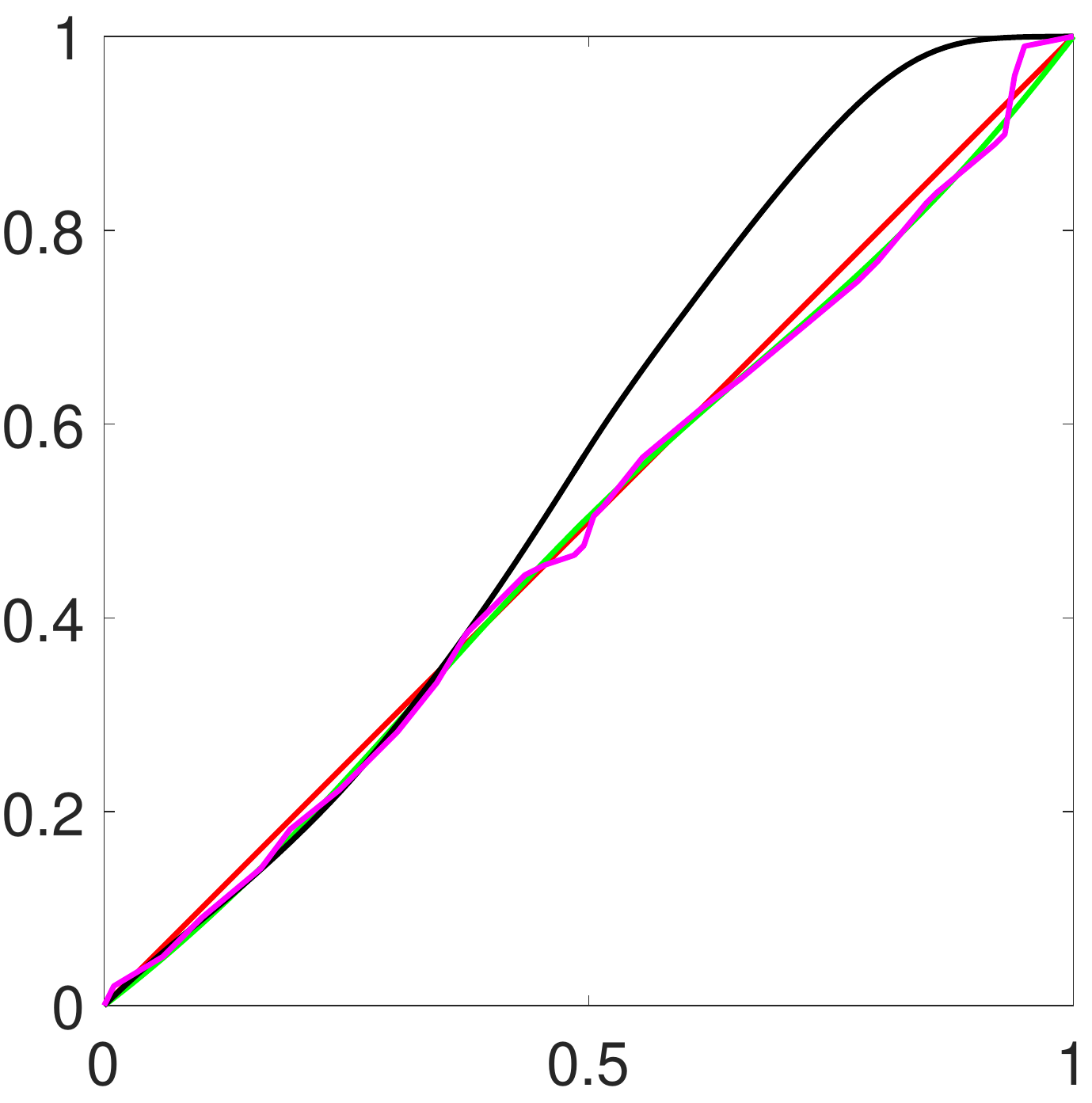}&\includegraphics[width=1.2in]{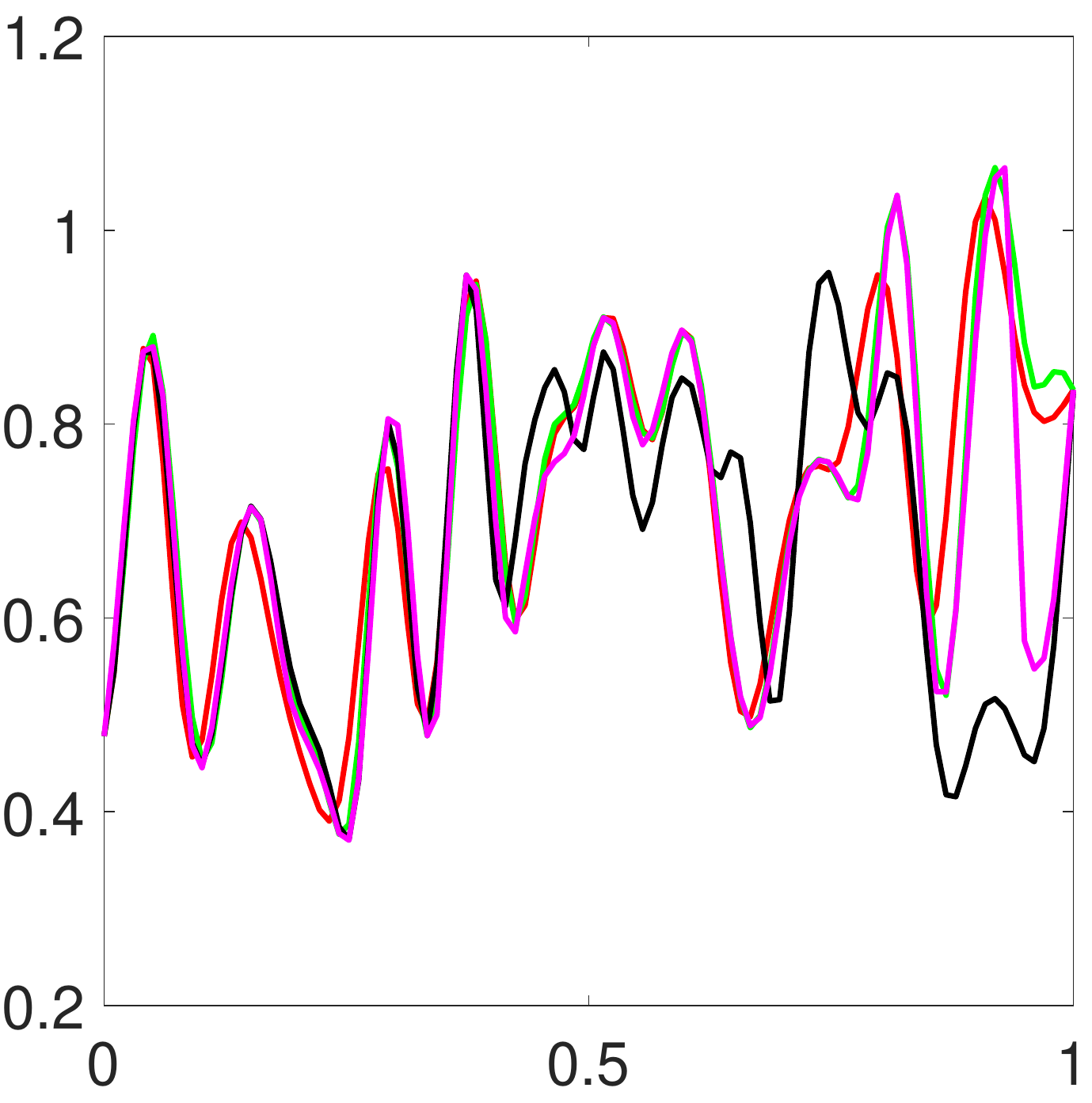}\\
\end{tabular}
\begin{tabular}{|c|c|c|c|}
\hline
(d)&(e)&(f)&(g)\\
\hline
\includegraphics[width=.9in]{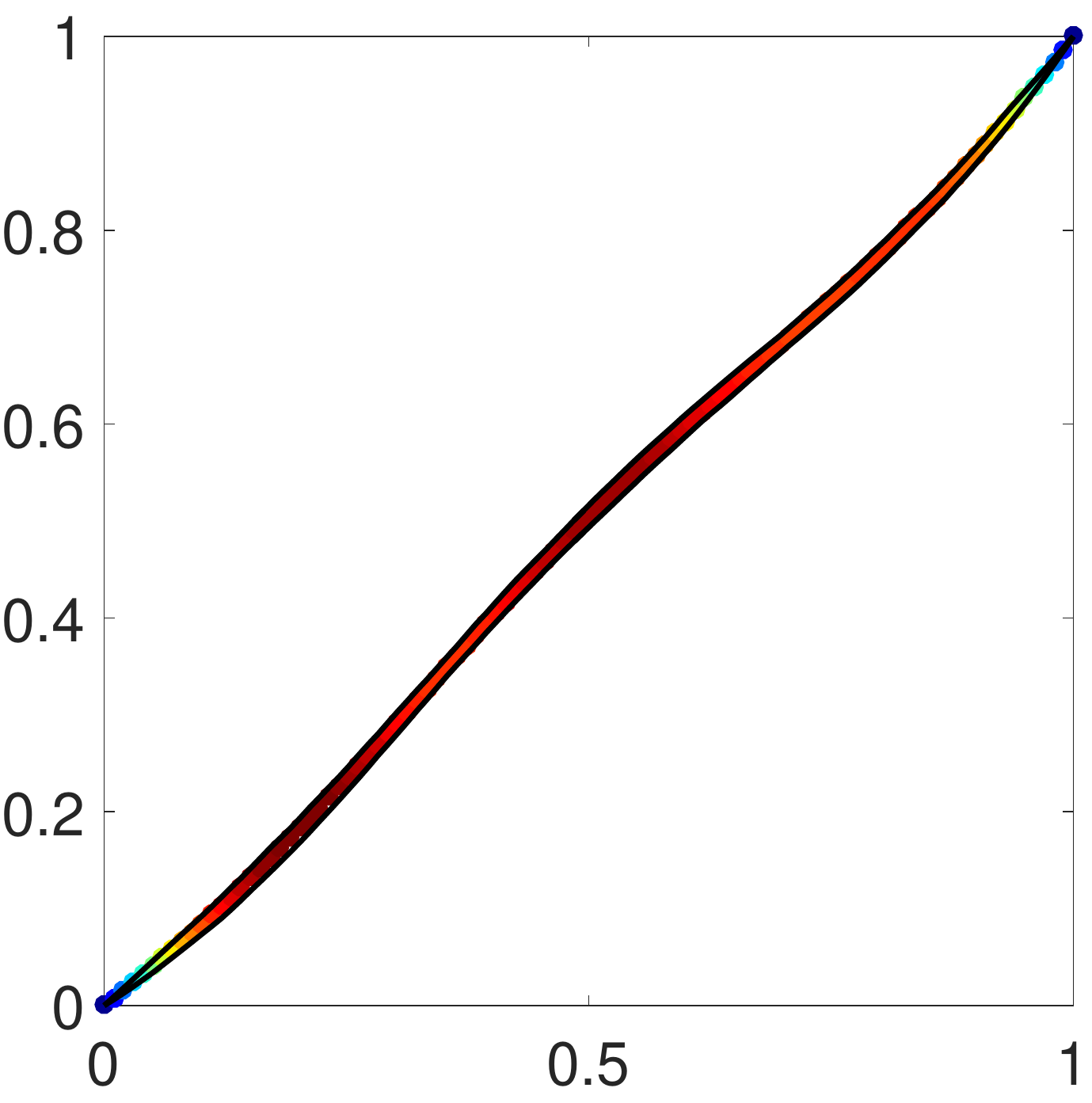}&\includegraphics[width=.9in]{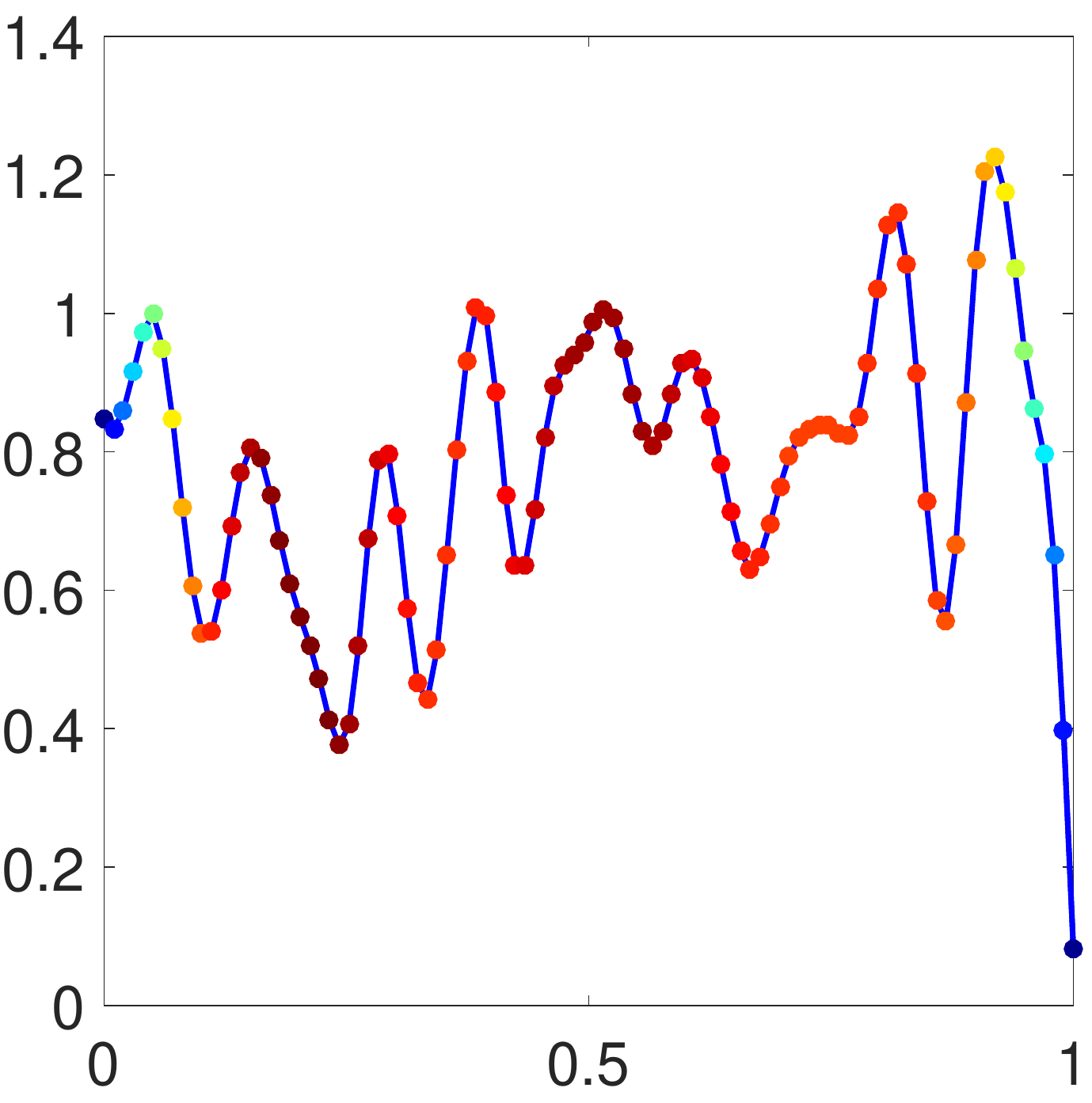}&\includegraphics[width=.9in]{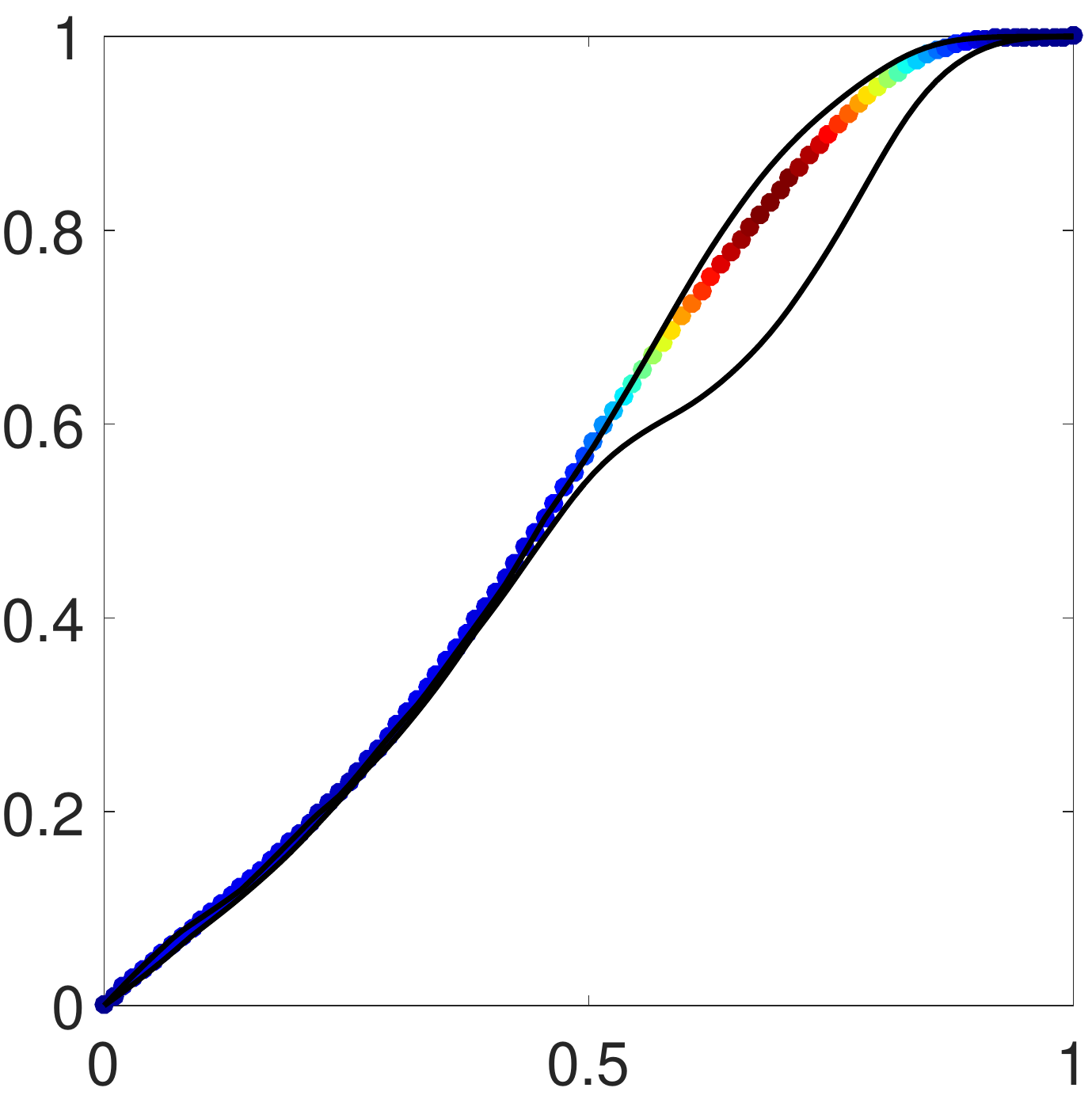}&\includegraphics[width=.9in]{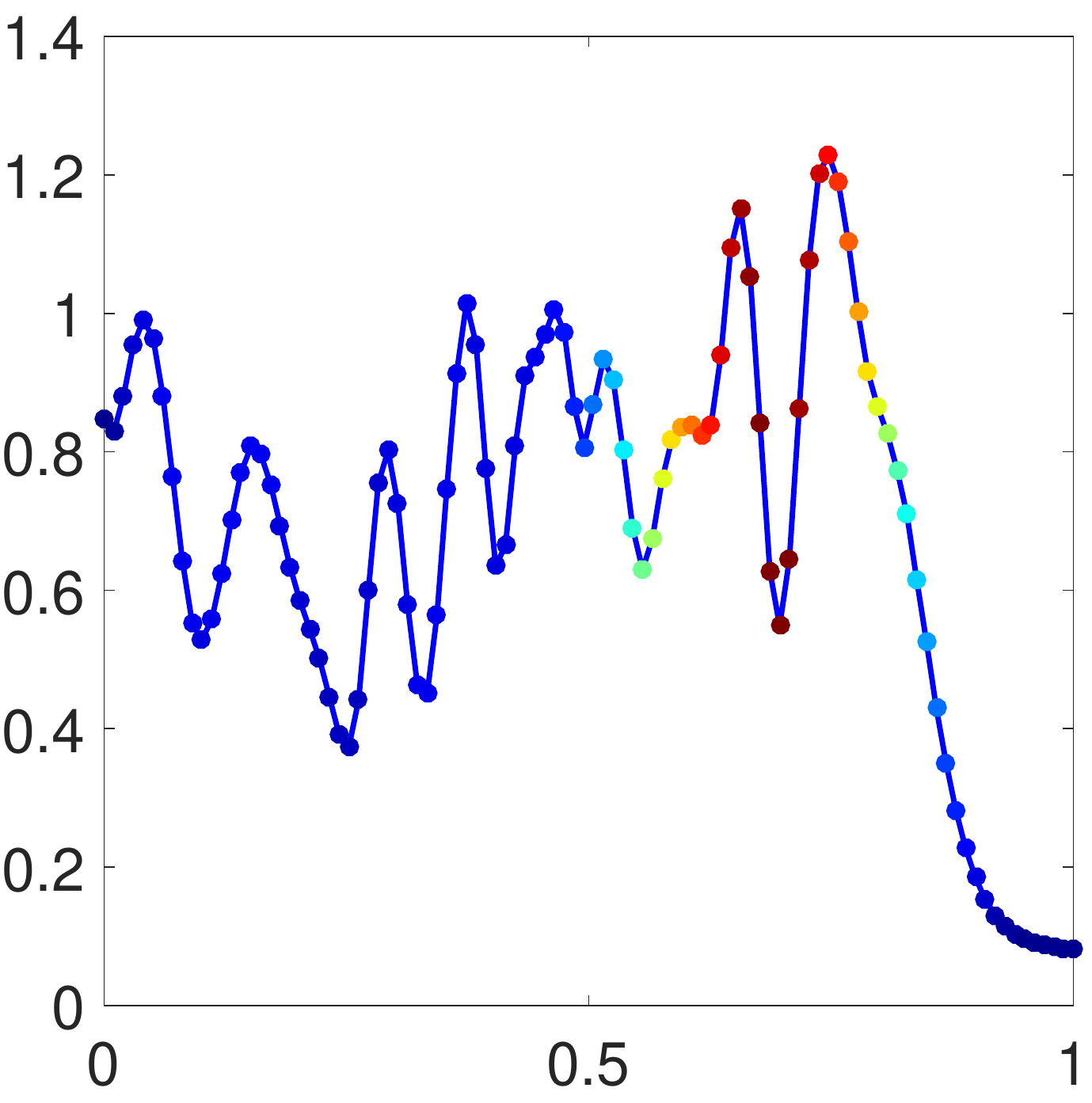}\\
\hline
\end{tabular}
\caption{Pairwise alignment of two signature acceleration functions. (a) Original functions $f_1$ and $f_2$ in blue and red, respectively; $f_2\circ\gamma_{DP}$ in magenta, $f_2\circ\bar{\gamma}_1$ in green (cluster 1) and $f_2\circ\bar{\gamma}_2$ in black (cluster 2). (b) $\gamma_{DP}$ in magenta, $\gamma_{id}$ in red, and $\bar{\gamma}_1$ and $\bar{\gamma}_2$ in green and black, respectively. (c) Pointwise average of $f_1$ and $f_2$ for each alignment result (colored in the same way as (a) and (b)). (d) Pointwise standard deviation (hot colors correspond to higher values) plotted on $\bar{\gamma}_1$, and the $95\%$ credible interval in black. (e) Pointwise standard deviation (hot colors correspond to higher values) plotted on $f_2\circ\bar{\gamma}_1$. (f)-(g) Same as (d)-(e) but for cluster 2.} \label{fig:ex1sig}
\end{center}
\end{figure}

\begin{figure}[!t]
\begin{center}
\begin{tabular}{|c|c|c|}
\hline
(a)&(b)&(c)\\
\hline
\includegraphics[width=1.2in]{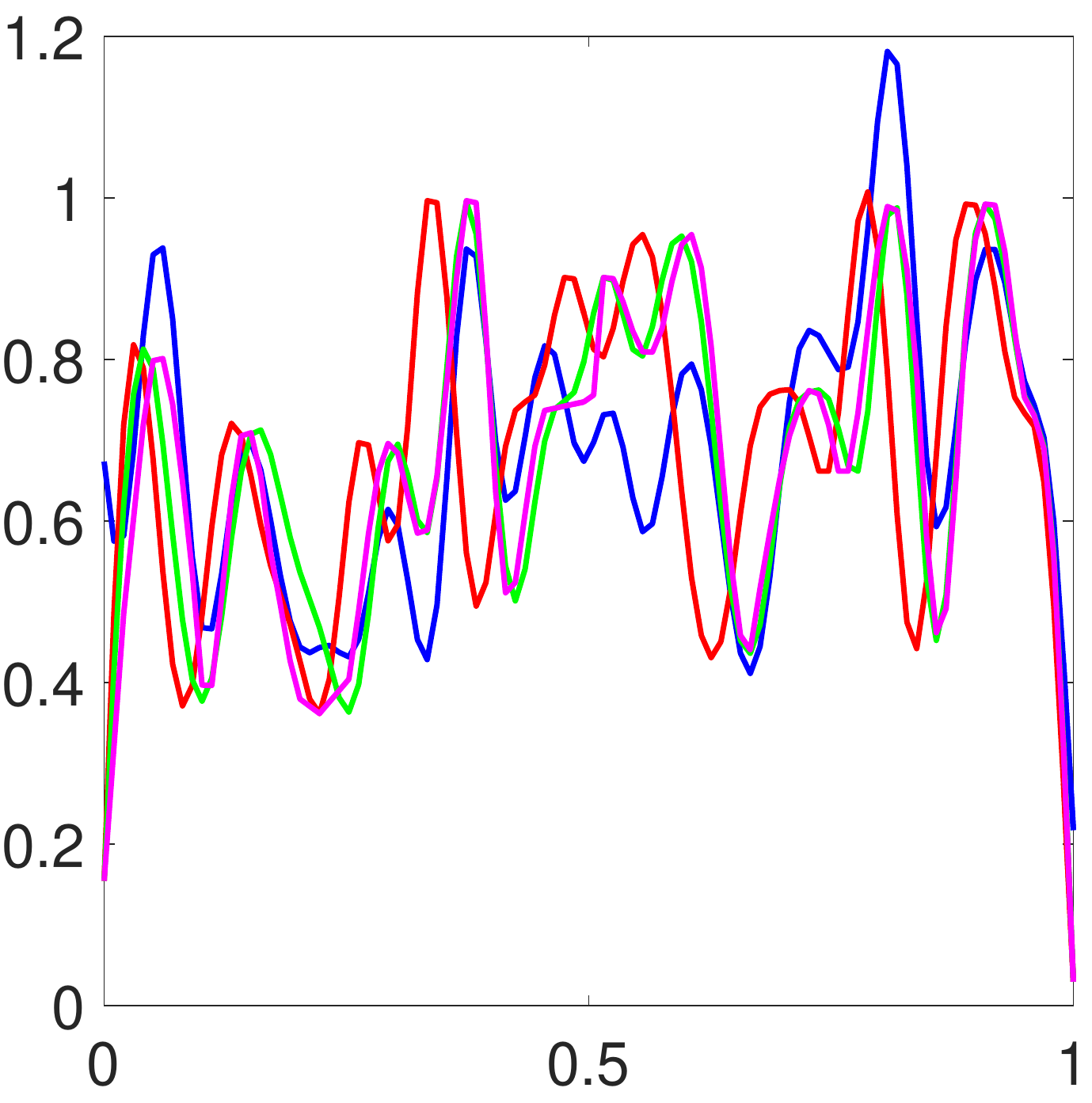}&\includegraphics[width=1.2in]{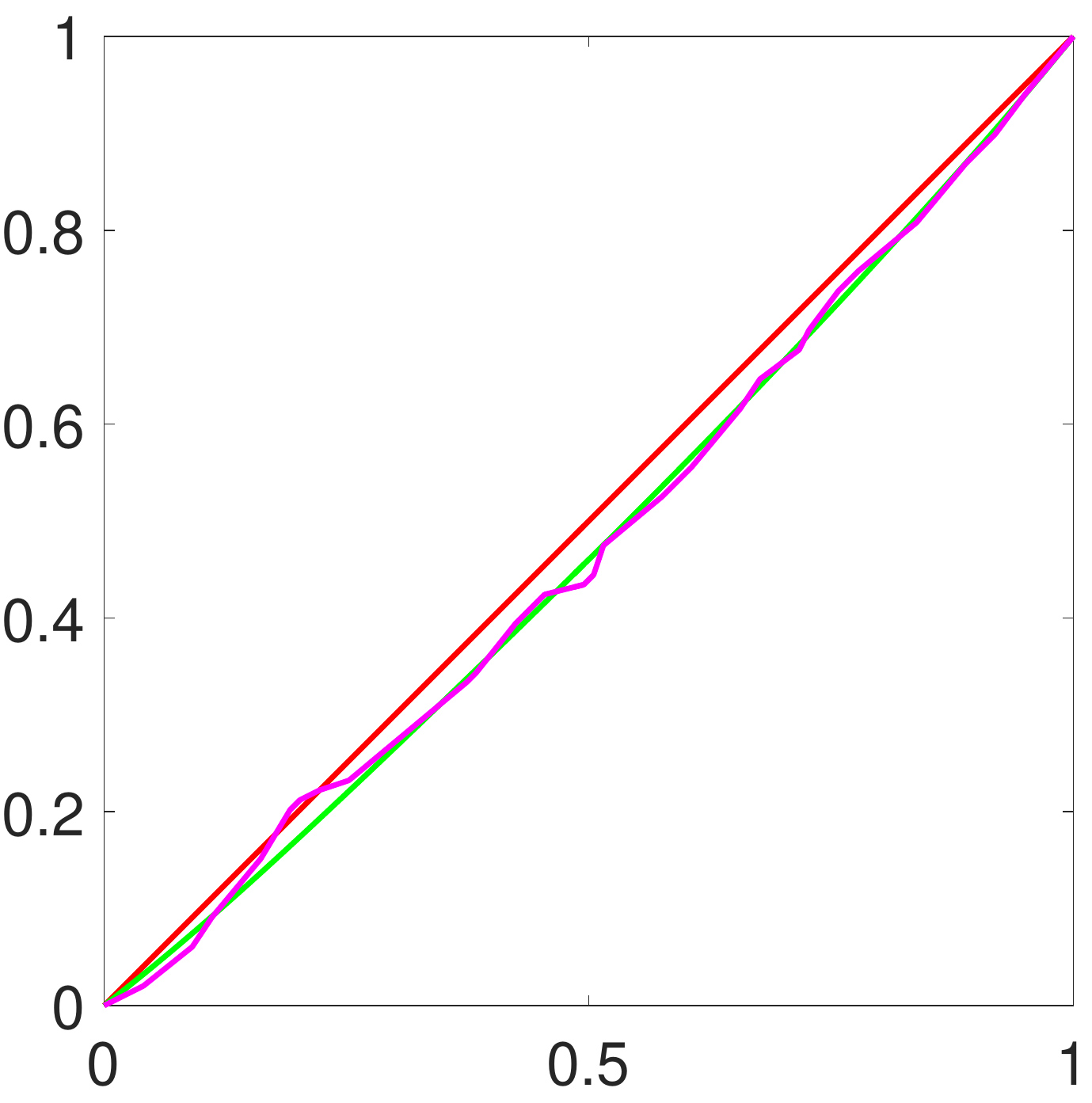}&\includegraphics[width=1.2in]{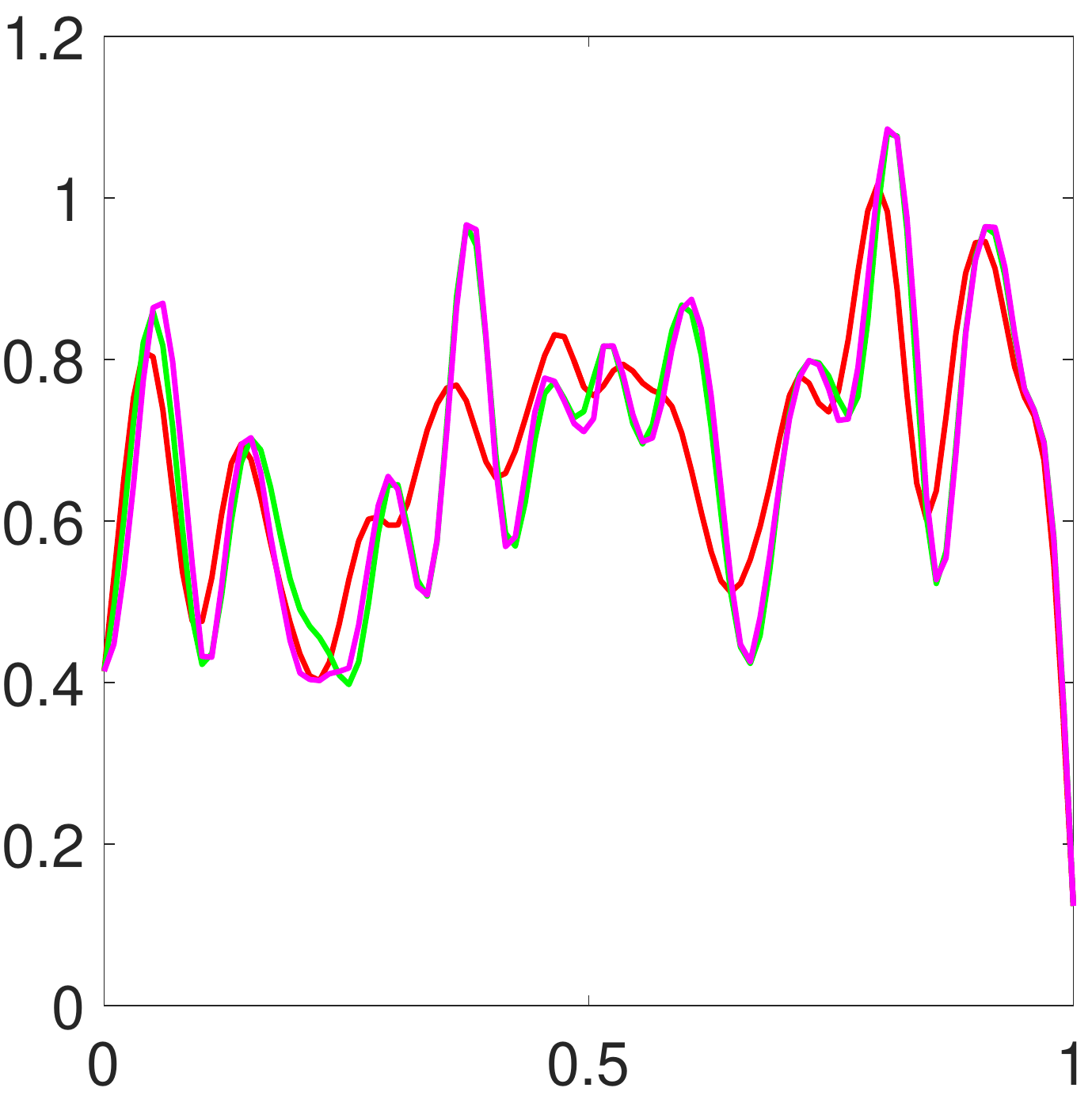}\\
\hline
\end{tabular}
\begin{tabular}{|c|c|}
(d)&(e)\\
\hline
\includegraphics[width=1.2in]{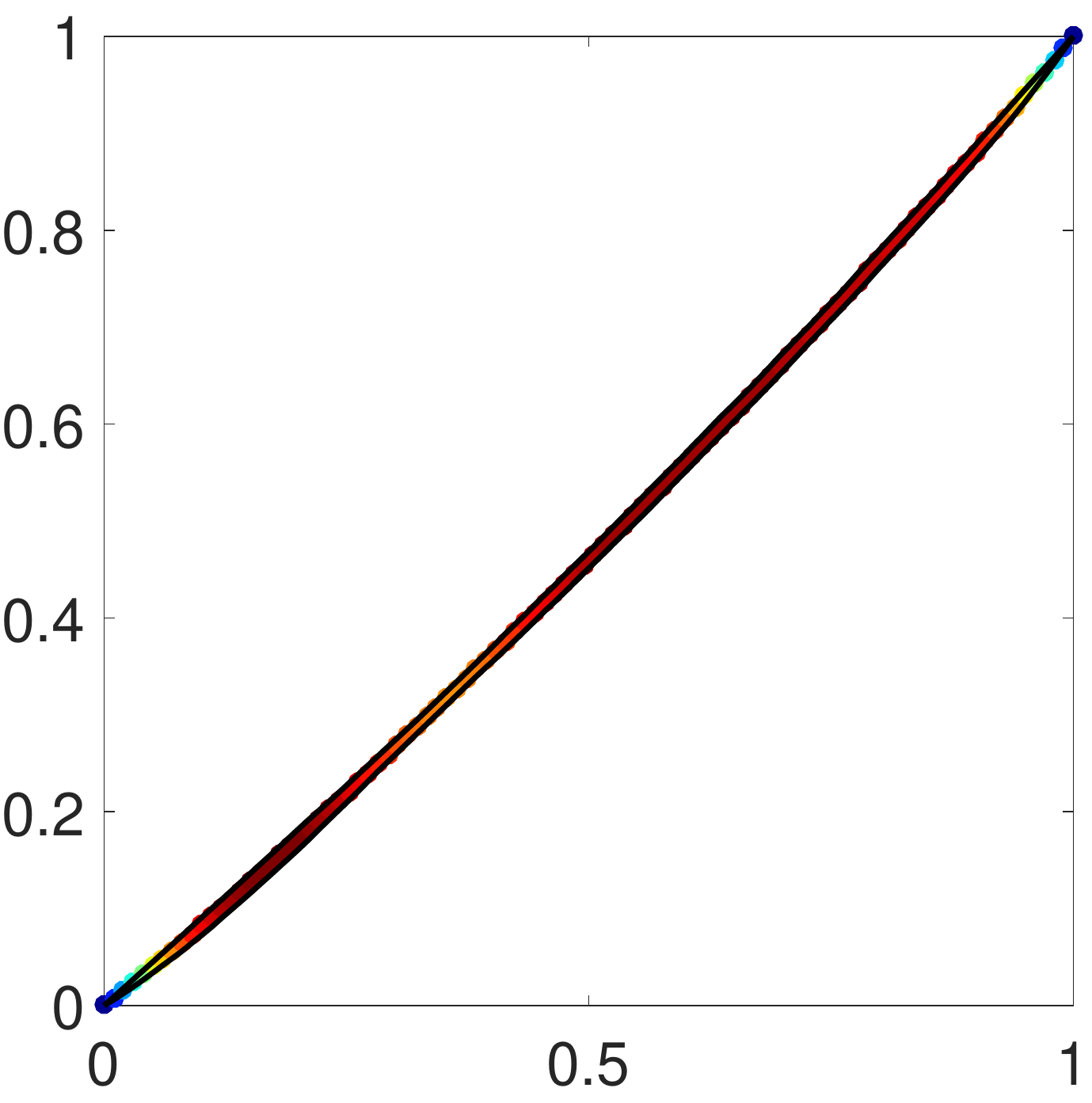}&\includegraphics[width=1.2in]{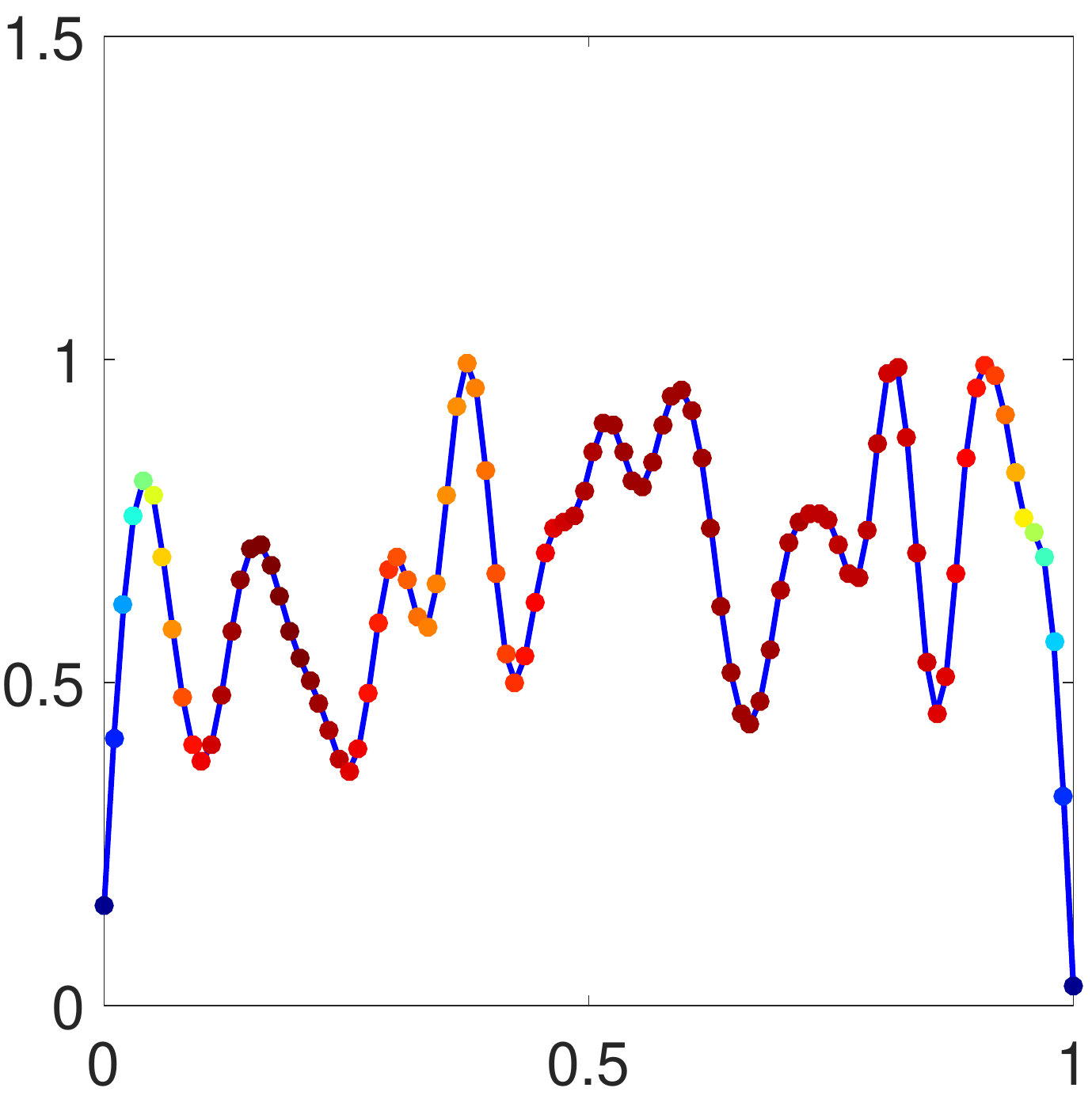}\\
\hline
\end{tabular}
\caption{Pairwise alignment of two signature acceleration functions. (a) Original functions $f_1$ and $f_2$ in blue and red, respectively; $f_2\circ\gamma_{DP}$ in magenta and $f_2\circ\bar{\gamma}$ in green. (b) $\gamma_{DP}$ in magenta, $\gamma_{id}$ in red, and $\bar{\gamma}$ in green. (c) Pointwise average of $f_1$ and $f_2$ for each alignment result (colored in the same way as (a) and (b)). (d) Pointwise standard deviation (hot colors correspond to higher values) plotted on $\bar{\gamma}$, and the $95\%$ credible interval in black. (e) Pointwise standard deviation (hot colors correspond to higher values) plotted on $f_2\circ\bar{\gamma}$.} \label{fig:ex2sig}
\end{center}
\end{figure}

\subsection{Signature Acceleration Functions}

The final application considers alignment of signature acceleration functions. As described in \cite{ramsay-silverman-2005,tucker1}, each planar signature curve is first summarized using its tangential acceleration. Comparison and modeling of such functions are important in understanding inter and intra-class signature variability, and for signature classification. A major difficulty that arises in the analysis pipeline is that the signature acceleration functions contain natural warping variability. Thus, in order to obtain satisfactory results, such variability must be accounted for. We present two different pairwise registration results in Figures \ref{fig:ex1sig} and \ref{fig:ex2sig}. In the first example, the posterior distribution of warping functions contains two different modes. The posterior mean alignment agrees for close to half of the time interval at which point the mean warping in cluster 2 (black) diverges from the identity warping. This results in two drastically different alignments between the two signatures (and potentially different inferences depending on which alignment is used). Another interesting feature is that there is a large amount of uncertainty in the region where the mean warping in cluster 2 diverges from the identity element; this indicates that the corresponding region of the two acceleration functions is difficult to align. The posterior distribution in the second example is unimodal, and the posterior mean is very close to the dynamic programming solution. Furthermore, perhaps surprisingly, there is very little uncertainty in the alignment.

\begin{figure}[!t]
\begin{center}
\begin{tabular}{|c|c|c|c|c|}
\hline
&(a)&(b)&(c)&(d)\\
\hline
(1)&\includegraphics[width=.9in]{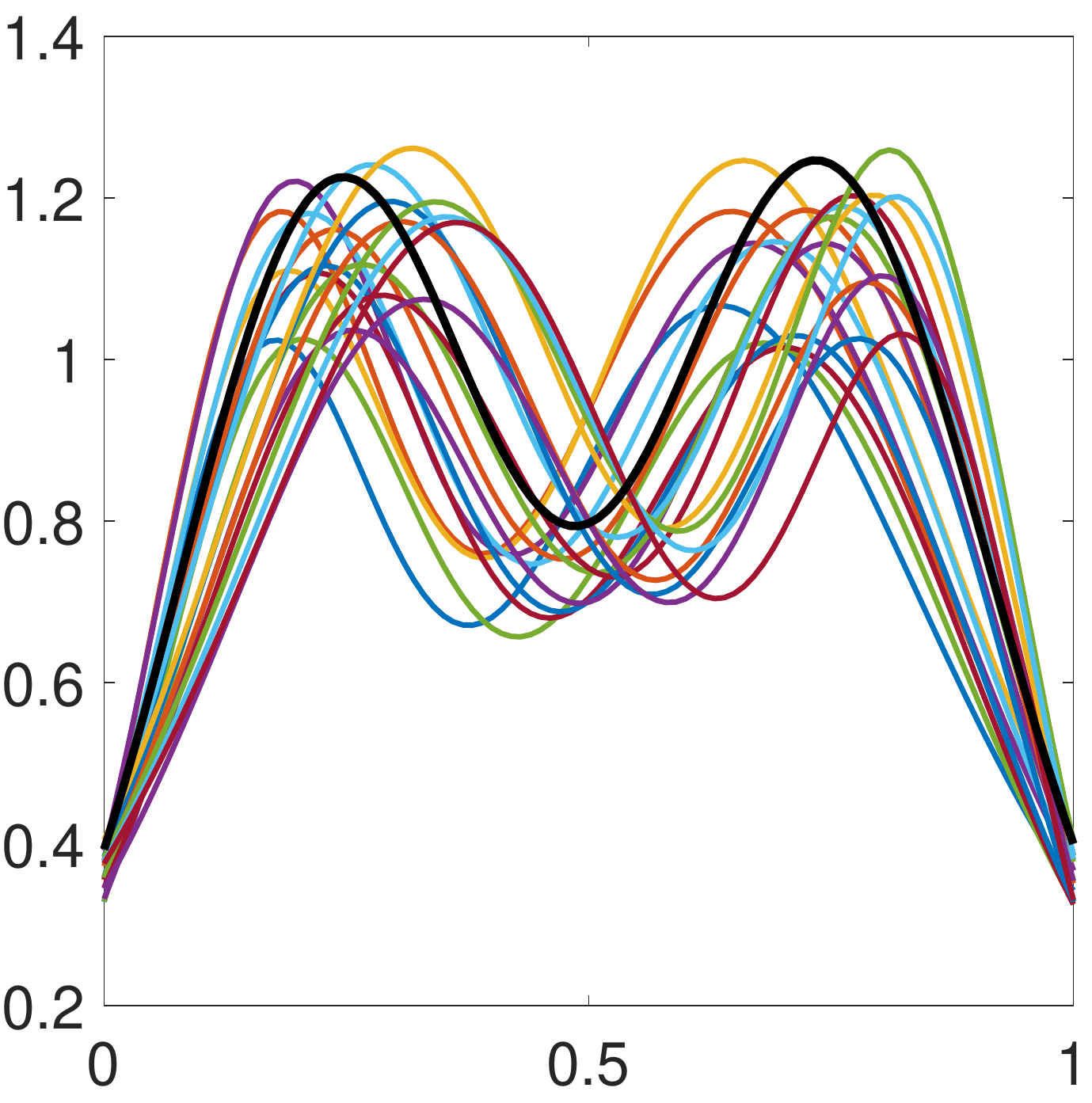}&\includegraphics[width=.9in]{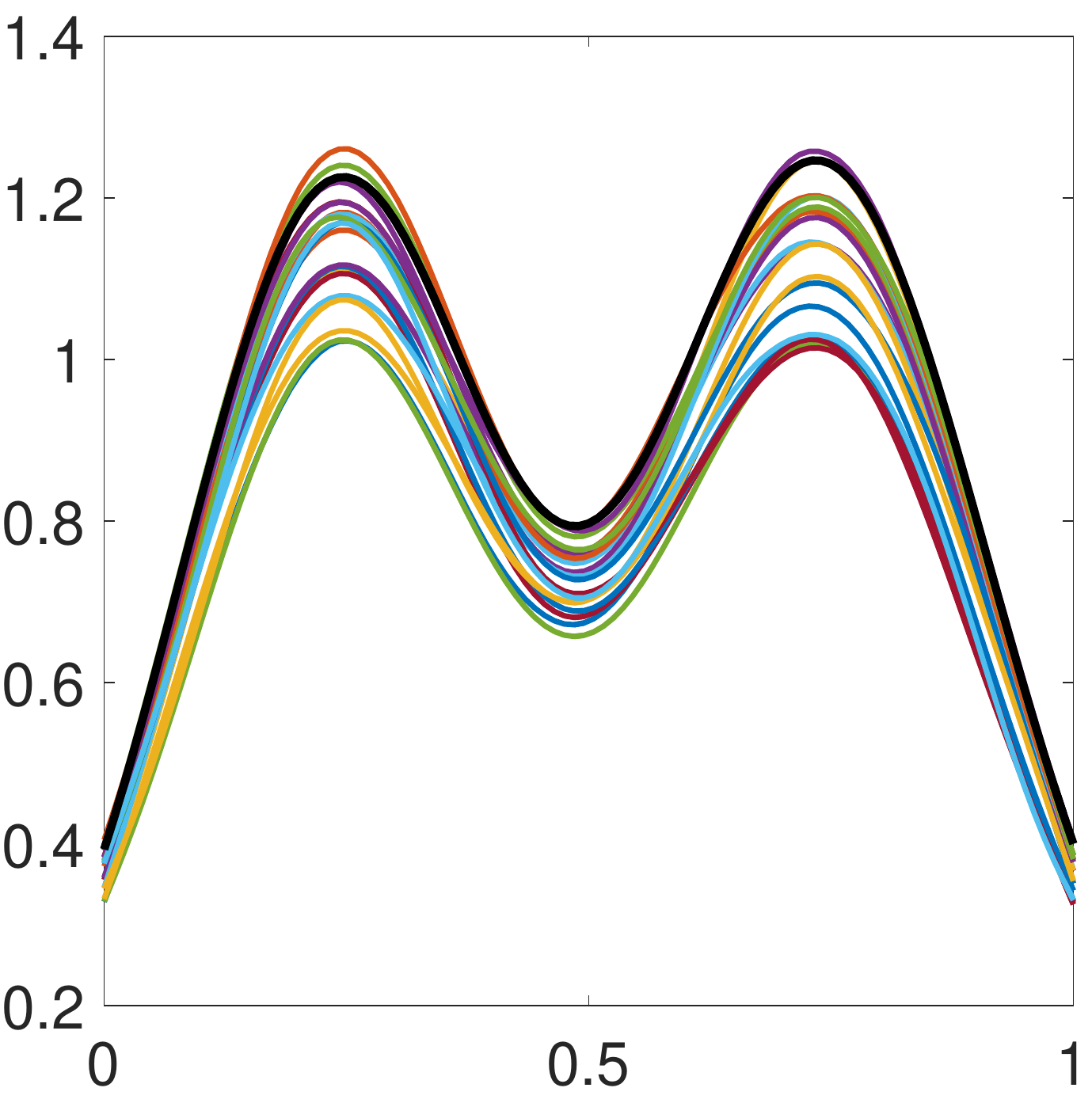}&\includegraphics[width=.9in]{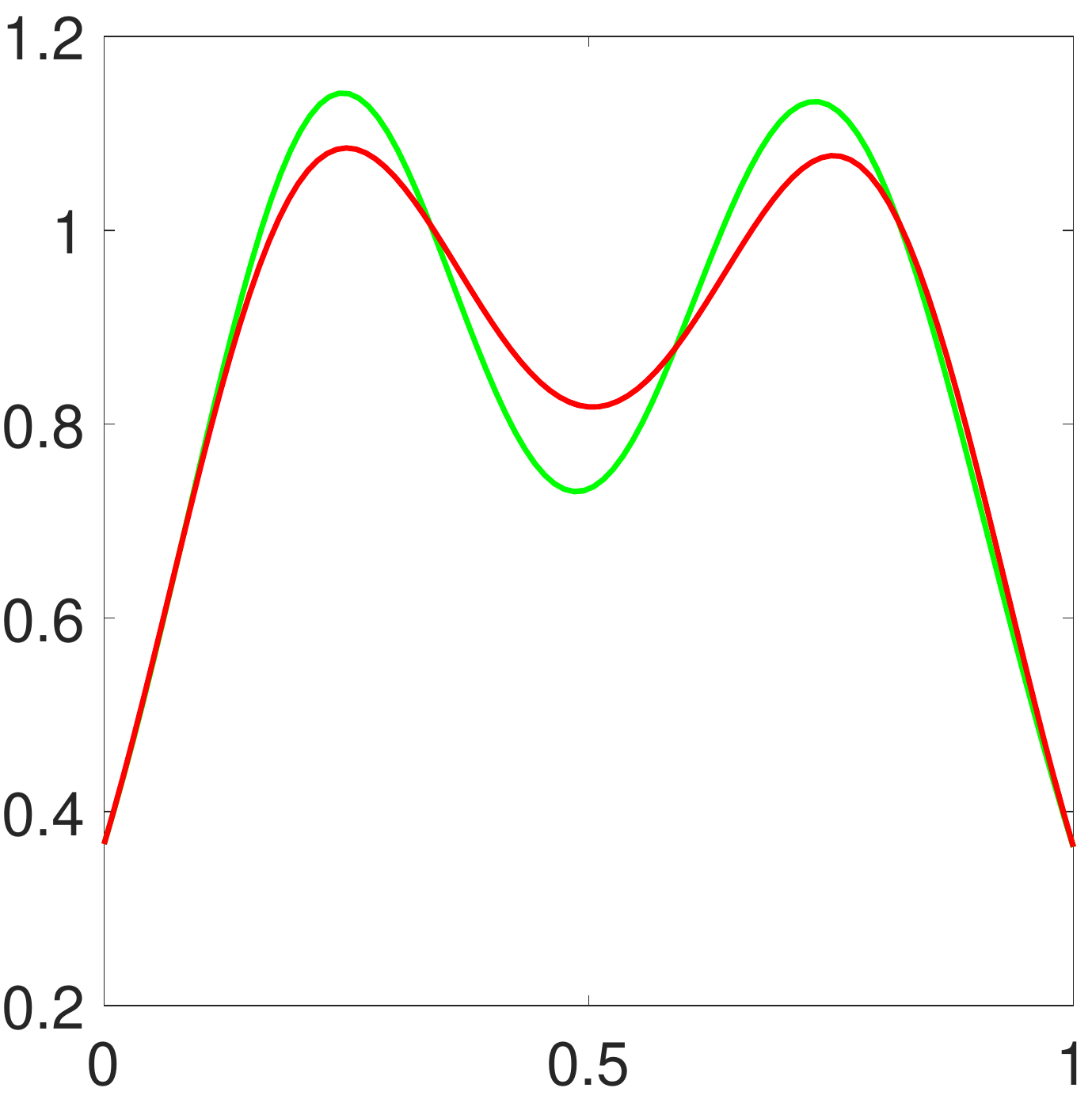}&\includegraphics[width=.9in]{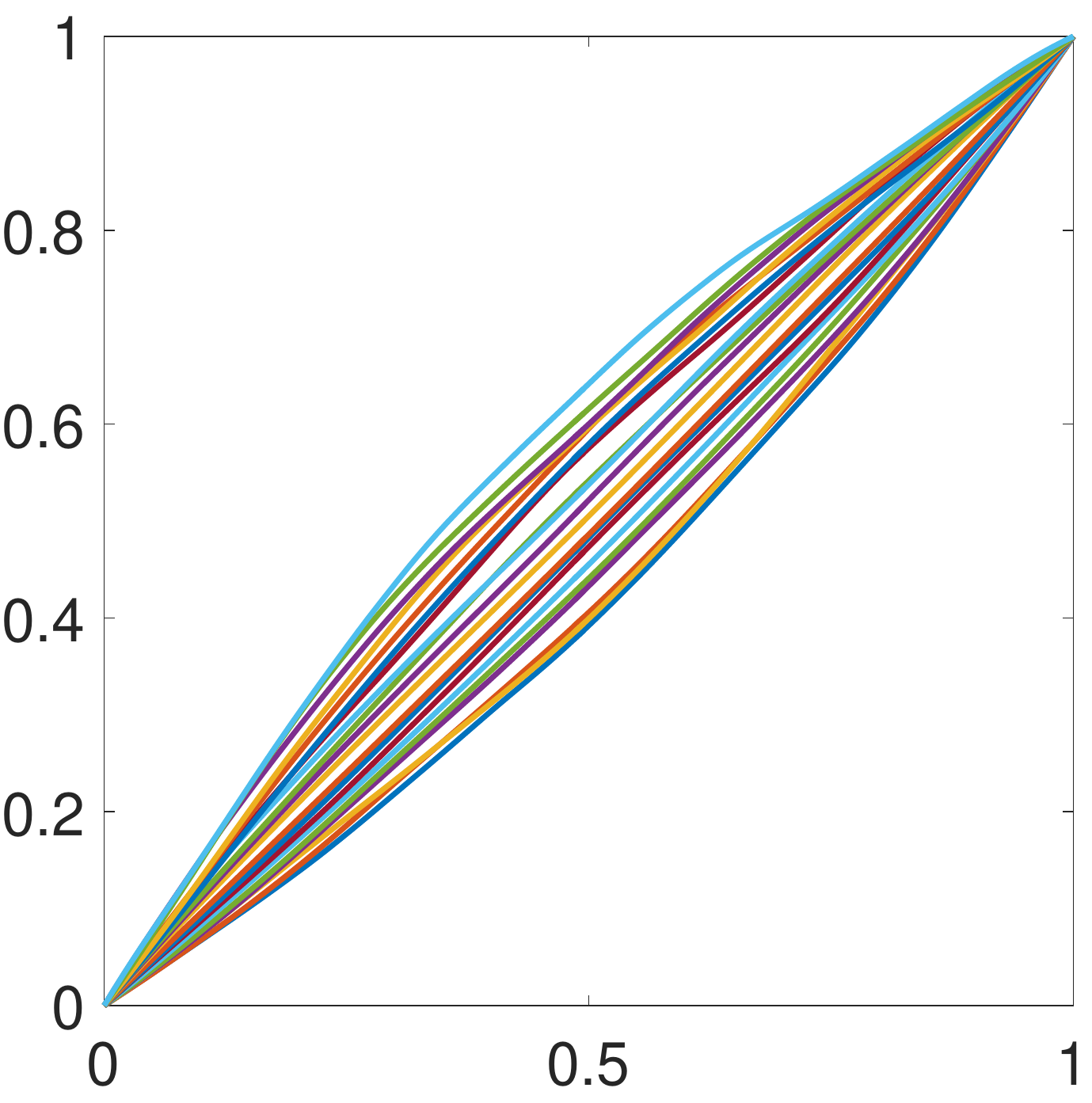}\\
\hline
(2)&\includegraphics[width=.9in]{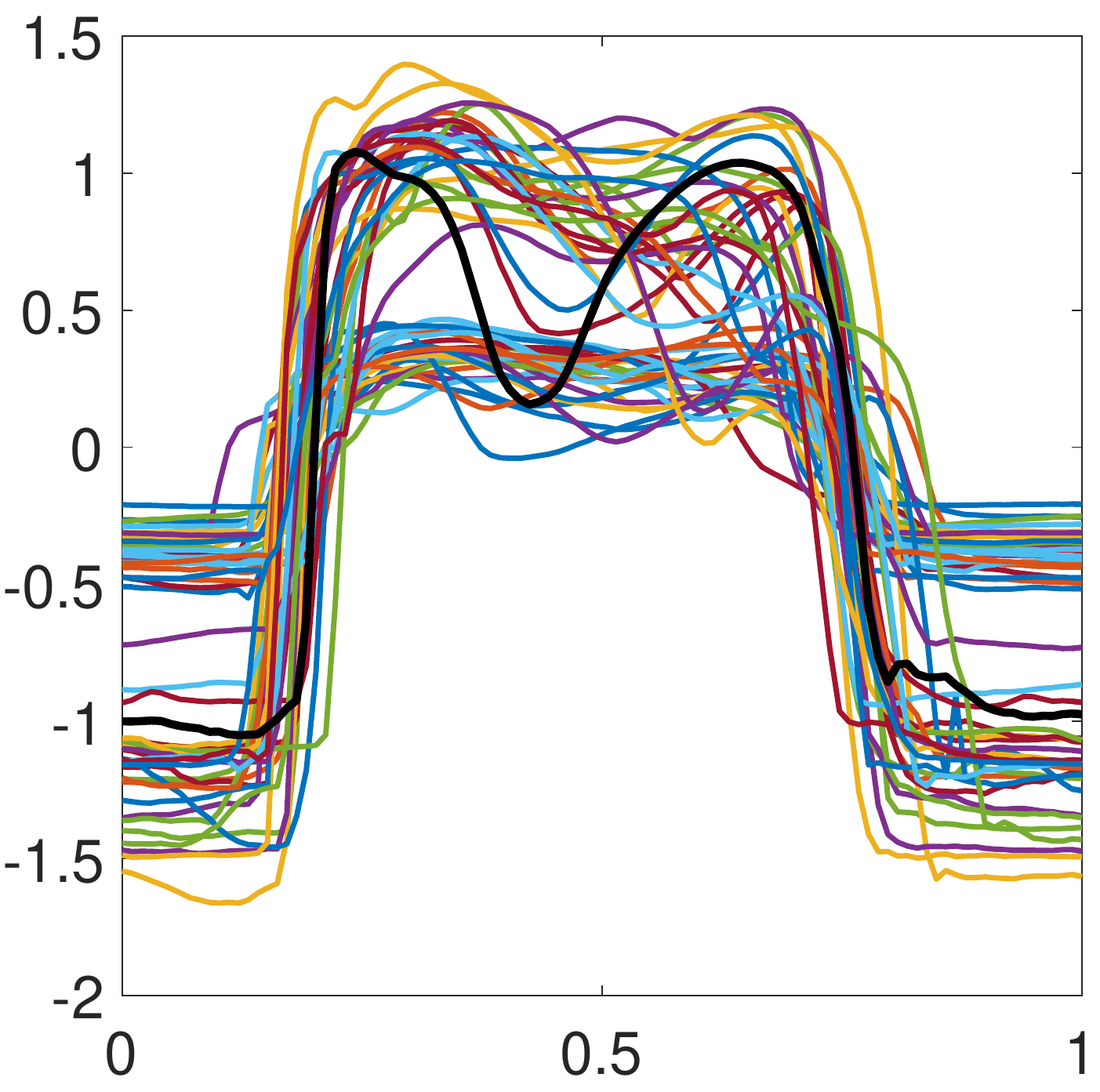}&\includegraphics[width=.9in]{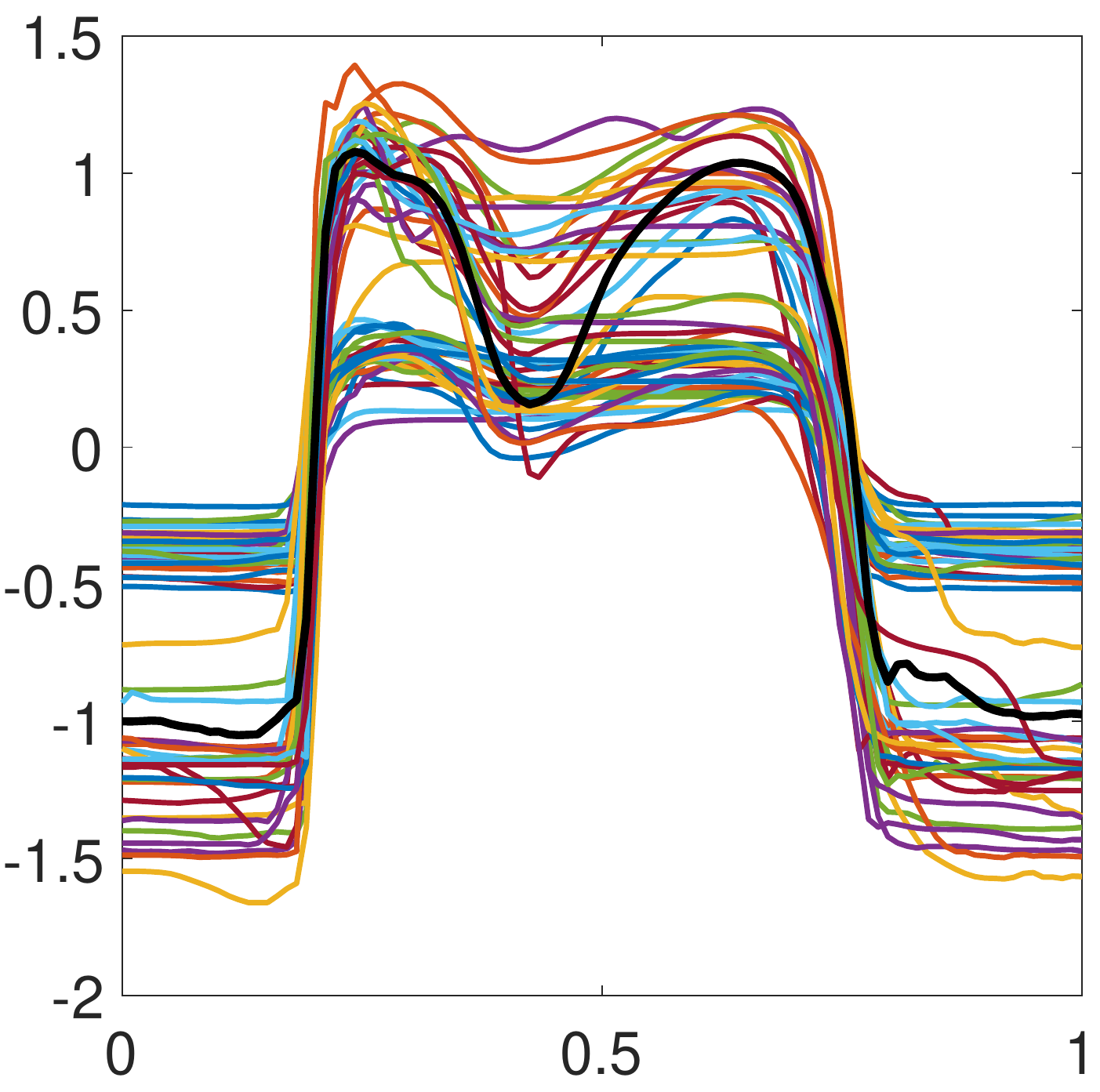}&\includegraphics[width=.9in]{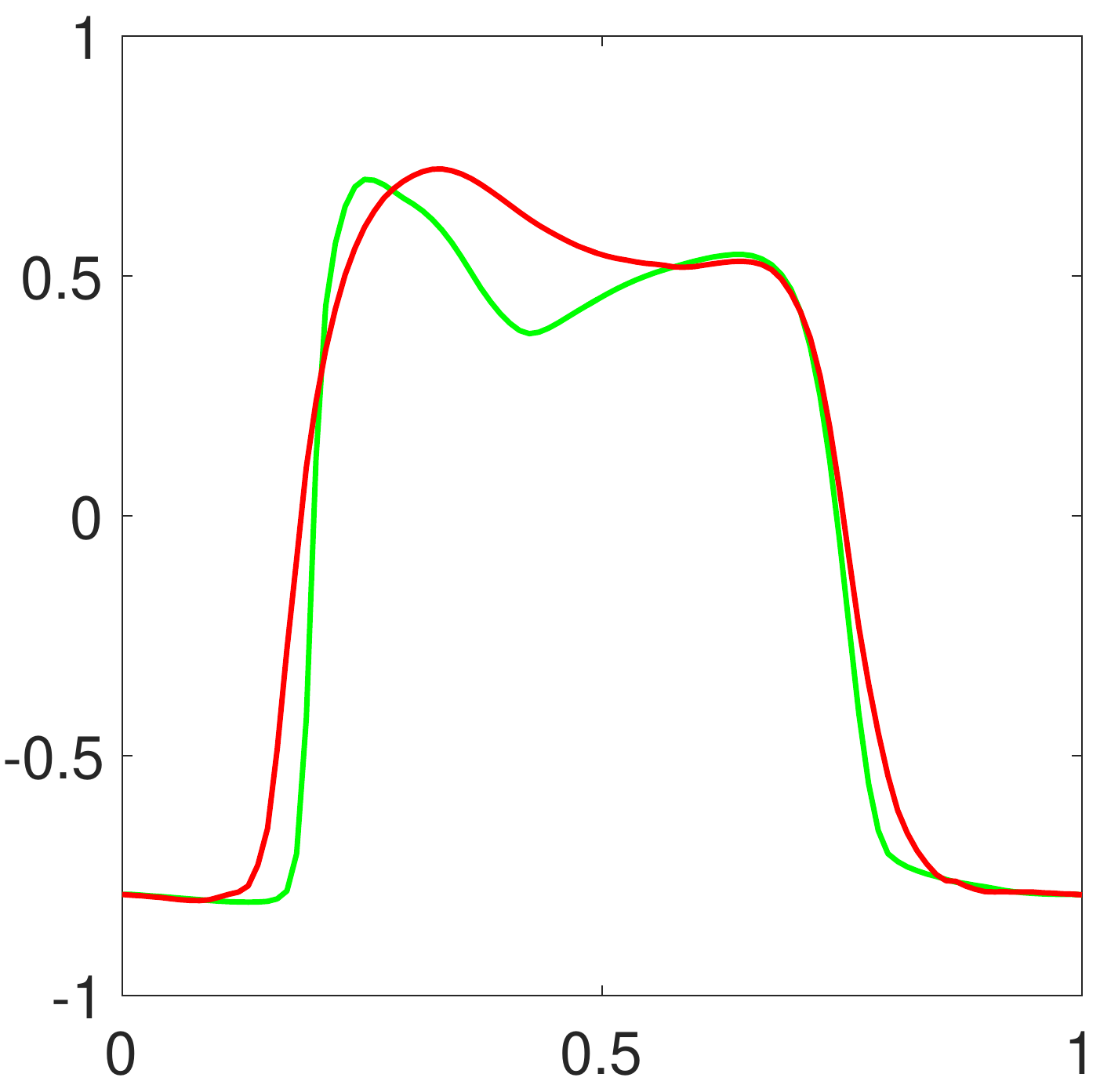}&\includegraphics[width=.9in]{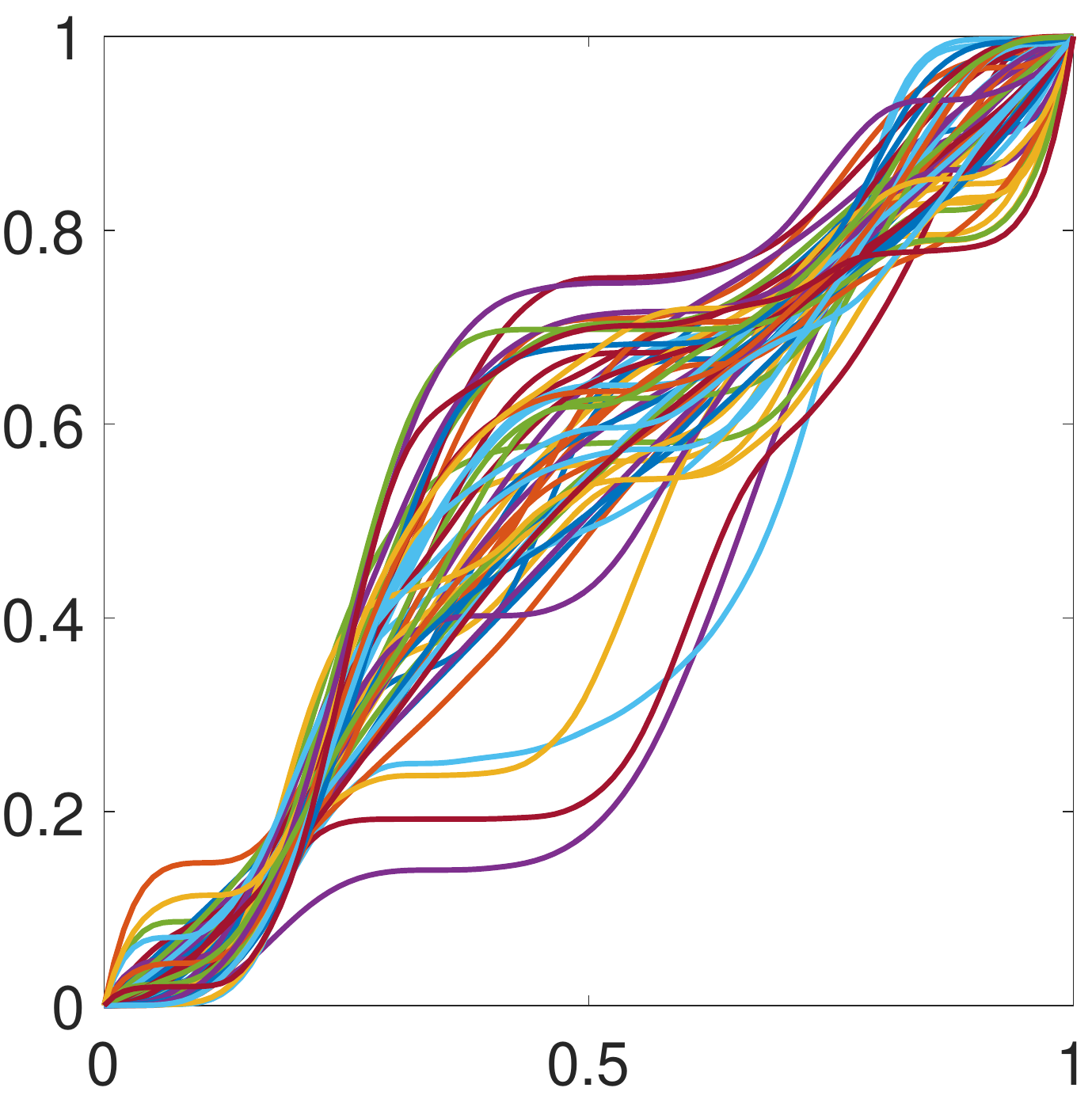}\\
\hline
(3)&\includegraphics[width=.9in]{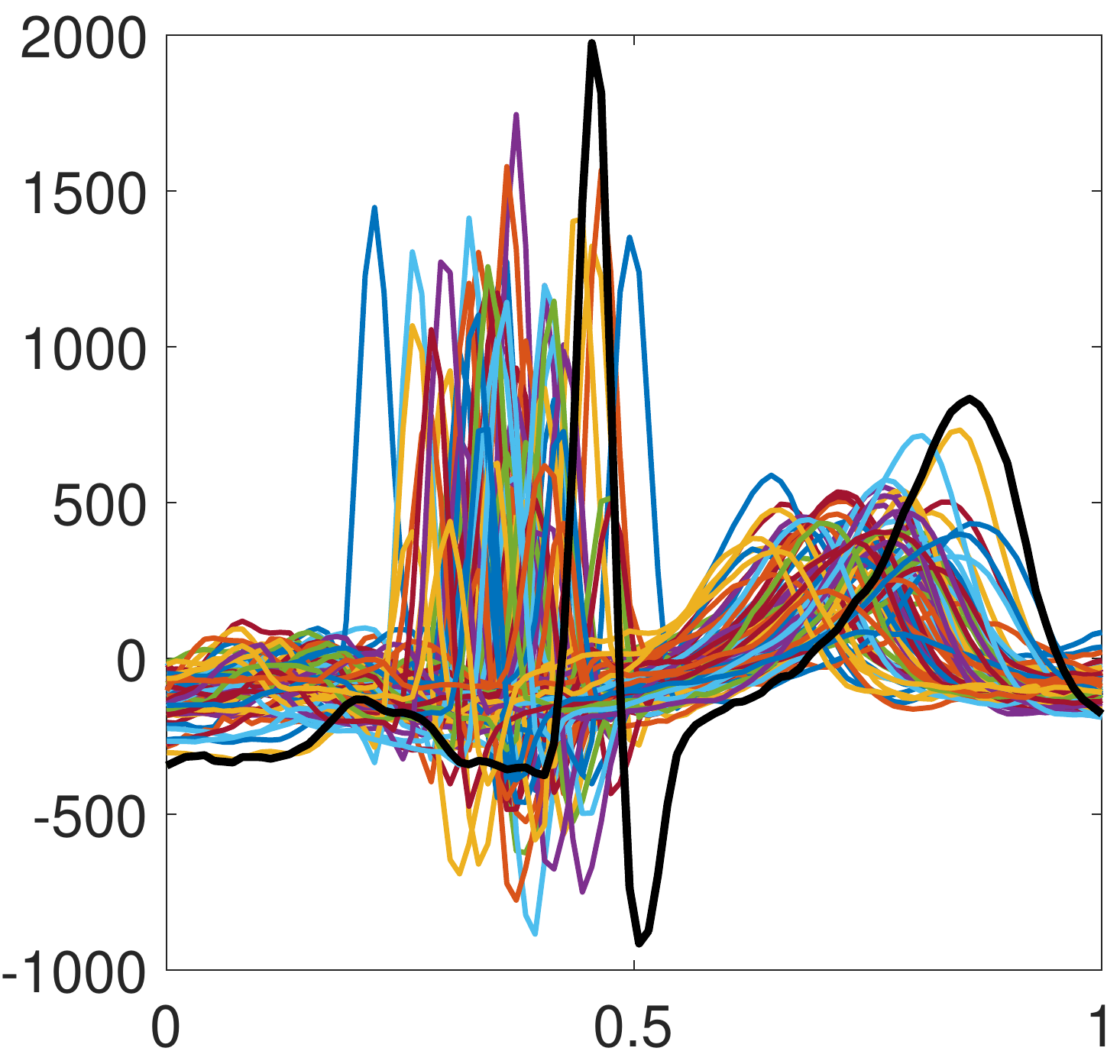}&\includegraphics[width=.9in]{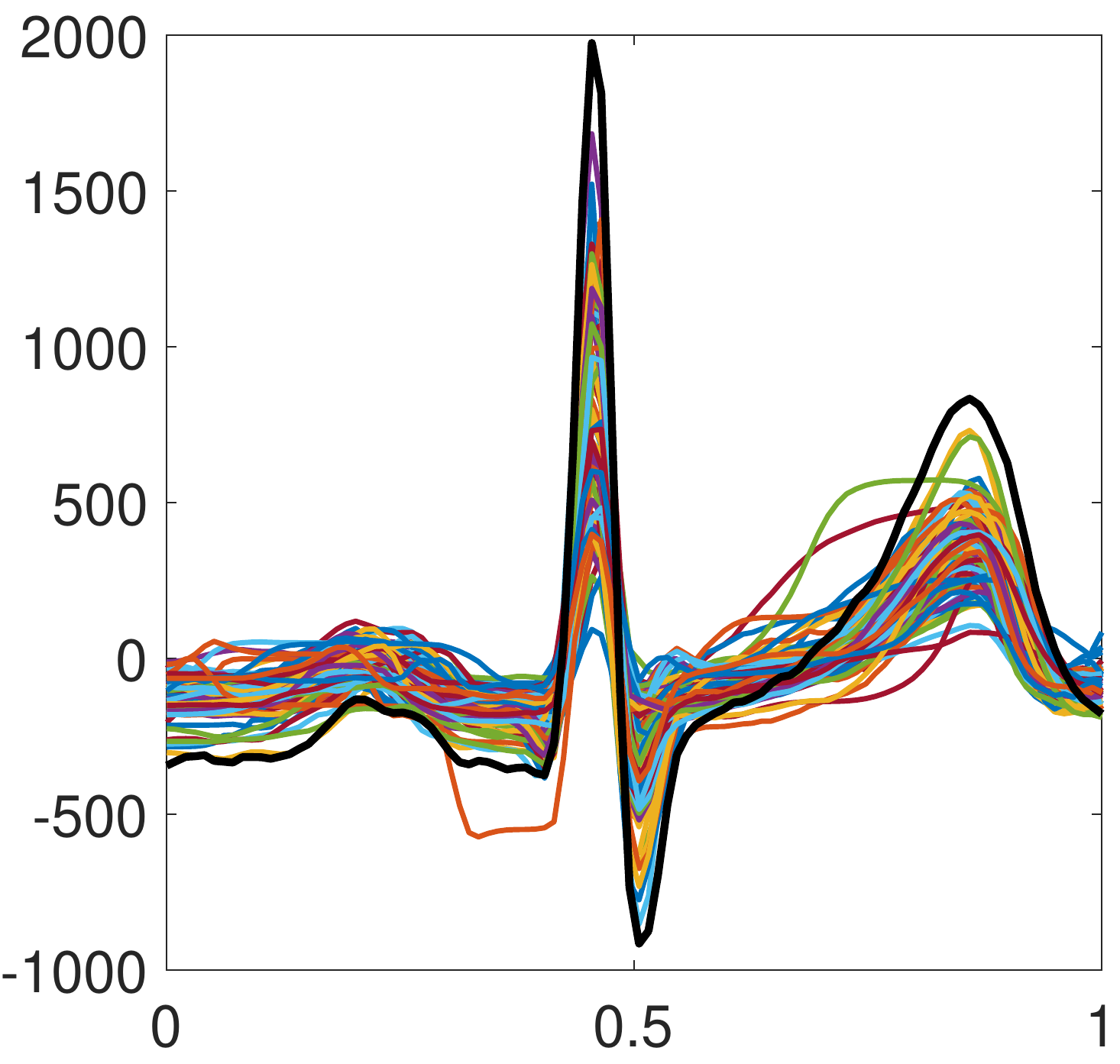}&\includegraphics[width=.9in]{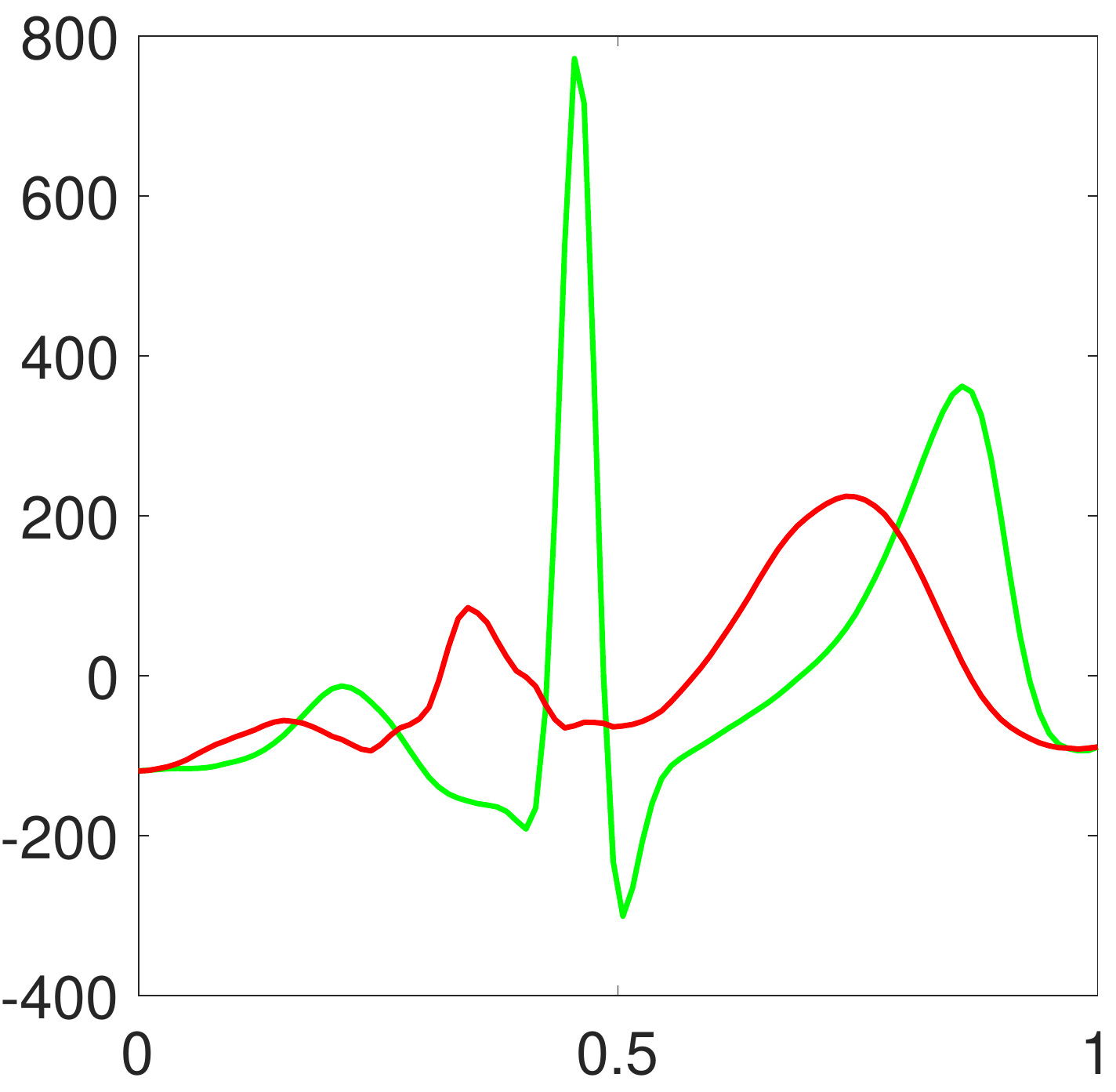}&\includegraphics[width=.9in]{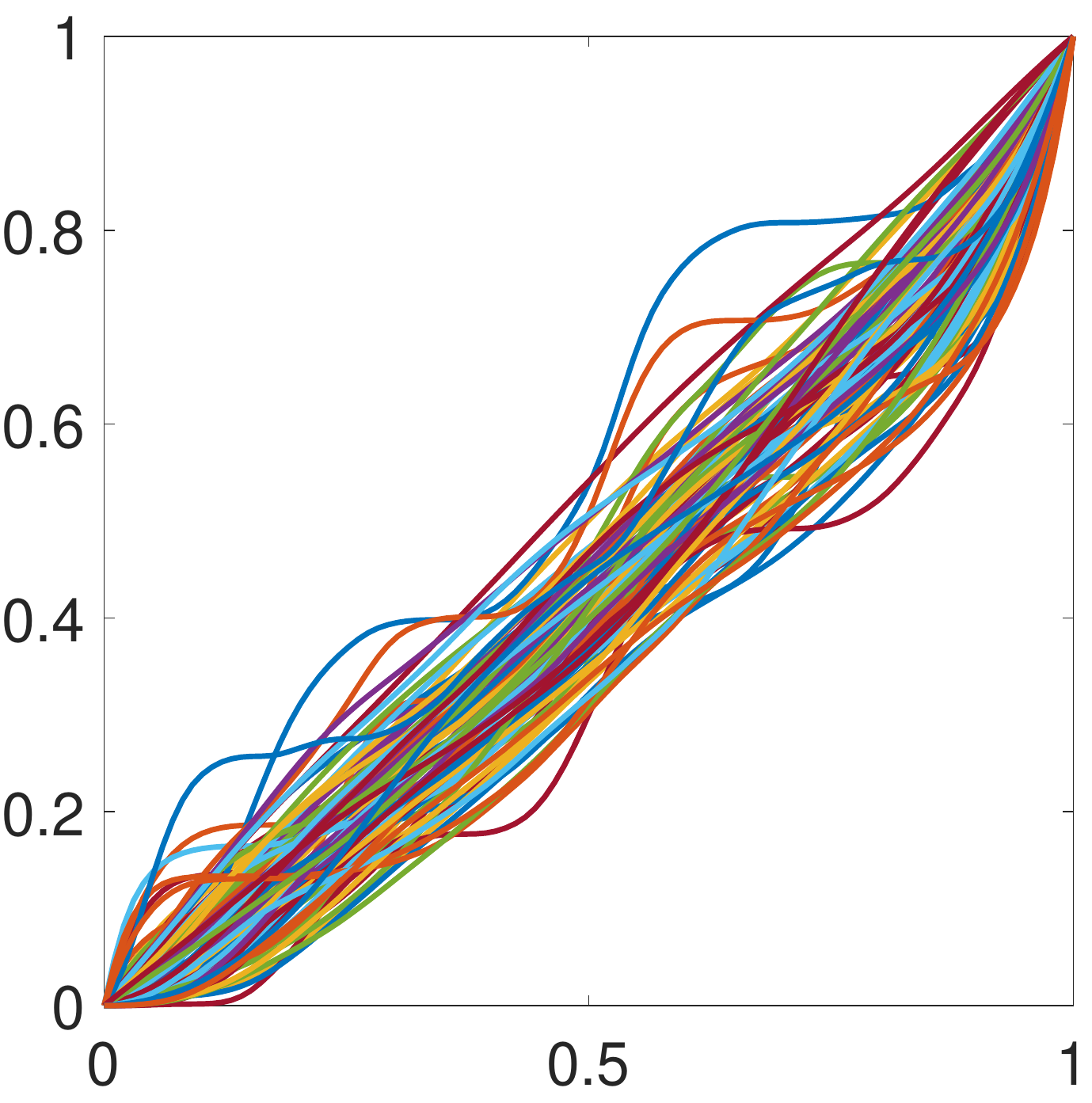}\\
\hline
(4)&\includegraphics[width=.9in]{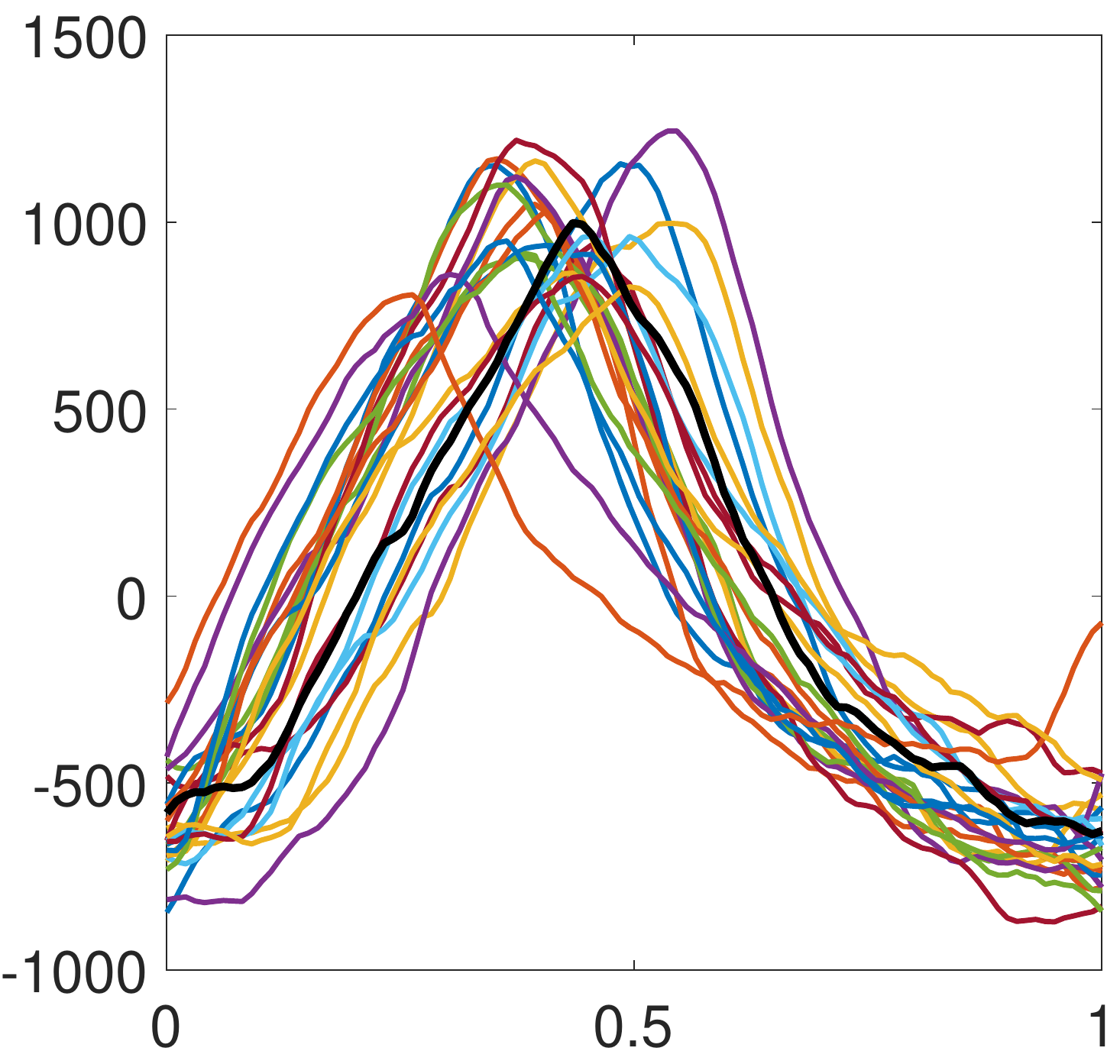}&\includegraphics[width=.9in]{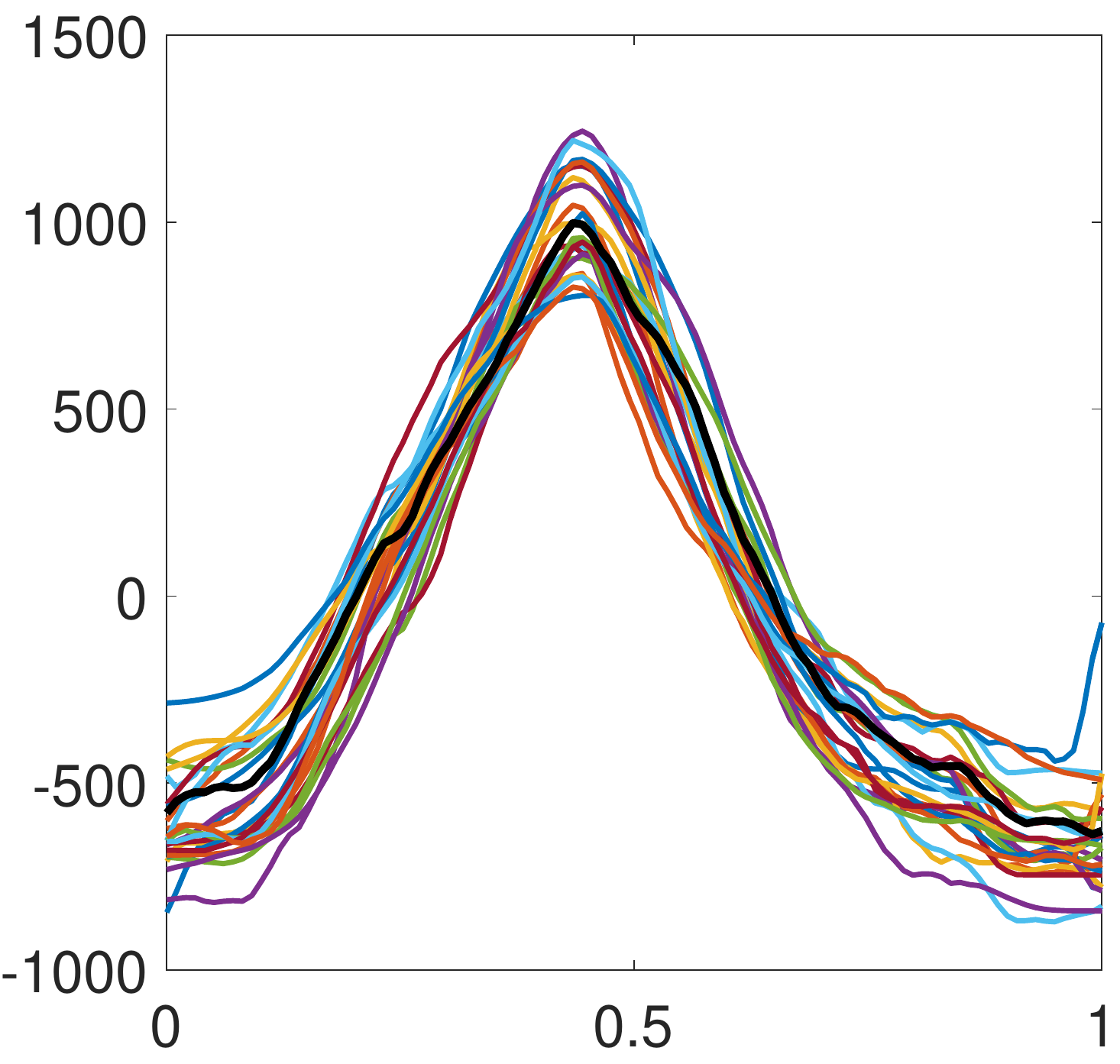}&\includegraphics[width=.9in]{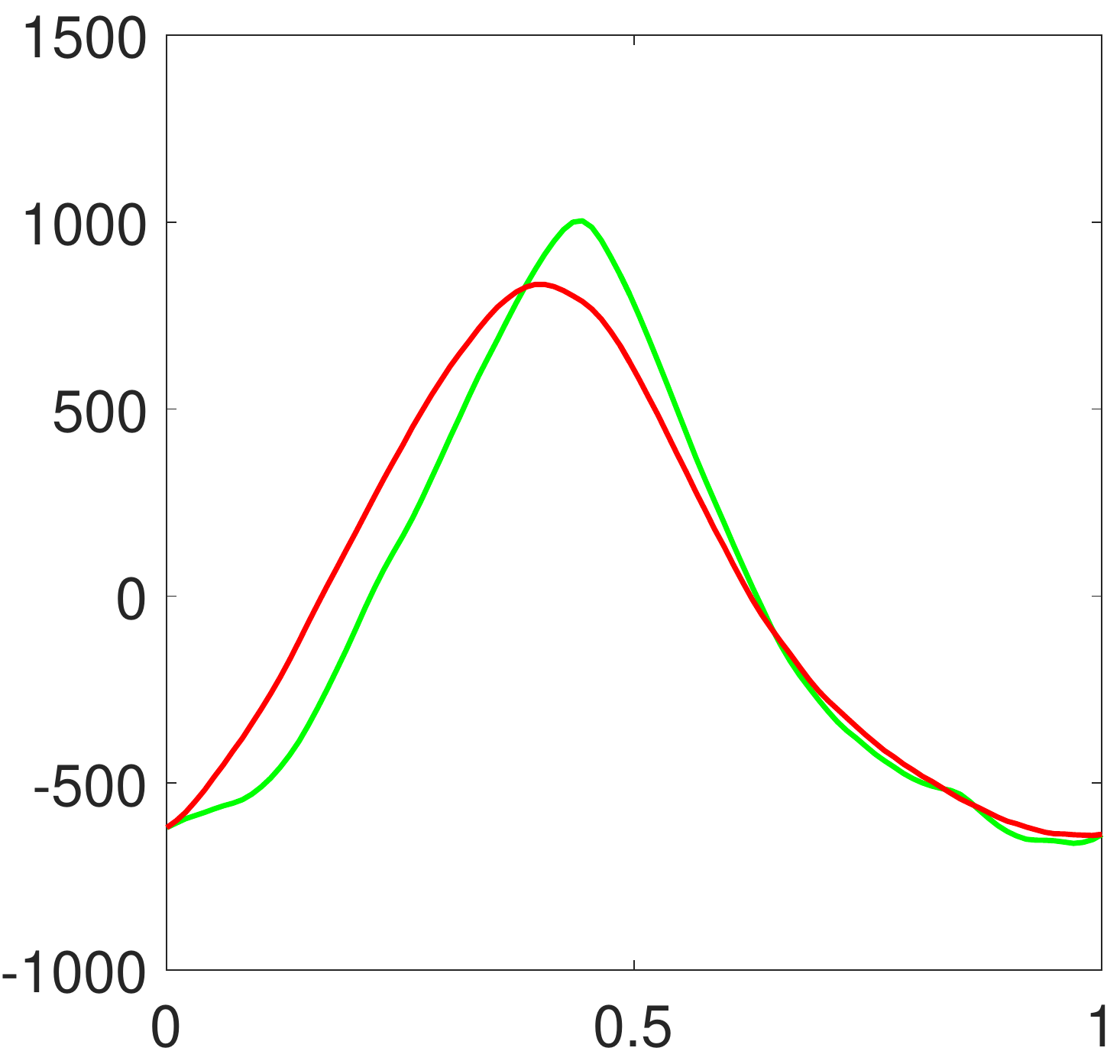}&\includegraphics[width=.9in]{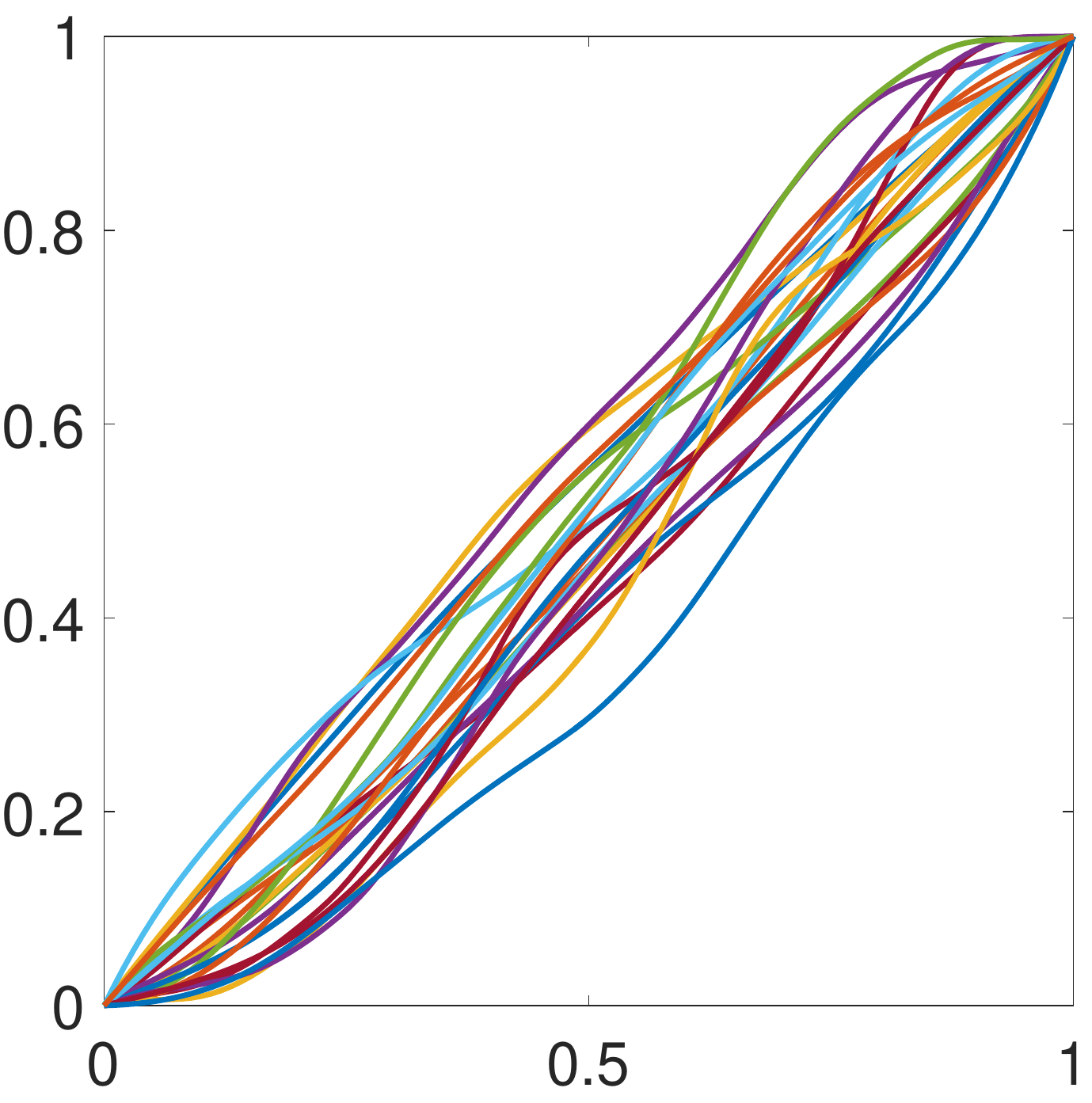}\\
\hline
\end{tabular}
\caption{Groupwise MAP alignment to a known template. (a) Original functions with the template in black. (b) Aligned functions ($f\circ\gamma_{MAP}$) with the template in black. (c) Pointwise average before (red) and after alignment (green). (d) Warping functions $\gamma_{MAP}$. (1) Simulated data. (2) Gait pressure functions. (3) PQRST complexes. (4) Respiration functions.} \label{fig:exma1}
\end{center}
\end{figure}

\begin{figure}[!t]
\begin{center}
\begin{tabular}{|c|c|c|c|c|}
\hline
&(a)&(b)&(c)&(d)\\
\hline
(5)&\includegraphics[width=.9in]{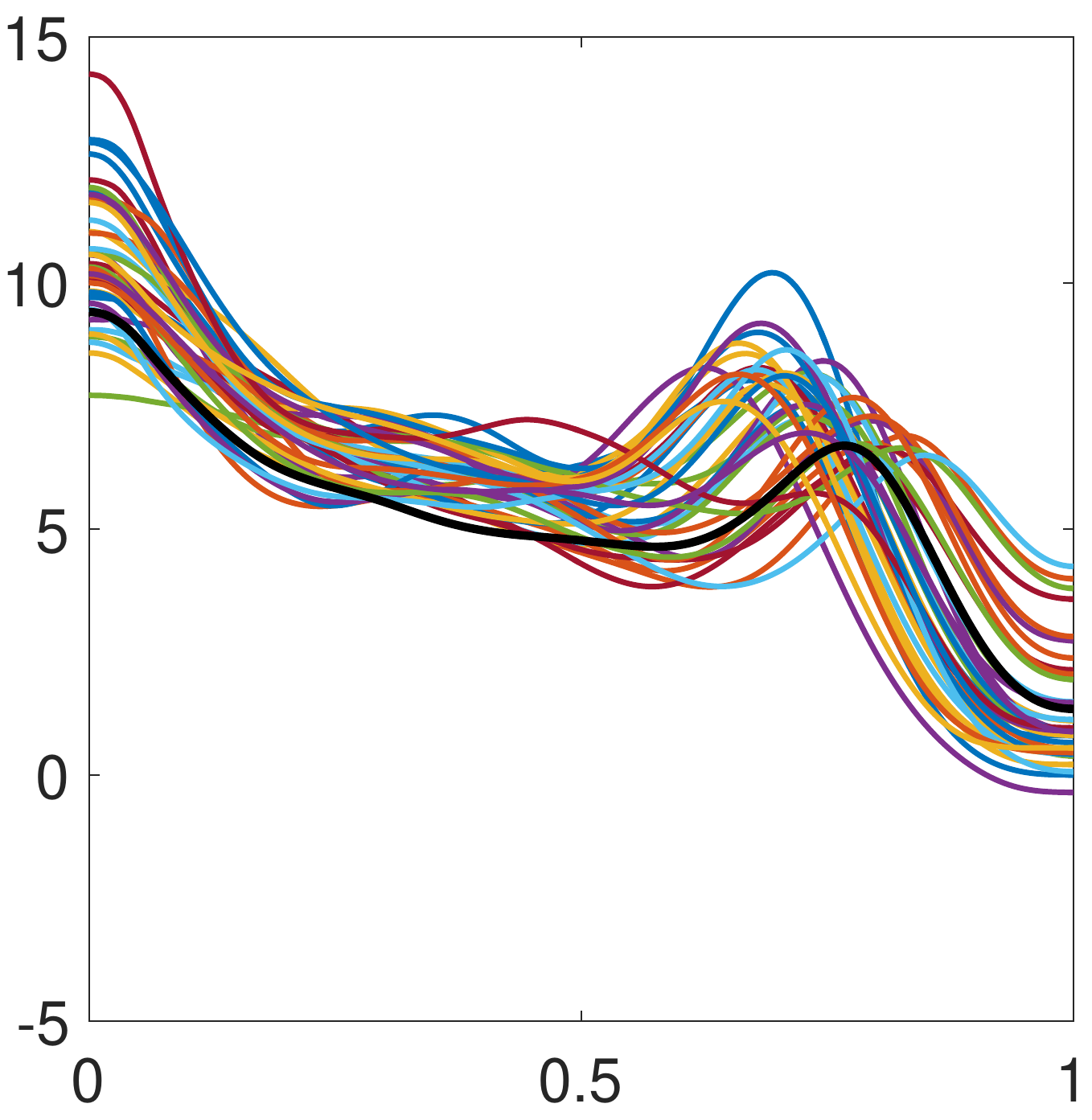}&\includegraphics[width=.9in]{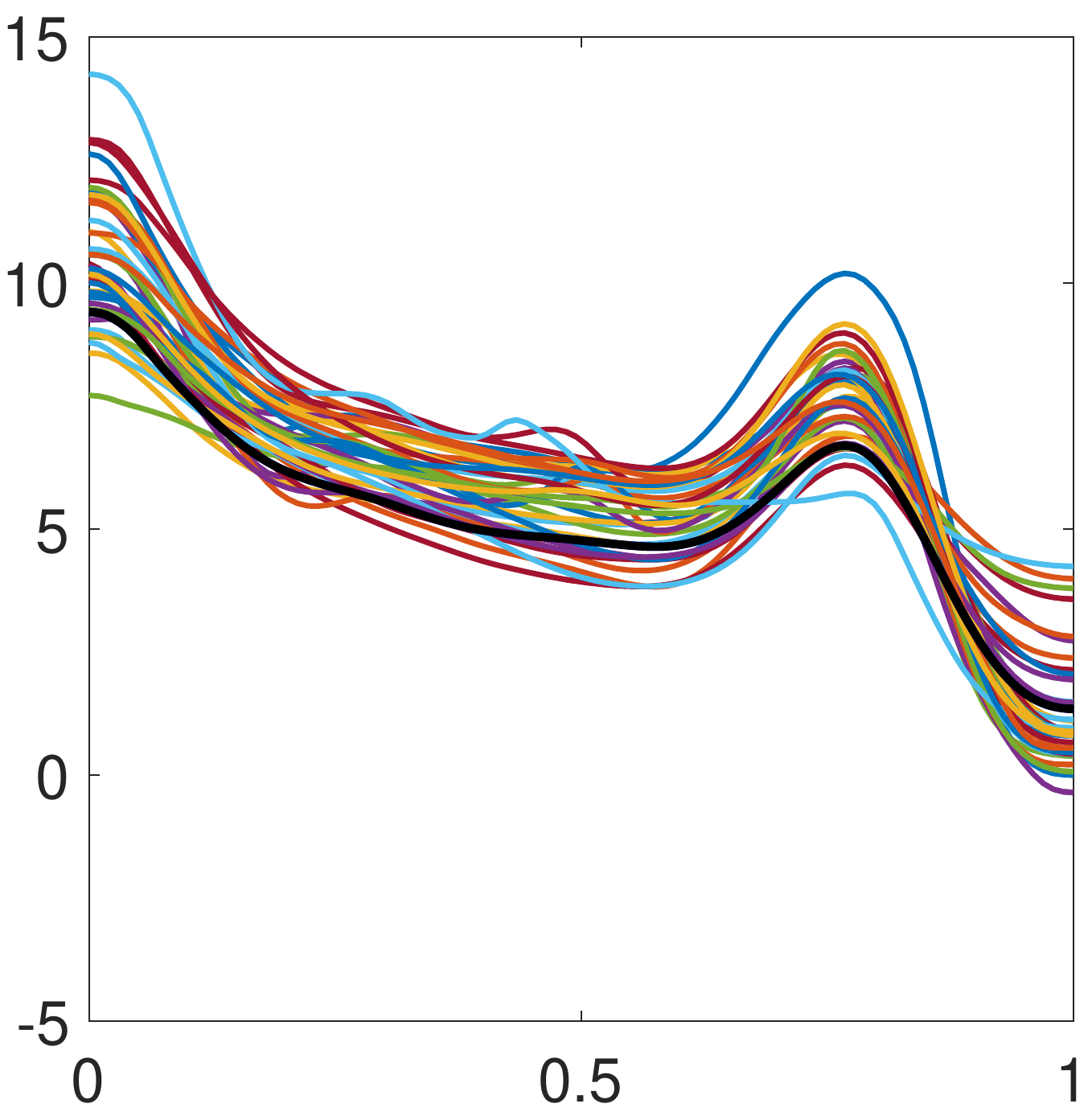}&\includegraphics[width=.9in]{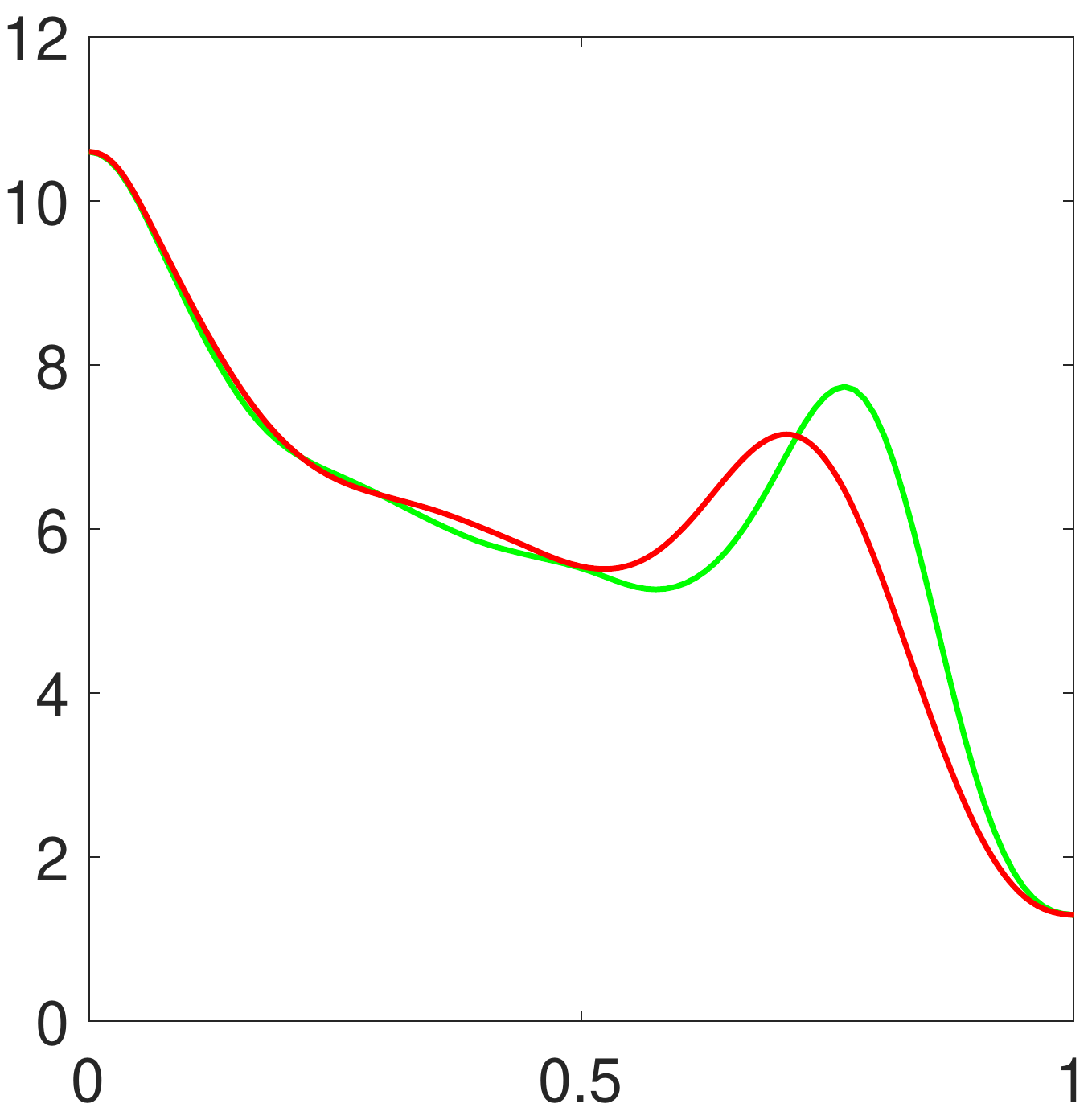}&\includegraphics[width=.9in]{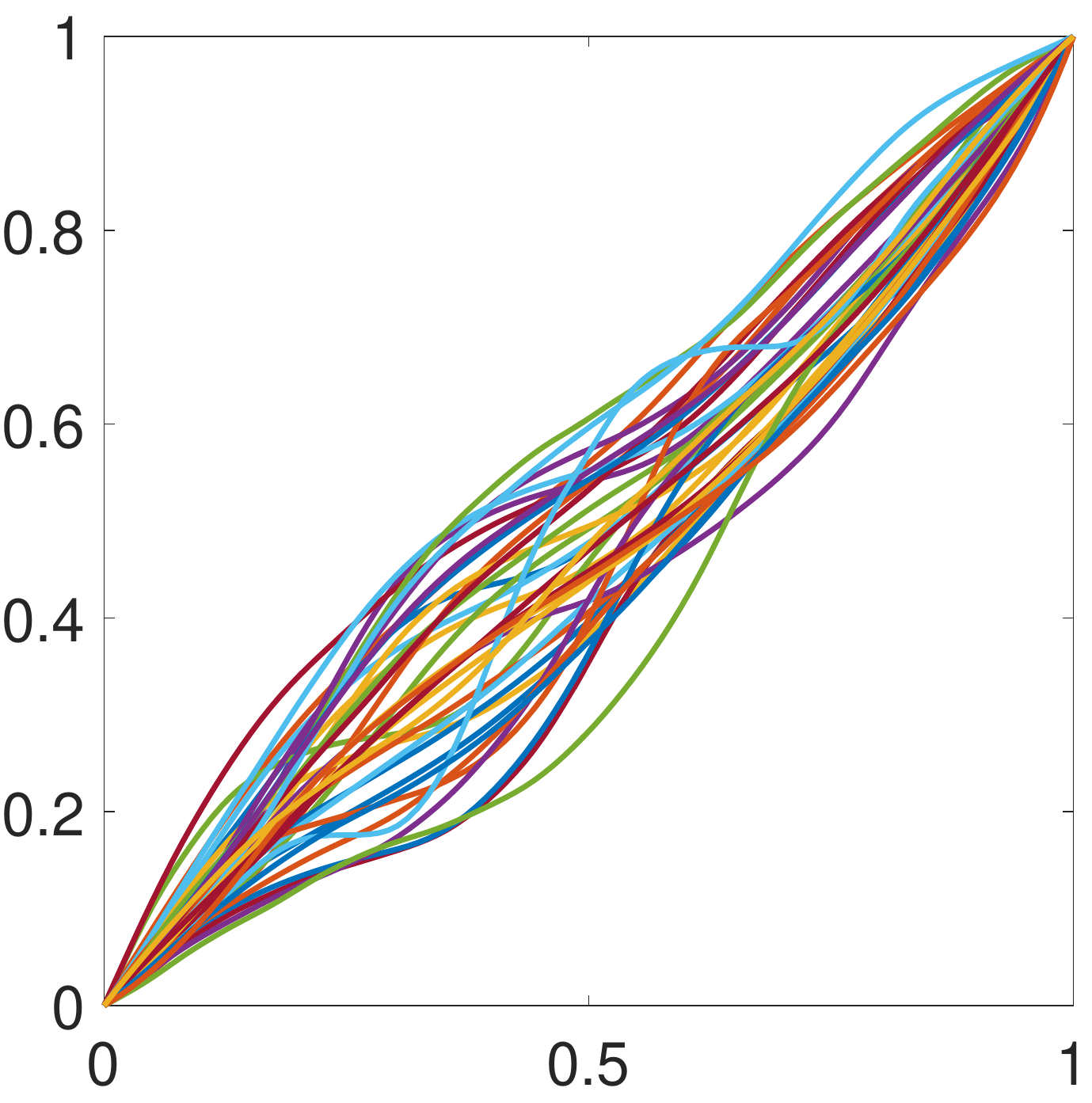}\\
\hline
(6)&\includegraphics[width=.9in]{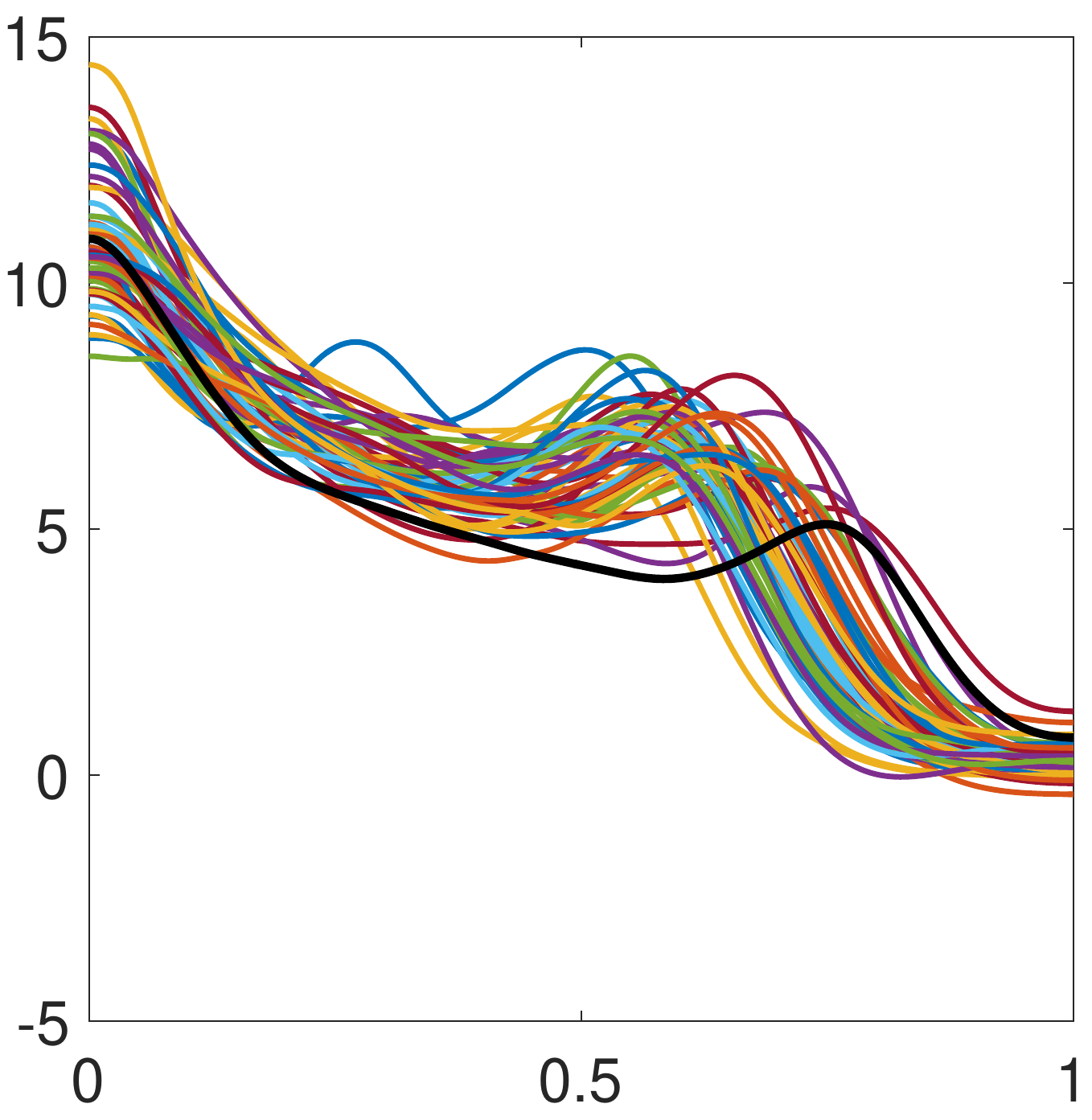}&\includegraphics[width=.9in]{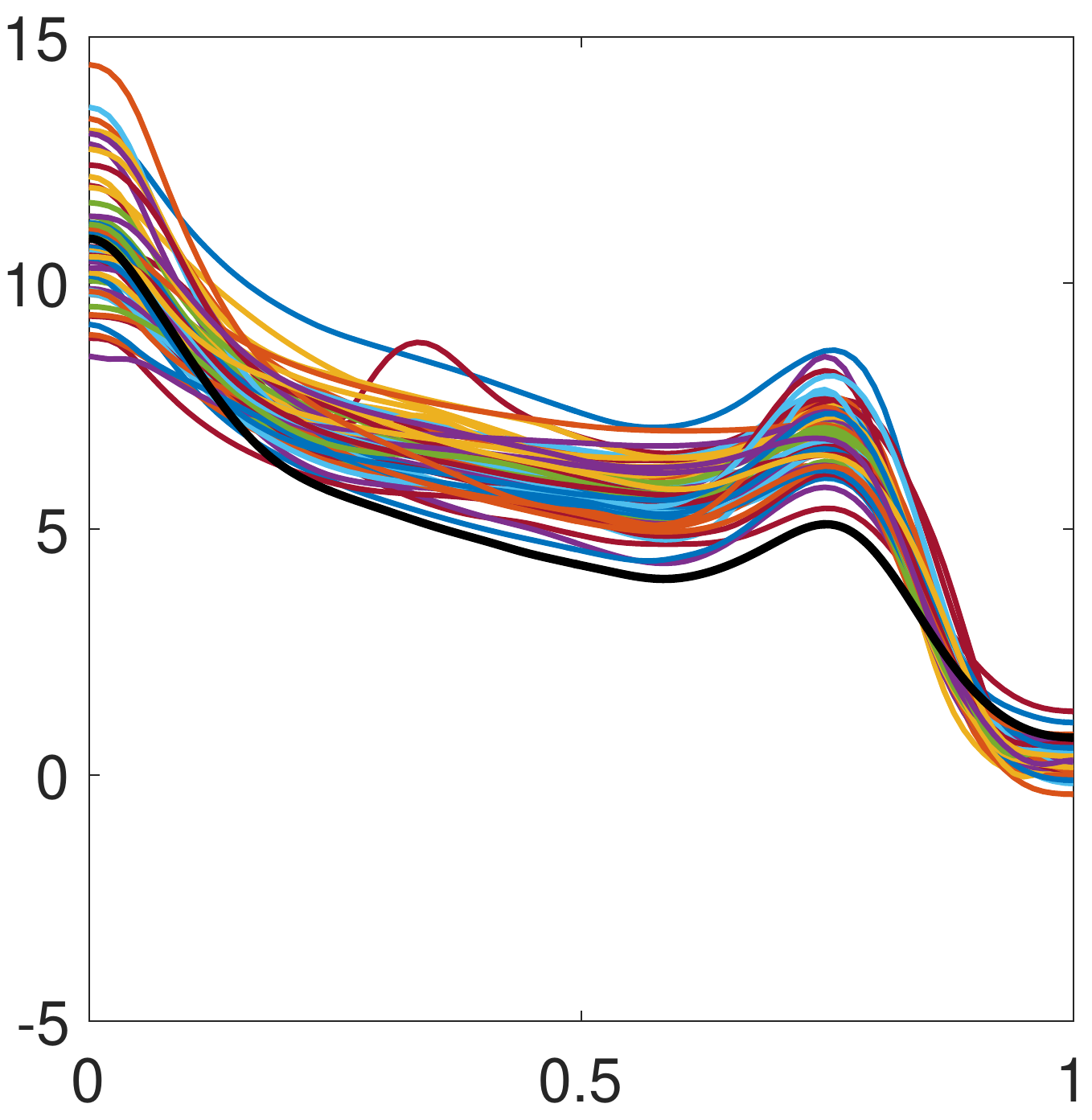}&\includegraphics[width=.9in]{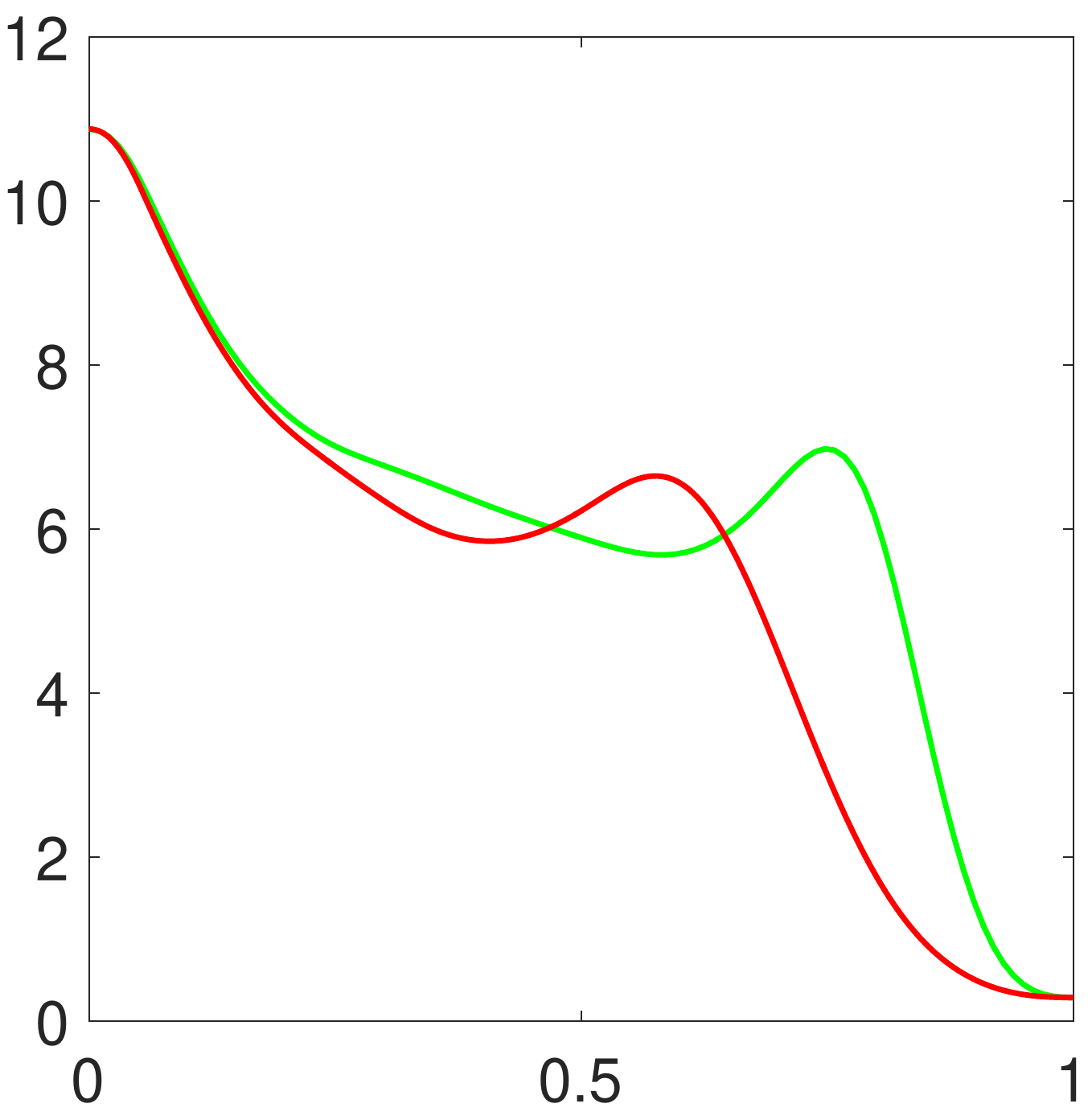}&\includegraphics[width=.9in]{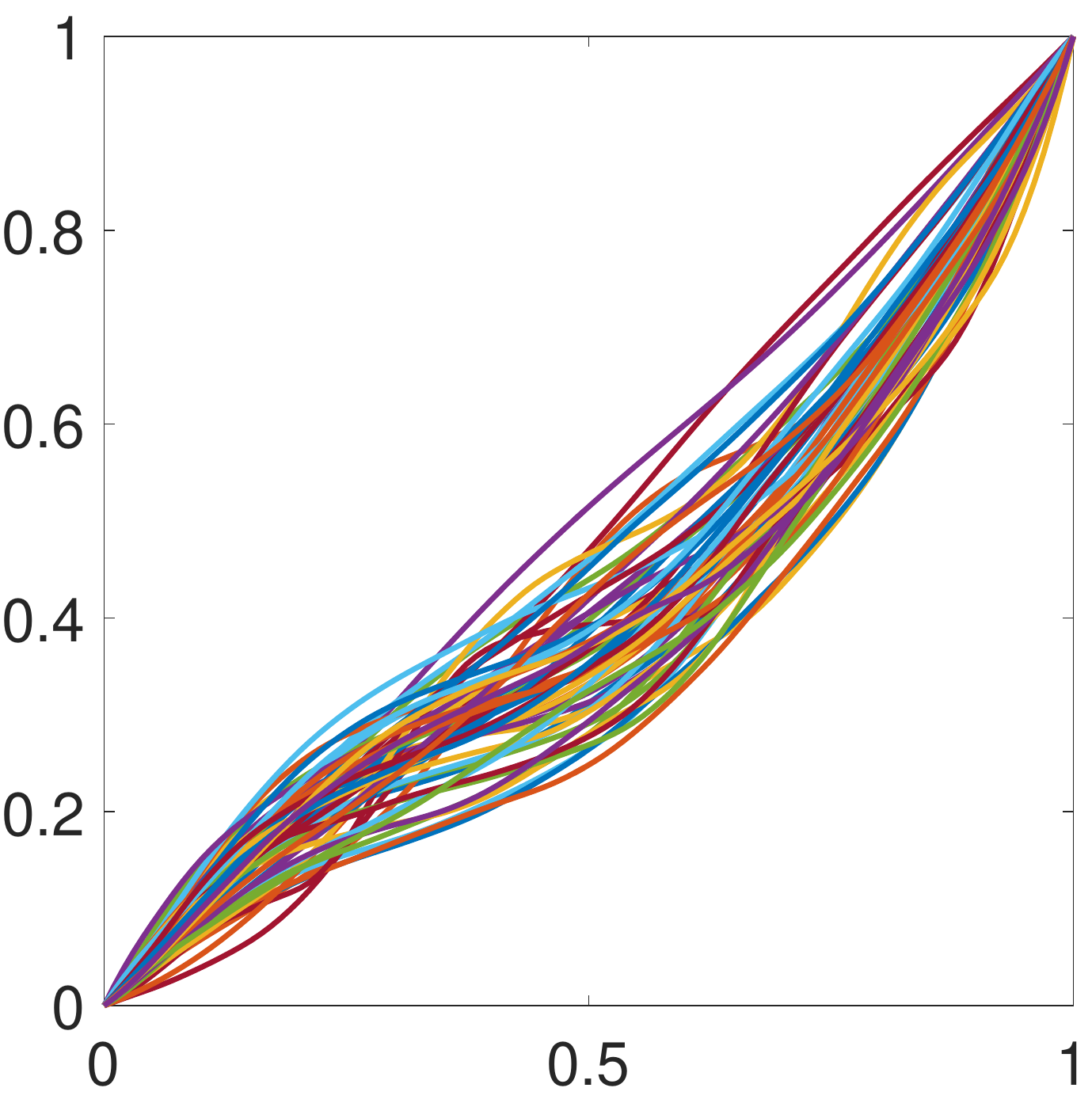}\\
\hline
(7)&\includegraphics[width=.9in]{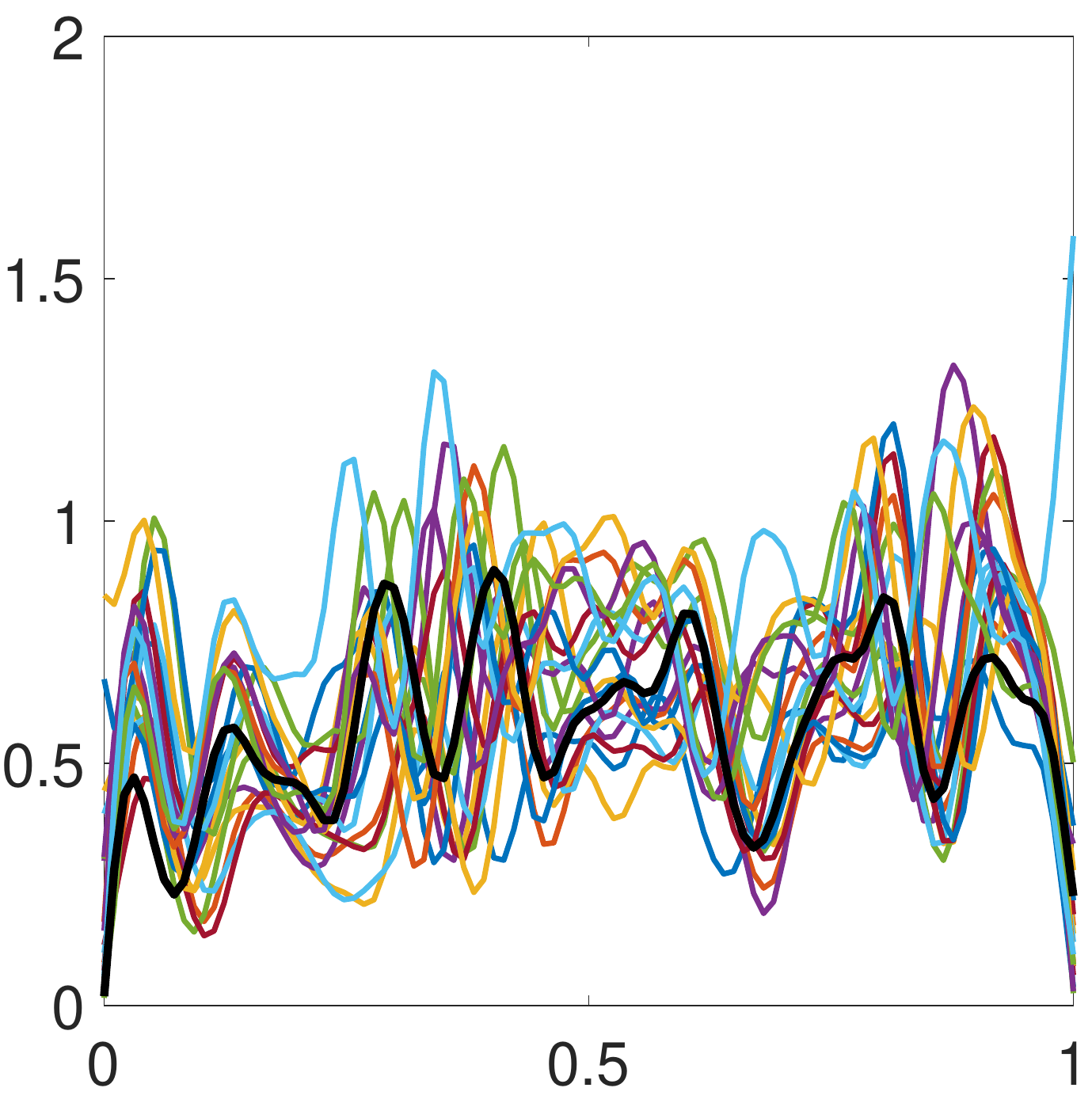}&\includegraphics[width=.9in]{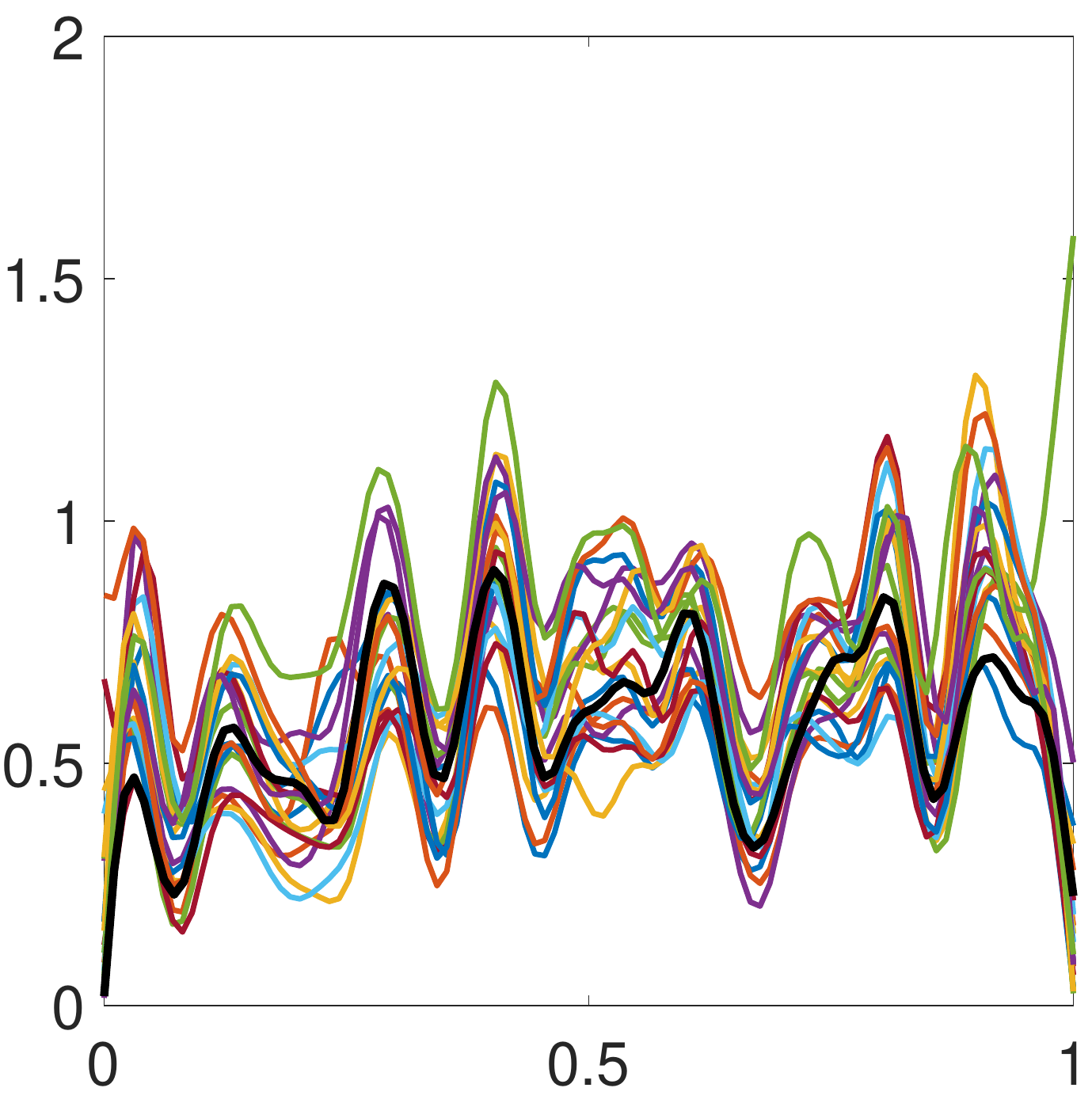}&\includegraphics[width=.9in]{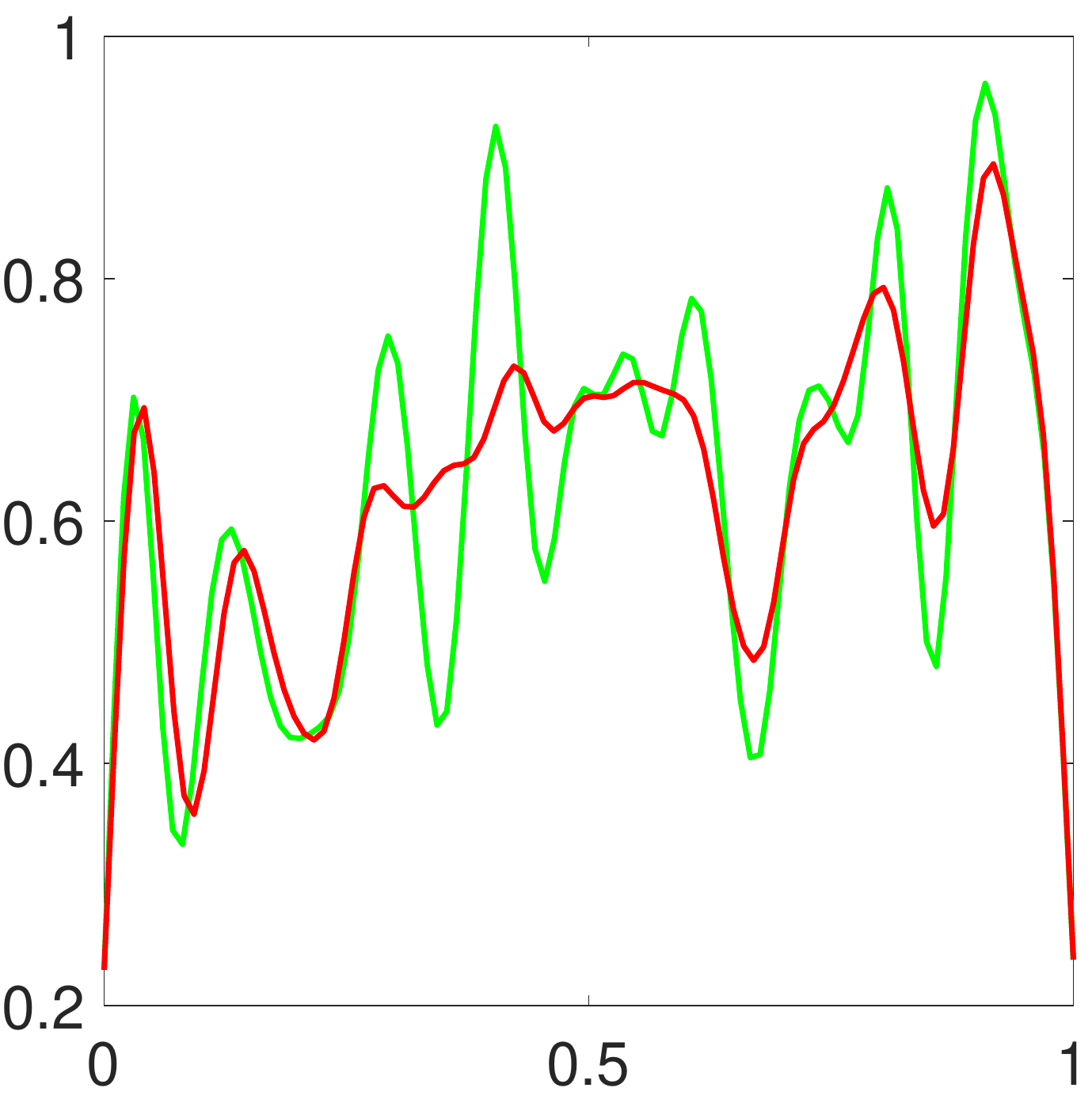}&\includegraphics[width=.9in]{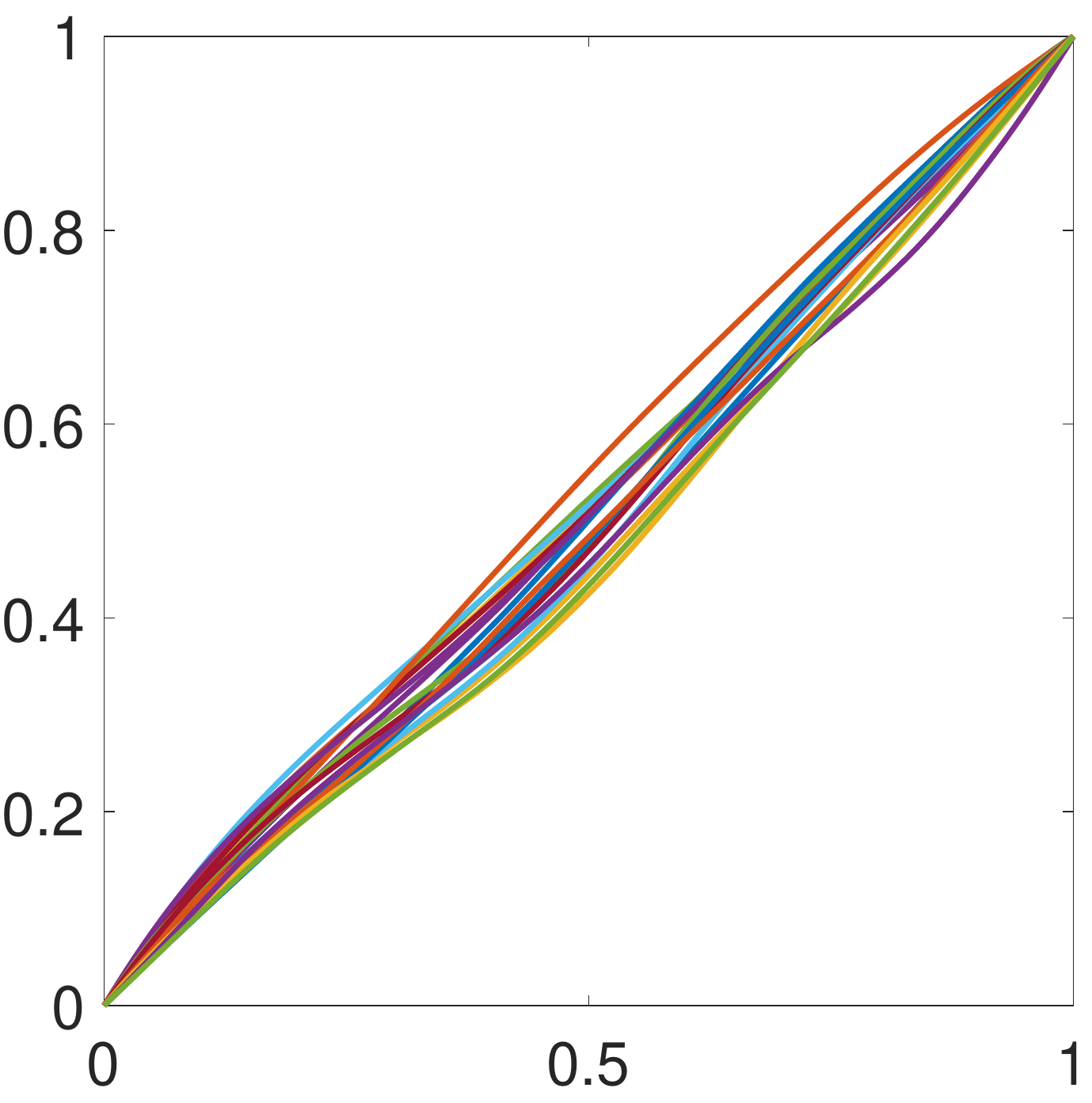}\\
\hline
\end{tabular}
\caption{Groupwise MAP alignment to a known template. (a)-(d) Same as in Figure \ref{fig:exma1}. (5) BGD for boys. (6) BGD for girls. (7) Signature acceleration data.} \label{fig:exma2}
\end{center}
\end{figure}

\subsection{Groupwise MAP Alignment to a Known Template}

We close the applications section with several examples of groupwise function alignment to a known template. For each of the datasets described above (and an additional simulated dataset), we randomly select one of the functions in the data as a template and align all functions in a pairwise manner to this template. In these examples, we do not account for multimodality in the posterior distribution and use the MAP warping ($\gamma_{MAP}$) for alignment. The results are presented in Figures \ref{fig:exma1} and \ref{fig:exma2}. For each example, we display the full original dataset with the template highlighted in black in panel (a). We show the aligned data in panel (b), the pointwise function averages before (red) and after (green) alignment in panel (c), and the estimated warping functions in panel (d). In all examples, we see a drastic improvement in function alignment using the proposed method, which directly translates to better data summaries such as the pointwise function averages. As a specific example, consider the PQRST complexes in row (3). The MAP alignment is able to correctly match the P waves, QRS complexes and T waves across all of the given data. This results in an accurate representation of the pointwise average, which shares all of the features present in the original data. On the other hand, the QRS complex is highly distorted in the unaligned pointwise average. Similar results are observed in the other examples with highest improvement for the gait pressure data and the signature acceleration data.

\section{Summary and Future Work}
\label{sec:conc}

We have presented a Bayesian model for pairwise registration of functional data. This model utilizes a convenient geometric representation of warping functions called the square-root density, which allows for efficient sampling from the posterior distribution via importance sampling. A main advantage of the proposed approach over previous optimization-based approaches is that it is possible to discover multiple plausible registrations, which are given by different modes in the posterior distribution. We present several simulated and real data examples that highlight these advantages. We use simulations to compare the results obtained using the proposed model to those obtained using a similar model with a Dirichlet process prior on the warping functions (which does not exploit the geometry of the space of warping functions).

There are multiple directions for future work. First, we will extend these methods to a groupwise registration model where the template function and the warping functions are estimated jointly. Second, we will extend these methods to a setting where soft landmark information is provided on the functions of interest. In such a case, one can incorporate this information into the prior distribution of the Bayesian model. Third, we will consider a more general problem of curve alignment for shape analysis where the curves are functions from a unit interval (open curves) or unit circle (closed curves) to $\real^n,\ n>1$. Shapes of objects are invariant to translation, scale, rotation and re-parameterization, and thus, the prior distributions in our Bayesian model will be defined on product spaces, whose geometric structure will play an important role. Finally, a major question relates to propagating the registration uncertainty to subsequent statistical inference problems. One example is template estimation in the presence of multiple plausible warping solutions.

\section*{Acknowledgements}

The author would like to thank the editor, associate editor and reviewers for providing valuable feedback. He would also like to thank Anuj Srivastava and Karthik Bharath for useful suggestions and comments. This research was partially supported by NSF DMS 1613054.

\bibliographystyle{imsart-number}
\bibliography{bibfile}

\end{document}